\RequirePackage{fix-cm}
\documentclass[a4paper]{article}                     
\usepackage{graphicx}
\usepackage{color}
\usepackage{amsmath}
\usepackage{amsfonts}
\usepackage[normal]{subfigure}
\usepackage{a4wide}
\usepackage{natbib}

\renewcommand\harvardyearright[1]{.} 
\newcommand{\Ub}{{\bf U}}
\newcommand{\vb}{{\bf v}}
\newcommand{\x}{{\bf x}}

\newcommand{\ie}{{\it i.e.}}
\newcommand{\eg}{{\it e.g.}}
\begin{document}

\title{Kinetic models with non-local sensing determining cell polarization and speed according to independent cues}


\author{Nadia Loy \and Luigi Preziosi}


\author{Nadia Loy \thanks{Department of Mathematical Sciences ``G. L. Lagrange'', Politecnico di Torino, Corso Duca degli Abruzzi 24, 10129 Torino, Italy, and Department of Mathematics ``G. Peano'', Via Carlo Alberto 10 ,10123 Torino, Italy
                (\texttt{nadia.loy@polito.it})}\and
        Luigi Preziosi \thanks{Department of Mathematical Sciences ``G. L. Lagrange'', Dipartimento di Eccellenza 2018-2022, Politecnico di Torino, Corso
                Duca degli Abruzzi 24, 10129 Torino, Italy
                (\texttt{luigi.preziosi@polito.it})}}
                		
\date{}


\maketitle
\begin{abstract}
Cells move by run and tumble, a kind of dynamics in which the cell alternates runs over straight lines and re-orientations. This erratic motion may be influenced by external factors, like chemicals, nutrients, the extra-cellular matrix, in the sense that the cell measures the external field and elaborates the signal eventually adapting its dynamics.\\
We propose a kinetic transport equation implementing a velocity-jump process in which the transition probability takes into account a double bias, which acts, respectively, on the choice of the direction of motion and of the speed. The double bias depends on two different non-local sensing cues coming from the external environment. We analyze how the size of the cell and the way of sensing the environment with respect to the variation of the external fields affect the cell population dynamics by recovering an appropriate macroscopic limit and directly integrating the kinetic transport equation. A comparison between the solutions of the transport equation and of the proper macroscopic limit is also performed.
\end{abstract}


\section{Introduction}\label{intro}

It is well known that a classical migration mode of bacteria consists in alternating runs or swims over straigth lines and tumbles \citep{Berg}. Also eukariotic cells alternate persistent crawlings along a polarization axis with reorientation phases in  which they re-polarize as a result of several external stimuli. These can be represented by nutrients, chemical factors, or noxious substances \citep{Berg_Block_Segall}, and also by environmental cues, such as density, stiffness, and structure of the extra-cellular matrix (ECM). In addition, the presence of other cells along the way can act in a two fold-way, either attractive due to the interaction of adhesion molecules (e.g, cadherin complexes) expressed on the cellular membrane, or repulsive when the region is becoming too overcrowded.

Cells measure all these signals by transmembrane receptors located on their protrusions that can extend up to several cell diameters. The captured chemical or physical signals activate in turn transduction pathways downstream that lead to the cell response. This includes in particular  the polarization of the cell with the formation of a ``head" and a ``tail" and  the activation of adhesion molecules and traction forces leading to motion \citep{Adler}. In the framework of kinetic models, in the present paper we will focus on how the environmental sensing over a finite radius can be translated in the choice of the direction of motion and of the speed of the cell. In fact, in order to do that we develop a kinetic model in which the distribution function depends on cell speed and orientation, respectively a scalar and a unit vector, in addition to the usual dependence on space and time. In the tumbling phase, then, the new orientation and subsequent speed are chosen as a result of a sensing over a finite neighbourhood of the cell, giving the kinetic model a non-local character. 

Focusing on mesoscopic models of cell migration and referring  to \cite{Hillen_Painter.08} for an extensive and more general review on PDE models of chemotaxis, we can observe that the run and tumble dynamics can be modeled by a stochastic process called velocity-jump process that was introduced by \cite{Stroock}. In this work, the author derived a linear transport equation from the velocity-jump process. Such an equation describes the movement of a single-particle distribution function like in the Boltzmann equation \citep{Cercignani}. The main elements of such a process are the tumbling frequency, the mean speed, and the transition probability that describes the probability of choosing a certain velocity after re-orientation. The mean speed, mean runtime (which is the inverse of the frequency), and the tumbling probability may be measured from individual patterns of members of the population. Another advantage of such models is that they admit finite propagation speed. \cite{Alt.80} and \cite{Alt.88} introduced the bias induced by an external stimuli in the linear transport equation. \cite{Dickinson} generalized the model to a transport equation in an anisotropic environment and applied to an environment with a gradient of a stimulus. 

\cite{Hillen.05} proposed a Boltzmann-like model taking into account of the interaction between  cells and ECM, also including the degradation of ECM fibers as a function of the angle between the ECM fiber and the cell velocity. In particular, the author recovered macroscopic limits (diffusive or hyperbolic) according to the structure of the matrix. \cite{Painter.09} used a similar model to describe ECM remodelling resulting from the migration of cells in a heterogeneous environment by integrating numerically the kinetic equation.

An extension of this Boltzmann-like model taking also into account of the interaction between cells and between cells and ECM is proposed  by \cite{Chauviere_Hillen_Preziosi.07} and \cite{Chauviere_Hillen_Preziosi.08}. A further extension is suggested by  \cite{Chauviere_Preziosi} where the operators describing interactions between cells and between cells and ECM include the bias due to an external cue. Surulescu and coworkers \citep{Engwer_Stinner_Surulescu.08, Engwer_Hillen_Surulescu.15, Engwer_Hunt_Surulescu.16, Engwer_Knappitsch_Surulescu.16, Stinner_Surulescu.15} 
instead discussed in more details cell-ECM adhesion phenomena, and, focusing on tumour growth, mainly gliomas, also included proliferation and therapies.

Many of the papers above determine for the proposed mesoscopic models the related macroscopic, usually diffusive, limits and implement them numerically. The diffusive limit of transport equations is deeply described by \cite{Othmer_Hillen.00}.
More in detail,  \cite{Othmer_Hillen.02} investigated various forms and orders of the bias which may be included in the transition probability. However, in some regimes a hyperbolic limit  is more suitable (for example in presence of formation of networks, as studied by \cite{Chalub_Markowich_Perthame_Schmeiser.04}, \cite{Hwang_Stevens.05}, \cite{Filbet_Perthame}).

Coming to the sensing strategies and then to the determination of cell re-orientation and speed, they have been considered in several models. In position jump processes, the transition probabilities model the strategy of sensing and may typically be (i) local, if the information at the present position is considered, (ii) neighbour based, if the information at the target jump site is taken into account, (iii) gradient based, if the local difference in information  
between the target and the local site is considered, (iv) local average, if the average of the information
between the present and target site as done by \cite{Othmer_Stevens.97} and by \cite{Painter_Sherratt.03}. \\
In general, a way of including cell sensing
 is to consider a non-local average of the external fields.  For what concerns bacteria and cells dynamics,  \cite{Othmer_Hillen.02} and \cite{Hillen_Painter_Schmeiser.06} introduced a finite sampling radius and they defined a non-local gradient as the average of the external field on a surface which represents the membrane of the cell.  In particular, the Authors derived a macroscopic diffusive model with a non-local gradient both from a position jump process and from a velocity jump process by postulating that the non-local sensing is a bias of higher order. 

The notion of non-local sensing was also used for cell adhesion and haptotaxis by \cite{Armstrong_Painter_Sherratt.06} yielding a macroscopic integro-differential equation. This model was recently derived from a space jump process by \cite{Butt}.  Similar equations were proposed in 2D set-ups by \cite{Col_Sci_Prez.17}, \cite{Col_Sci_Tos.15},  and applied also to model crowd dynamics and traffic flow by \cite{Tosin_Frasca}.

\cite{Eftimie2} and \cite{Eftimie} proposed non-local kinetic models including repulsion, alignement and attraction. They distinguish cells going along the two directions in the one-dimensional set-up with the possibility of switching between the two directions with a constant speed. The one-dimensional kinetic model takes then a discrete structure that is then integrated directly. 

As already stated, in this paper we want to include the sensing strategy in a velocity-jump process and to model the tactic and kinetic response of a cell as a consequence of the non-local sensing of the environment. In particular, we focus on the fact that different fields can influence respectively cell direction and speed. Because of these characteristics we name the model as a double-bias non-local kinetic model. To be specific, a chemical factor may influence the orientation of a cell (taxis), while the cell population density or the matrix density may influence the speed of the cell once the direction is chosen. For instance, a volume filling effect can hamper the motion in a certain direction due to cell overcrowding. Similarly, fiber networks with too narrow pores can hamper or even completely inhibit migration, a phenomenon called physical limit of migration by \cite{Boekhorst} and \cite{Wolf_Friedl.13}. 

Therefore, cell response is related to the choice of the transition probability evaluated in a non-local way over the neighborhood of the cell.  Several examples are given, some in order to show the link with existing models \citep{Butt, Hillen_Painter_Schmeiser.06,  Painter_Sherratt.2010, Painter.Gerish.15, Painter_Hillen.02}, others to point out the novelties of the proposed model. In particular, from the numerical point of view, we simulate directly the kinetic equation taking into account of different phenomena, $\eg$, chemotaxis,  ECM steryc hindrance, adhesion induced aggregation, durotaxis. 

We also discuss how the choices made on the transition probability and on the size of the sampling volume determine the type of  macroscopic limit that can be performed. Then, we compare the outcome of the numerical integration of the transport equation with the proper macroscopic limits and the related analytical solution that can be obtained for the stationary case. This  shows a satisfactory agreement when the hypothesis necessary to perform the limits are satisfied. 

The plan of the article is as follows. After a section introducing the general and preliminary concepts, Section 3 introduces the general structure of the double-bias non-local kinetic model. Then, in Section 4 the macroscopic limits are discussed, distinguishing when it is possible to perform a parabolic scaling or a hyperbolic scaling according to the modelling assumptions. In Section 5 we start with a simpler single-bias model in which there is no cue affecting the speed, that is determined through a given distribution function. Several sensing strategies determining cell orientation are discussed, starting from local responses to non-local averages and strategies comparing the level of chemicals in different points of the neighbourhood, also including the possible dependence of the encounter rate from taxis factors. In Section 6 the double-bias model is discussed. Again, for clarity we first focus on the simpler case in which there is no cue influencing cell orientation, so that it is randomly chosen. We finally give an example for the complete model, already presented in its general form in Section 3.

\section{Preliminaries}

The statistical description of the cell population is performed through the distribution density $p = p(t,\x, \vb_p)$ which is parametrized by the time $t>0$, the position $\x \in \Omega \subseteq \mathbb{R}^d$ and the velocity pair $\vb_p = (v,\hat{\vb}) \in V_p=[0,U]\times\mathbb{S}^{d-1}$, being  $\mathbb{S}^{d-1}$ the unit sphere boundary in $\mathbb{R}^d$ and $U$ the maximal cell speed, so that $\hat{\vb}$ is the unit vector and $v$ is the speed, $\ie$, the modulus of the velocity vector $\vb=v \hat{\vb}$. The choice of representing the distribution function $p$ depending on the direction and on the speed of the velocity, instead of the velocity vector $\vb$, lies in the need of separating the mechanisms governing cell polarization and motility, for instance in response of chemotaxis and in presence of other influencing factors \citep{Alt.80}.
The mesoscopic model consists in the transport equation for the cell distribution:
\begin{equation}\label{transport.general}
\dfrac{\partial p}{\partial t}(t,\x,\vb_p) + \vb\cdot \nabla p(t,\x,\vb_p) =  \mathcal{J} [p] (t,\x,\vb_p)
\end{equation}
where the operator $\nabla$ denotes the spatial gradient.
The term $\mathcal{J}[p](t,\x,\vb_p)$, named {\it turning operator},
is an  integral operator that describes the change in velocity which is not due to the free-particle transport. It may describe the classical run and tumble behaviors, contact guidance phenomena, or cell-cell interactions. For the moment we will consider the classical run and tumble, $\eg$, random re-orientations, which, however, may be biased  by external cues. Therefore, our turning operator will be the implementation of a velocity-jump process in a kinetic transport equation as introduced by \cite{Alt.80}.

A macroscopic description for the cell population can be classically recovered  through the definition of moments of the distribution function $p$:

\noindent - the cell number density
\begin{equation}\label{def_rho}
\rho(t,\x) = \int_{V_p} p(t,\x,\vb_p) \,d\vb_p\,;
\end{equation}


\noindent - the cell mean velocity
\begin{equation}\label{mean.U}
\Ub(t,\x) = \dfrac{1}{\rho(t,\x)}\int_{V_p} p(t,\x,\vb_p)
\,\vb\, d\vb_p\,;
\end{equation}

\noindent - the cell variance-covariance matrix 
\begin{equation}\label{mean.PP}
\mathbb{P}(t,\x) = \int_{V_p} (\vb -\Ub) \otimes (\vb-\Ub) p(t,\x,\vb_p)
\, d\vb_p\,;
\end{equation}

\noindent - the cell speed variance 
\begin{equation}\label{mean.E}
E(t,\x) = \int_{V_p} \frac{|\vb -\Ub|^2}{2}p(t,\x,\vb_p)
\, d\vb_p\,.
\end{equation}
We remark that, because of the definition of $V_p$, the integrals over the velocity space are
\begin{equation}
\int_{V_p} f(v,\hat{\vb}) \, d\vb_p= \int_{0}^{U} \int_{\mathbb{S}^{d-1}} f(v,\hat{\vb})  \, d\hat{\vb} \, dv.
\end{equation}
In particular, if the dependence of $f$ on $v$ and $\hat{\vb}$ can be factorized, $\ie \  f(v,\hat{\vb})=f^1(v) f^2(\hat{\vb})$, we have that
\begin{equation}
\int_{V_p} f(v,\hat{\vb}) \, d\vb_p= \int_{0}^{U} f^1(v) \, dv \int_{\mathbb{S}^{d-1}} f^2(\hat{\vb}) \, d\hat{\vb}.
\end{equation}
The integral over $\mathbb{S}^{d-1}$, which we remind to be the boundary of the unit sphere, is not a surface integral. It has to be interpreted as
$$\int_{\mathbb{S}^{d-1}} f^2(\hat{\vb}) \, d\hat{\vb}=\int_{0}^{2\pi} f^2(\cos (\theta), \sin (\theta)) d\theta$$ in 2D and similarly in 3D.

Computing the moments of the transport equation \eqref{transport.general} allows to obtain evolution equations for the moments above.
Macroscopic limits can then be achieved by classical procedures. In particular, the diffusive limit of transport equations with velocity jump processes is deeply treated by \cite{Othmer_Hillen.00}, \cite{Othmer_Hillen.02}, and  \cite{Hillen.05}, where the hyperbolic scaling is treated as well. 
The interest here is focused on when, according to the hypothesis done on the transition probability and on the sensing radius, it is possible to perform either a parabolic or a hyperbolic scaling, leading to the related diffusive and hyperbolic limits.
  
\section{The turning operator}\label{sec2}
In taxis processes cells are capable of detecting and measuring external signals  through membrane receptors located along cell protrusions that can extend over a finite radius. Information of both mechanical and chemical origin are then transduced and act as control factors for the dynamics of cells. Therefore, in the turning operator of the kinetic model we will include the evaluation of  mean fields 
 in a neighborhood of the re-orientation position. In particular,  we will consider two different control factors $\mathcal{S}$ and $\mathcal{S}'$ which are positive quantities defined on $\Omega$. They respectively bias cell polarization (and therefore orientation) and speed, once cell orientation is set. Hence, we will have a transition probability and a frequency of tumbling which depend on both $\mathcal{S}$ and $\mathcal{S}'$.

The general form of the turning operator which implements a velocity jump processes is
\begin{equation}\label{turning.operator}
\mathcal{J}[p](\x,\vb_p) = \mathcal{G}[p](\x,\vb_p)-\mathcal{L}[p](\x,\vb_p)\,,
\end{equation}
where the gain term is
\begin{equation}\label{gain.turning.operator}
\mathcal{G}[p](\x,\vb_p) = \int_{V_p} 
\mu(\x,\vb_p') T[\mathcal{S},\mathcal{S}'](\x,\vb_p|\vb'_p)p(t,\x,\vb_p')\, d\vb_p' \,,
\end{equation}
the loss term is 
\begin{equation}\label{loss.turning.operator}
\mathcal{L}[p](\x,\vb_p) = \int_{V_p} 
\mu(\x,\vb_p) T[\mathcal{S},\mathcal{S}'](\x,\vb_p\!\!\!'{}'|\vb_p)p(t,\x,\vb_p)\, d\vb_p\!\!\!'{}' \,,
\end{equation}
and $\vb_p'$ is the pre-turning velocity of the gain term and $\vb_p\!\!\!'{}'$ is the post-turning velocity of the loss term.
The so-called {\it turning kernel} $T[\mathcal{S},\mathcal{S}'](\x,\vb_p|\vb_p')$  is the probability for a cell in $\x$ of choosing the velocity $\vb_p$ after a re-orientation biased by the external fields $\mathcal{S}$ and $\mathcal{S}'$ given the pre-turning velocity $\vb_p'$.
Being a transition probability, it satisfies 
\begin{equation}\label{normalization.T}
\int_{V_p} T[\mathcal{S},\mathcal{S}'](\x,\vb_p\!\!\!'{}'|\vb_p) d\vb_p\!\!\!'{}' =1 \,,\quad \ \forall \x \, \in \, \Omega, \, \, \forall \vb_p \, \in \, V_p. 
\end{equation}
Also the {\it turning frequency} $\mu$ may depend on the external signal and/or on its gradient, and on the microscopic orientation, $\eg$ through its polarization, typically through $\vb\cdot \nabla \mathcal{S}$ (see for example the work by \cite{Othmer_Hillen.02}). In fact, it is proved that the speed and the turning rates of individuals
depend not only on the magnitude of an external signal $\mathcal{S}$ but also on its
temporal and spatial variations as highlighted by \cite{Berg}, \cite{Fisher}, \cite{Koshland}, \cite{Soll}.\

As done by \cite{Chauviere_Hillen_Preziosi.07} and \cite{Chauviere_Hillen_Preziosi.08}, in the following, we will assume that cells retain no memory of their velocity prior to the re-orientation, $\ie$, $T=T[\mathcal{S},\mathcal{S}'](\x,\vb_p)$.
The independence from the pre-tumbling velocity lies in the fact that the choice of the new velocity is linked to the slow interaction process also related to cell ruffling and sensing which is responsible for the biased re-orientation. However, the assumption might be restrictive in some cases, as it does not include, for instance, the case in which the sensing region depends on the incoming velocity through a polarization-dependent expression of transmembrane proteins. It also excludes  persistence effects in which the re-orientation direction depends on the pre-tumbling polarization of the cell. 
Under such an assumption, the operator (\ref{turning.operator}) simplifies to
\begin{equation}\label{J_r.mu.S}
\mathcal{J}[p](\x,\vb_p) =   \int_{V_p}\mu(\x,\vb_p')p(t,\x,\vb_p') \, d\vb_p' \,T[\mathcal{S},\mathcal{S}'](\x,\vb_p) - \mu(\x,\vb_p)p(t,\x,\vb_p) \,,
\end{equation}
where $\int_{V_p}\mu(\x,\vb_p')p(t,\x,\vb_p') \, d\vb_p'$ represents the fraction of re-orientating cells per time unit in $\x$. If also the frequency $\mu(\x)$ does not depend on the  microscopic velocity, then
we have that
\begin{equation}\label{J_r.S}
 \mathcal{J}[p](t,\x,\vb_p) = \mu(\x) \, \Big( \rho(t,\x)  T[\mathcal{S},\mathcal{S}'](\x,\vb_p) - p(t,\x,\vb_p) \Big) \,.
\end{equation}
However, in the following, we will also consider a 'weak' dependence of the frequency on the incoming velocity and in particular on cell polarization.

We observe that the distribution function nullifying the operator in (\ref{J_r.S}) is simply
\begin{equation}\label{stationary.eq}
p(t,\x,\vb_p)=\rho(t,\x)  T[\mathcal{S},\mathcal{S}'](\x,\vb_p) ,
\end{equation}
which, in general, is time dependent, but it may also be a stationary equilibrium state. 

\noindent As a consequence of (\ref{normalization.T}), for any function $T[\mathcal{S},\mathcal{S}']$ the operator $\mathcal{J}[p]$ satisfies the property
\begin{equation}
\int_{V_p} \mathcal{J}[p](t,\x,\vb_p)\, d\vb_p = 0 \,,
\end{equation}
that expresses mass conservation. On the other hand, the evaluation of the first moment of $\mathcal{J}[p]$ with respect to the velocity variable yields
\begin{equation}\label{momentum_r}
\int_{V_p} \mathcal{J}[p](t,\x,\vb_p)\,\vb \, d\vb_p = \mu (\x) \rho(t,\x) \big({\bf U}_{\scriptscriptstyle\mathcal{S},\mathcal{S}'}(\x)- \Ub(t,\x)\big) \,
\end{equation}
that indicates no preservation of momentum in general. Similarly, also energy is not preserved. In (\ref{momentum_r}) the vector $\Ub$ is the mean velocity of the cells defined by (\ref{mean.U}), while ${\bf U}_{\scriptscriptstyle\mathcal{S},\mathcal{S}'}$ is a macroscopic velocity due to the variations of $\mathcal{S}, \, \mathcal{S}'$, defined by
\begin{equation}\label{mean.Vr1}
{\bf U}_{\scriptscriptstyle
\mathcal{S},\mathcal{S}'
 }(\x)  = \int_{V_p} T[\mathcal{S},\mathcal{S}'](\x,\vb_p) \vb \ d\vb_p \,.
\end{equation}
It is the mean outgoing velocity after a re-orientation biased by $\mathcal{S}$ and $\mathcal{S}'$.
Another important quantity is the variance-covariance matrix of $T$, which is defined as
\begin{equation}
\mathbb{D}_{\scriptscriptstyle\mathcal{S},\mathcal{S}'}(t,\x) = \int_{V_p} (\vb -\Ub_{\scriptscriptstyle\mathcal{S},\mathcal{S}'}) \otimes (\vb-\Ub_{\scriptscriptstyle\mathcal{S},\mathcal{S}'}) T[\mathcal{S},\mathcal{S}'](\x,\vb_p)
\, d\vb_p\,.
\end{equation}

\cite{Bisi.Carrillo.Lods} and \cite{Petterson} deal with an operator which has a structure similar to (\ref{J_r.S}). They prove that, provided that the probability distribution has initially a finite mass and energy and non-absorbing boundary conditions hold, the function (\ref{stationary.eq}), which makes (\ref{J_r.S}) vanish, is a stable asymptotic equilibrium state. It is stationary, which means that $\rho$ is stationary. We shall consider no-flux boundary conditions ($\ie$ \eqref{noflux}) which are non-absorbing boundary conditions.
Moreover, ${\bf U}_{\scriptscriptstyle
\mathcal{S},\mathcal{S}'
 }$ and $\mathbb{D}_{\scriptscriptstyle\mathcal{S},\mathcal{S}'}$ are the mean velocity and variance-covariance tensor of the cells at the equilibrium.

\subsection{Structure of the turning probability $T$}

From the physiological point of view, the decision process determining the new cell velocity can be split in two parts. First the cell decides where to go and polarizes, forming the so-called cell ``head" and ``tail". Then, it starts to move in that direction with a certain speed.
These two processes are mainly independent and might be influenced by different chemical and mechanical cues.
For instance, environment microstructure, availability of intracellular space and of adhesion sites, expression of integrins, activity of motor proteins, calcium influx, availability of ATP all influence cell speed, while gradients of free and bound chemical factors, of ECM components and ECM stiffness all influence cell polarization.
A cell can even be polarized but unable to move. 
This, for instance, occurs if the cell is treated with latrunculin as shown by 
\cite{Devreotes.03}, or because of overcrowding due to the presence of too many cells in the direction where it would like to move,  giving rise to the so-called volume filling effect (see the work by \cite{Painter_Hillen.02} and references therein), or because of a too dense ECM, giving rise to the so-called physical limit of migration
\citep{Wolf_Friedl.13, Arduino_Preziosi.15, Giverso.18, Scianna_Preziosi.13, Scianna_Preziosi.14,Scianna_Preziosi.13.2}.

As already stated, the choice of the new velocity is a result of a sensing activity of the cell in its neighborhood performed to  monitor the quantity of two generally different fields $\mathcal{S}$ and $\mathcal{S}'$, respectively determining cell polarization and speed. 
This information is then weighted by two functions 
$$\gamma_{\scriptscriptstyle \mathcal{S}}:\mathbb{R}_+\longmapsto\mathbb{R}_+\quad {\rm and}\quad
   \gamma'_{\scriptscriptstyle \mathcal{S}'}:\mathbb{R}_+\longmapsto\mathbb{R}_+\,,$$
both with compact support that is related to the finite size of the influencing neighborhood related respectively to the fields 
$\mathcal{S}$ and $\mathcal{S}'$ around the position $\x$ and weighting the information according to the distance from the cell center. They may be 
\begin{itemize}
\item Dirac deltas, $\eg$, $\gamma_{\scriptscriptstyle \mathcal{S}}(\lambda)=\delta(\lambda-R_{\scriptscriptstyle\mathcal{S}})$, if the cells only measure the information perceived on a surface of given radius $R_{\scriptscriptstyle\mathcal{S}}$,
\item Heaviside functions, $\eg$, $\gamma_{\scriptscriptstyle \mathcal{S}}(\lambda)=H (R_{\scriptscriptstyle\mathcal{S}}-\lambda)$ if the cells explore the whole volume of the sphere centered in $\x$ with radius $R_{\scriptscriptstyle\mathcal{S}}$ and weight the information uniformly, or 
\item decreasing functions of the distance $\lambda$ from the cell center $\x$ taking for instance into account that the probability of making longer protrusion decreases with the distance, so the sensing of closer regions is more accurate. 
\end{itemize}
The distance $R_{\scriptscriptstyle\mathcal{S}}$ is, therefore, a measure of the capability of sensing and detecting of the cell. It may be the measure of cell protrusions, or it may be bounded by $\frac{ U}{\mu}$ which is the mean linear tract travelled between two consecutive re-orientations. \cite{Berg_Purcell.77} have shown that the sampling volume in which the signal is significantly detected depends on how rapidly the receptors on the cell's membrane process the signal.  The same considerations are valid for $\gamma_{\scriptscriptstyle \mathcal{S}'}$ with a radius $R_{\scriptscriptstyle \mathcal{S}'}$ which in general, might be different from $R_{\scriptscriptstyle\mathcal{S}}$.\\ 
The above considerations justify the following splitting
\begin{equation}\label{distribution.g.r}
T[\mathcal{S},\mathcal{S}'](\x, \vb_p)=c(\x)\int_{\mathbb{R}_+}\gamma_{\scriptscriptstyle \mathcal{S}}(\lambda) \mathcal{T}_{\lambda}^{\hat{\vb}}[\mathcal{S}](\x) d\lambda \,  \int_{\mathbb{R}_+}\gamma_{\scriptscriptstyle \mathcal{S}'}(\lambda') \psi(\x,v|\mathcal{S}'(\x+\lambda'\hat{\vb})) \, d\lambda '.
\end{equation}
The quantity $\mathcal{T}_{\lambda}^{\hat{\vb}}[\mathcal{S}](\x)$ is a functional which acts on $\mathcal{S}$ and describes the way the cell measures the quantity $\mathcal{S}$ around $\x$ along the direction $\hat{\vb}$ and, therefore, the bias intensity in the direction $\hat{\vb}$.
In some cases, we shall write for instance
\begin{equation}
\mathcal{T}_{\lambda}^{\hat{\vb}}[\mathcal{S}](\x) = b\big(\mathcal{S}(\x + \lambda\hat{\vb})\big)\,,
\end{equation}
where $b$ is a quantity depending on the value of $\mathcal{S}$ in a point at a distance $\lambda$ from $\x$ along the direction $\hat\vb$.  So, if $b$ is an increasing function of $\mathcal{S}$ and the signal is stronger in the direction $\hat\vb$, then there will be a higher probability for the cell to move along $\hat\vb$ than along $-\hat\vb$. We shall also consider other forms for $\mathcal{T}_{\lambda}^{\hat{\vb}}[\mathcal{S}](\x)$ as well.\\
The quantity $\psi(\x,v|\mathcal{S}'(\x+\lambda'\hat{\vb}))$ is a conditional probability distribution function of the cell speed given the value of $\mathcal{S}'$ in a neighborhood of $\x$ in the direction $\hat\vb$ that is then weighted by $\gamma_{\scriptscriptstyle \mathcal{S}'}$. Again, we shall discuss this further in Section \ref{double.bias}. Finally, the factor $c(\x)$ is a normalizing factor which is given by
\begin{equation}\label{norm.c}
c(\x)=\dfrac{1}{\displaystyle \int_{\mathbb{S}^{d-1}} \Gamma_0 '\int_{\mathbb{R}_+}\gamma_{\scriptscriptstyle \mathcal{S}}(\lambda) \mathcal{T}_{\lambda}^{\hat{\vb}}[\mathcal{S}](\x) d\lambda \, d\hat{\vb} }
\end{equation}
where $\Gamma_0'=\displaystyle \int_{\mathbb{R}_+}\gamma_{\scriptscriptstyle \mathcal{S}'}(\lambda') \, d\lambda'$, so that (\ref{normalization.T}) is satisfied.

\section{Macroscopic limits}
In order to highlight the driving macroscopic phenomenon which may be diffusion or convection, we will discuss the appropriate macroscopic limits of the kinetic equations. A way to do that is to perform, respectively, a parabolic scaling and then a diffusive limit, or a hyperbolic scaling and then a hydrodynamic limit. In these approaches it is assumed that there is a small parameter $\epsilon\ll 1$ that allows to suitably rescale the transport equation.  The two scalings are, respectively, 
$$
\begin{array}{lr}
\tau=\epsilon^2 t, \quad {\bf \boldsymbol{\xi}} =\epsilon \x, \\[8pt]
\tau=\epsilon t, \quad {\bf \boldsymbol{\xi}} =\epsilon \x.
\end{array}
$$
Following the work by \cite{Othmer_Hillen.02}, we also assume that $T$ can be expanded as
\begin{equation}\label{T.expanded}
T[\mathcal{S},\mathcal{S}'](\boldsymbol{\xi}, \vb_p)=T[\mathcal{S},\mathcal{S}']_0(\boldsymbol{\xi}, \vb_p)+\epsilon T[\mathcal{S},\mathcal{S}']_1(\boldsymbol{\xi}, \vb_p)+\mathcal{O}(\epsilon^2) 
\end{equation}
This means that there are different orders of bias as will be better illustrated in the examples to follow. If we assume that $\mu=\mathcal{O}(1)$, we may denote $\mathcal{J}^0$ and $\mathcal{J}^1$ the related operators defined by $T_0$ and $T_1$ respectively, and
we assume that
\begin{subequations}
\begin{equation}\label{diff.cond.1}
\displaystyle \int_{V_p} T[\mathcal{S},\mathcal{S}']_0(\boldsymbol{\xi}, \vb_p)  \, d\vb_p =1,
\end{equation}
and
\begin{equation}\label{diff.cond.2}
\displaystyle \int_{V_p} T[\mathcal{S},\mathcal{S}']_i(\boldsymbol{\xi}, \vb_p)  \, d\vb_p =0 \quad \forall i \geq 1 .
\end{equation}
\end{subequations}
Coherently, the velocity and tensor of the second moments of the transition probability may be written as
\begin{equation}
\Ub_{\scriptscriptstyle\mathcal{S},\mathcal{S'}}=\Ub_{\scriptscriptstyle\mathcal{S},\mathcal{S'}}^0+\epsilon \Ub_{\scriptscriptstyle\mathcal{S},\mathcal{S'}}^1 +\mathcal{O}(\epsilon^2)
\end{equation} 
and
\begin{equation}
\mathbb{D}_{\scriptscriptstyle\mathcal{S},\mathcal{S'}}=\mathbb{D}_{\scriptscriptstyle\mathcal{S},\mathcal{S'}}^0+\epsilon \mathbb{D}_{\scriptscriptstyle\mathcal{S},\mathcal{S'}}^1+\mathcal{O}(\epsilon^2)
\end{equation}
where $$\Ub_{\scriptscriptstyle\mathcal{S},\mathcal{S'}}^i=\displaystyle\int_{V_p}T[\mathcal{S},\mathcal{S}']_i \vb \, d\vb_p$$ and $$\mathbb{D}_{\scriptscriptstyle\mathcal{S},\mathcal{S'}}^i=\displaystyle\int_{V_p} T[\mathcal{S},\mathcal{S}']_i (\vb-\Ub_{\scriptscriptstyle\mathcal{S},\mathcal{S'}}^i) \otimes (\vb-\Ub_{\scriptscriptstyle\mathcal{S},\mathcal{S'}}^i) \, d\vb_p$$ for $i\geq 1$.
In both limits we consider an expansion of the distribution function $p$ in the form
\begin{equation}
p=p_0+\epsilon p_1 +\mathcal{O}(\epsilon^2) \, ,
\end{equation}
where the function $p_0$ is the equilibrium function given by the solution to the leading order part of the turning operator
\begin{equation}\label{Jr0}
\mathcal{J}^0[p_0](\vb_p)=0 \, .
\end{equation}
As $\mathcal{J}^0[p_0](\vb_p)=\mu(\rho_0T[\mathcal{S},\mathcal{S}']_0(\vb_p)-p_0(\vb_p))$, Eq.\eqref{Jr0} is satisfied if and only if 
\begin{equation}\label{p0}
p_0(\vb_p)=\rho_0T[\mathcal{S},\mathcal{S}']_0 (\vb_p).
\end{equation}

\subsection{The diffusive limit}\label{The diffusive limit}
The diffusive limit of velocity jump processes is deeply treated by \cite{Othmer_Hillen.00} and \cite{Othmer_Hillen.02}. In particular, in \citep{Othmer_Hillen.02} macroscopic models for chemosensitive movement are systematically derived for different orders of $\epsilon$ of the turning probability. \noindent In the same spirit as in the works above, because of the conservation of mass, we have that
\begin{itemize}
\item all the mass is in $p_0$, $\ie$,
\begin{equation}\label{rho0}
\rho_0=\rho, \quad \rho_i=0 \quad \forall   i \geq 1 \, ,
\end{equation}
where $\displaystyle{\rho_i=\int_{V_p}p_i d\vb}$;
\item \begin{equation}\label{rhoM0}
\displaystyle{\int_{V_p} p_i \vb d\vb_p =0 } \quad \forall i \geq 2
\end{equation}.
\end{itemize}
In the works by \cite{Othmer_Hillen.00} and \cite{Othmer_Hillen.02}   the turning probability depends on the pre-tumbling velocity, $\ie, T=T(\vb_p|\vb_p')$ and it is required that
\begin{equation}\label{T4}
\int_{V_p} T(\vb_p|\vb_p') d\vb_p'=1.
\end{equation}
Thanks to this property and to other regularity assumptions on $T$ \cite{Othmer_Hillen.00} show that the function identically equal to $1$ is an eigenfunction of the turning operator with eigenvalue 0 that is a simple eigenvalue. In the present work there is no dependence on the pre-tumbling velocity and \eqref{T4} does not hold true. Furthermore, because of the structure of the turning operator, the density $p_0$ making the leading part of the turning operator vanish \eqref{p0} is an eigenfunction of the velocity jump operator and it depends on the velocity. Moreover, as all the functions nullifying the turning operators are proportional to $T$ up to a multiplicative constant, we have that the eigenvalue 0 is simple and the subspace of $L^2(V_p)$ defined by
\[
\langle T[\mathcal{S},\mathcal{S}']_0(\boldsymbol{\xi}, \vb_p)\rangle
\]
is the kernel of the turning operator and it has dimension equal to one.
Therefore, the appropriate scalar product allowing to determine uniquely the solvability condition  is the one proposed by \cite{Hillen.05}
\begin{equation}\label{scalar.product}
\langle f,g\rangle=\int_{V_p} f g T^{-1} d\vb_p.
\end{equation}
Whence the equilibrium function $p_0$ belongs to the subspace generated by the eigenfunction of $\mathcal{J}^0$ with eigenvalue $0$, $\ie$, $\langle T[\mathcal{S},\mathcal{S}']_0(\boldsymbol{\xi}, \vb_p)\rangle$. This means that the equilibrium function has mean and the variance-covariance tensor equal to $\Ub_{\scriptscriptstyle\mathcal{S},\mathcal{S'}}^0$ and $\mathbb{D}_{\scriptscriptstyle\mathcal{S},\mathcal{S'}}^0$ respectively.
\noindent For $i\geq 1$, the functions $p_i \in \langle T[\mathcal{S},\mathcal{S}']_0(\boldsymbol{\xi}, \vb_p)\rangle^{\bot} \subseteq L^2(V_p)$ for $a.e. \, \boldsymbol{\xi}$, where 
$\langle  T[\mathcal{S},\mathcal{S}']_0(\boldsymbol{\xi}, \vb_p)\rangle^{\bot}$ is the orthogonal subspace to the subspace generated by 
$T[\mathcal{S},\mathcal{S'}]_0(\boldsymbol{\xi}, \vb_p)$. Thanks to the scalar product \eqref{scalar.product} we have that $g\in \langle  T[\mathcal{S},\mathcal{S}']_0(\boldsymbol{\xi}, \vb_p)\rangle^{\bot}$ if and only if $\int_{V_p}g(\vb_p) d\vb_p=0$. This justifies the assumptions \eqref{rho0}.

The dependence of $p_0$ on $\vb_p$ has also consequences on the conditions for the isotropy of $T$ and of the diffusion tensor are not the same as the ones considered by \cite{Othmer_Hillen.00} and \cite{Othmer_Hillen.02}. The case in which $T$ does not depend on the pre-tumbling velocity is treated by \cite{Hillen.05} where it is assumed that the turning probability $T$ is of order zero in $\epsilon$. Therefore, we illustrate the diffusive limit procedure in the case in which the turning probability does not depend on the pre-tumbling velocity and it can be expressed as in (\ref{T.expanded}), and we also illustrate the necessary conditions which are needed in order to perform it.\\
The discriminating factor is whether $T$ is such that
\begin{equation}\label{even}
\Ub_{\scriptscriptstyle \mathcal{S},\mathcal{S}'}^0 =\displaystyle\int_{V_p}T[\mathcal{S},\mathcal{S}']_0 \vb \, d\vb_p =\boldsymbol{0} 
\quad \textit{for}\quad a.e. \quad \boldsymbol{\xi},
\end{equation}
$\ie$, the leading order macroscopic velocity vanishes, or not.
In this case we are led to perform a parabolic scaling, $\ie$,
$\tau=\epsilon^2 t$, ${\bf \boldsymbol{\xi}} =\epsilon \x$, $\epsilon \ll 1$. \\
The rescaled equation is
\begin{equation}
\begin{aligned}
\epsilon^2\dfrac{\partial p}{\partial \tau}(\tau,{\bf \boldsymbol{\xi}},\vb_p) + \epsilon\vb\cdot \nabla p(\tau,{\bf \boldsymbol{\xi}},\vb_p)=&\mathcal{J}^0[p]+\epsilon\mathcal{J}^1[p] +\mathcal{O}(\epsilon^2) \, . 
\end{aligned} 
\end{equation}
We suppose that $\mu \sim \mathcal{O}(1)$. 
From now on, to simplify the notation, we will drop the dependencies  on $\tau$ and $\boldsymbol{\xi}$.
Comparing terms of equal order in $\epsilon$, we then obtain the following system of equations:

In $\epsilon^{0}$:
\begin{equation}
\mathcal{J}^0[p_{0}](\vb_p)\equiv  \mu  \Big( \rho_0 T[\mathcal{S},\mathcal{S'}]_0(\vb_p)-p_{0}(\vb_p)  \Big)=0
\label{eps0q}
\end{equation}

In $\epsilon^{1}$:
\begin{equation}
\begin{aligned}
\nabla\cdot\big( p_{0}(\vb_p)\vb\big)&=\mathcal{J}^0[p_1](\vb_p) +\mathcal{J}^1[p_0](\vb_p)=\\[8pt]
&=\mu  \big(\rho_1 T[\mathcal{S},\mathcal{S'}]_0(\vb_p) -p_1(\vb_p)\big)
+ \mu  \rho_0 T[\mathcal{S},\mathcal{S'}]_1(\vb_p) 
\end{aligned}
\label{eps1q}
\end{equation}

In $\epsilon^{2}$:
\begin{equation}
\begin{array}{lclr}
\dfrac{\partial }{\partial \tau}p_{0}(\vb_p)+\nabla\cdot\big(p_{1}(\vb_p)\vb \big)=\mathcal{J}^0[p_2](\vb_p) +\mathcal{J}^1[p_1](\vb_p)+\mathcal{J}^2[p_0](\vb_p).
\label{eps2q}
\end{array}
\end{equation}
As already discussed, Eq. ($\ref{eps0q}$) implies (\ref{p0}),
that is the equilibrium state of order zero.
From Eq. ($\ref{eps1q}$),  by inverting $\mathcal{J}^0$, we get
\begin{equation}
p_{1}(\vb_p)=-\dfrac{1}{\mu}  \nabla \cdot \big(\vb p_{0}\big)+  \rho_0 T[\mathcal{S},\mathcal{S'}]_1(\vb_p) \, .
\end{equation}
The solvability condition for inverting $\mathcal{J}^0$ is that $p_1 \in \langle T[\mathcal{S},\mathcal{S'}]_0(\vb_p)\rangle^{\bot}$. Because of \eqref{scalar.product} this means that
\begin{equation}\label{solv.1}
\int_{V_p}\left[- \nabla \cdot \dfrac{1}{\mu} \big(\vb p_{0} \big)+ \rho_0 T[\mathcal{S},\mathcal{S'}]_1(\vb_p)\right]d\vb_p =0 \!\!\!\!\qquad \textit{for}\!\!\!\!\qquad a.e. \!\!\!\!\qquad \boldsymbol{\xi}.
\end{equation}
Equation (\ref{solv.1}) is satisfied because 
\begin{equation}\label{solv.2}
\int_{V_p} \vb p_{0}(\vb_p)d\vb_p =\rho_0\int_{V_p}T[\mathcal{S},\mathcal{S'}]_0(\vb_p)\vb \, d\vb_p=0 \quad \textit{for} \quad a.e. \quad \boldsymbol{\xi} ,
\end{equation}
that is ($\ref{even}$),
and
\begin{equation}\label{solv.3}
\int_{V_p} T[\mathcal{S},\mathcal{S'}]_1(\vb_p)d\vb_p =0\quad \textit{for} \quad a.e. \quad \boldsymbol{\xi},
\end{equation}
because of (\ref{diff.cond.2}).

\noindent The evolution equation for $\rho \, (=\rho_0)$ is obtained from the solvability condition at $\mathcal{O}(\epsilon^2)$, $\ie$, the integral on $V_p$ of Eq. ($\ref{eps2q}$) which is given by
\begin{equation}\label{macro.diff}
\dfrac{\partial \rho}{\partial \tau}+\nabla \cdot \Big(\rho \Ub_{\scriptscriptstyle\mathcal{S},\mathcal{S'}}^1 \Big)=\nabla \cdot \left( \dfrac{1}{\mu}\nabla \cdot \big(\mathbb{D}_{\scriptscriptstyle\mathcal{S},\mathcal{S'}}^0\rho \big)\right)\, 
\end{equation}
where we used \eqref{rhoM0}.
The term $\Ub_{\scriptscriptstyle\mathcal{S},\mathcal{S'}}^1$ is the chemotactic velocity, which is present if there is a bias of order $\epsilon$ as in the work by \cite{Othmer_Hillen.02}. In the following sections, we will show that under appropriate conditions of sensing of the cells, microscopic rules may allow to recover the chemotaxis equation in which the chemotactic velocity is proportional to the gradient of $\mathcal{S}$ and the chemotactic sensitivity depends on the sensitivity function, which captures the cell capability of sensing and detecting the signal.

\subsection{Diffusive limit for encounter rates weakly depeding on the incoming velocity}\label{diff.limit.weak.dep}
We now generalize the above procedure in the case it is possible to write $\mu(\x,\vb_p)= \mu_0(\x) +\epsilon \mu_1(\x,\vb_p)+\mathcal{O}(\epsilon^2)$.
In this case the rescaled transport equation reads
\begin{equation}
\begin{array}{lr}
\epsilon^2\dfrac{\partial p}{\partial \tau}(\vb_p) + \epsilon\vb\cdot \nabla p(\vb_p) =
  \mu_0  \,  \Big( \rho  \big( T[\mathcal{S},\mathcal{S'}]_0(\vb_p)+\epsilon T[\mathcal{S},\mathcal{S'}]_1(\vb_p) \big) - p(\vb_p) \Big) \\[8pt]
  +\displaystyle\int_{V_p} \epsilon \mu_1(\vb_p') p(\vb_p') d\vb_p' \Big( T[\mathcal{S},\mathcal{S'}]_0(\vb_p)+\epsilon T[\mathcal{S},\mathcal{S'}]_1(\vb_p) \Big)-  \epsilon \mu_1(\vb_p)p(\vb_p)\\[8pt]
+\mu_0 \, \epsilon^2\rho T[\mathcal{S},\mathcal{S'}]_2(\vb_p)+\epsilon^2 T[\mathcal{S},\mathcal{S'}]_0(\vb_p) \displaystyle \int_{V_p}\mu_2(\vb_p')p(\vb_p') \, d\vb_p' -\epsilon^2 \mu_2(\vb_p)p(\vb_p)+\mathcal{O}(\epsilon^3)\, ,  
\end{array}
\end{equation}
where also $p$ has to be intended as expanded in terms of $\epsilon$.
Comparing terms of equal order in $\epsilon$, we obtain the same equation as before in $\epsilon^0$, while at the first order
\begin{equation}
\begin{array}{lclr}
\nabla\cdot\big( p_{0}(\vb_p)\vb\big)=\mu_0  \big(\rho_1 T[\mathcal{S},\mathcal{S'}]_0(\vb_p) -p_1(\vb_p)\big)+ \mu_0 \rho_0 T[\mathcal{S},\mathcal{S'}]_1(\vb_p)\\[8pt]
+\displaystyle\int_{V_p} \mu_1(\vb_p') p_0(\vb_p') \, d\vb_p'T[\mathcal{S},\mathcal{S'}]_0(\vb_p)-\mu_1(\vb_p)p_0(\vb_p) ,
\end{array}
\label{eps1q.t}
\end{equation}
and at second order
\begin{equation}
\begin{array}{lclr}
\dfrac{\partial }{\partial \tau}p_{0}(\vb_p)+\nabla\cdot\big(p_{1}(\vb_p)\vb \big)=\mu_0 \big(\rho_2 T[\mathcal{S},\mathcal{S'}]_0(\vb_p) -p_2(\vb_p)\big)\\[8pt]
+\mu_0 \rho_1 T[\mathcal{S},\mathcal{S'}]_1(\vb_p)
+\displaystyle\int_{V_p} \mu_1(\vb_p') p_1(\vb_p') \, d\vb_p'T[\mathcal{S},\mathcal{S'}]_0(\vb_p)\\[8pt]
+\displaystyle\int_{V_p} \mu_1(\vb_p') p_0(\vb_p') \, d\vb_p'T[\mathcal{S},\mathcal{S'}]_1(\vb_p)-\mu_1(\vb_p)p_1(\vb_p)+\mu_0 \rho_0 T[\mathcal{S},\mathcal{S'}]_2(\vb_p)\\[8pt]
 + \displaystyle \int_{V_p}\mu_2(\vb_p')p_0(\vb_p') \, d\vb_p' T[\mathcal{S},\mathcal{S'}]_0(\vb_p)-\mu_2(\vb_p)p_0(\vb_p) .
\label{eps2q.t}
\end{array}
\end{equation}
From equation ($\ref{eps1q.t}$), we get
\begin{equation}
\begin{array}{lr}
p_{1}(\vb_p)=-\dfrac{1}{\mu_0}  \nabla \cdot \big(\vb p_{0}(\vb_p)\big)+  \rho_0 T[\mathcal{S},\mathcal{S'}]_1(\vb_p)
+\dfrac{\rho_0}{\mu_0}T[\mathcal{S},\mathcal{S'}]_0(\vb_p)\big(\bar{\mu}_1 -\mu_1(\vb_p) \big)\, ,
\end{array}
\end{equation}
where
\begin{equation}\label{barmu}
\bar{\mu}_1=\displaystyle\int_{V_p}\mu_1(\vb_p') T[\mathcal{S},\mathcal{S'}]_0(\vb_p') \, d\vb_p'.
\end{equation}
In this case the solvability condition for inverting $\mathcal{J}^0$ becomes
\begin{equation}\label{solv.1.t}
\begin{array}{lr}
\displaystyle \int_{V_p}\left[-\dfrac{1}{\mu_0}\nabla\cdot\big(\vb p_{0} \big)+ \rho_0 T[\mathcal{S},\mathcal{S'}]_1(\vb_p)
+\dfrac{\rho_0}{\mu_0}T[\mathcal{S},\mathcal{S'}]_0(\vb_p)\big(\bar{\mu}_1 -\mu_1(\vb_p)\big)\right]d\vb_p =0 \, ,
\end{array}
\end{equation}
that is again satisfied because
\begin{equation}\label{solv.2.t}
\int_{V_p} \vb p_{0}(\vb_p)d\vb_p =0 \,\,\,\,\textit{for} \, \,\,\, a.e. \,\,\,\, \boldsymbol{\xi},
\end{equation}
\begin{equation}\label{solv.3.t}
\int_{V_p} T[\mathcal{S},\mathcal{S'}]_1(\vb_p)d\vb_p =0 \,\,\,\,\textit{for} \, \,\,\, a.e. \,\,\,\, \boldsymbol{\xi}  
\end{equation}
and
\begin{equation}\label{solv.4.t}
\int_{V_p}T[\mathcal{S},\mathcal{S'}]_0(\vb_p)\big(\bar{\mu}_1 -\mu_1(\vb_p)\big)d\vb_p =0  \,\,\,\,\textit{for} \, \,\,\, a.e. \,\,\,\, \boldsymbol{\xi}. 
\end{equation}
The first two equations vanish as proved in the preceding section,
while recalling the definition (\ref{barmu}) and (\ref{diff.cond.1}), we have that (\ref{solv.4.t}) is trivially satisfied.\\
As in the previous section, the solvability condition at $\mathcal{O}(\epsilon^2)$, $\ie$, the integral on $V_p$ of \eqref{eps2q.t} gives the evolution equation for $\rho$ 
\begin{equation}\label{macro.diff.t}
\begin{array}{lr}
\dfrac{\partial \rho}{\partial \tau} +\nabla \cdot\Big(\rho \big(\Ub_{\scriptscriptstyle\mathcal{S},\mathcal{S}'}^1
+\Ub_{\scriptscriptstyle\mathcal{S},\mathcal{S}'}^{1,\mu_1}\big)\Big)=\nabla \cdot\left( \dfrac{1}{\mu_0} \nabla \cdot \big(  \mathbb{D}_{\scriptscriptstyle\mathcal{S},\mathcal{S}'}\rho\big) \right)
\, 
\end{array}
\end{equation}
where $\Ub_{\scriptscriptstyle\mathcal{S},\mathcal{S}'}^1= \displaystyle \int_{V_p} T[\mathcal{S},\mathcal{S'}]_1\vb  \, d\vb_p$ and
\begin{equation}
\begin{aligned}
\Ub_{\scriptscriptstyle\mathcal{S},\mathcal{S}'}^{1,\mu_1}=&\dfrac{1}{\mu_0}\displaystyle\int_{V_p}\left\lbrace T[\mathcal{S},\mathcal{S'}]_0(\vb_p)[\bar{\mu}_1-\mu_1(\vb_p)]\right\rbrace \vb d\vb_p\\[8pt]
=&-\dfrac{1}{\mu_0}\displaystyle\int_{V_p} T[\mathcal{S},\mathcal{S'}]_0(\vb_p)\mu_1(\vb_p) \vb d\vb_p,
\end{aligned}
\end{equation}
where the first term vanished because of (\ref{even}).
So, the chemotactic velocity presents a correction to the term $\Ub_{\scriptscriptstyle\mathcal{S},\mathcal{S}'}^1$  given by $\Ub_{\scriptscriptstyle\mathcal{S},\mathcal{S}'}^{1,\mu_1}$.
In the particular case in which  $\mu_1$ is independent of $\vb_p$ we obviously recover (\ref{macro.diff}) as $\Ub_{\scriptscriptstyle\mathcal{S},\mathcal{S}'}^{1,\mu_1}= {\bf 0}$.

\subsection{The hyperbolic limit}
The discriminating factor on whether a diffusive or a hyperbolic limit can be performed lies in the properties of the chosen turning probability $T$ and specifically on  the satisfaction of \eqref{even}. So, at variance with the previous section, we here assume  that $T$ does not satisfy (\ref{even}). In this case, a hyperbolic scaling, $\ie$ $\tau=\epsilon, {\bf \boldsymbol{\xi}} =\epsilon \x$, $\epsilon \ll 1$ can be performed. The rescaled transport equation is
\begin{equation}\label{hyp.resc}
\begin{array}{lr}
\epsilon \dfrac{\partial p}{\partial \tau}(\vb_p) + \epsilon\vb\cdot \nabla p(\vb_p) =  
\\[8pt]
\mu \,  \left[ \rho \big(T[\mathcal{S},\mathcal{S'}]_0(\vb_p)+\epsilon T[\mathcal{S},\mathcal{S'}]_1(\vb_p)\big)  - p(\vb_p) \right] \, .  
\end{array}
\end{equation}
As the equilibrium state is the same as before, we consider a Chapman-Enskog expansion of $p$ in the form
\begin{equation}\label{chap-ens}
p(\vb_p)= \rho_{0} T[\mathcal{S},\mathcal{S'}]_0(\vb_p) + \epsilon g +\mathcal{O}(\epsilon^2).
\end{equation}
where $g \in \langle T[\mathcal{S},\mathcal{S'}]_0(\vb_p)\rangle^{\bot} \subseteq L^2(V_p)$ with the same scalar product as before. Substituting ($\ref{chap-ens}$) in ($\ref{hyp.resc}$) and integrating on $V_p$ the equation at the order $\epsilon^{1}$, we obtain 
\begin{equation}\label{macro.drift}
\dfrac{\partial \rho}{\partial \tau}+\nabla \cdot \big( \rho \Ub_{\scriptscriptstyle\mathcal{S},\mathcal{S}'}^0 \big)=0 \, ,
\end{equation}
thanks to (\ref{rho0}),
where (with the dependencies)
\begin{equation}\label{drift}
\Ub_{\scriptscriptstyle\mathcal{S},\mathcal{S}'}^0(\boldsymbol{\xi})= \displaystyle\int_{V_p} T[\mathcal{S},\mathcal{S'}]_0(\boldsymbol{\xi},\vb_p)\vb  \, d\vb_p.
\end{equation}
In this case  we have then a drift-driven phenomenon.

If we consider a frequency weakly depending on the velocity, the rescaled transport equation  modifies into
\begin{equation}
\begin{array}{lr}
\epsilon \dfrac{\partial p}{\partial \tau}(\vb_p) + \epsilon\vb\cdot \nabla p(\vb_p) =  
\\[8pt]
\mu_0\,  \left[ \rho \big( T[\mathcal{S},\mathcal{S'}]_0(\vb_p)+\epsilon T[\mathcal{S},\mathcal{S'}]_1(\vb_p)\big)  - p(\vb_p) \right]\\[8pt]
+\epsilon T[\mathcal{S},\mathcal{S'}]_0(\vb_p)\displaystyle\int_{V_p}\mu_1(\vb_p')p(\vb_p') d\vb_p'-\mu_1(\vb_p)p(\vb_p) \, +\mathcal{O}(\epsilon^2),  
\end{array}
\end{equation}
and, thanks to (\ref{rho0}) and (\ref{barmu}),
we recover, again, (\ref{macro.drift}) with $\Ub_{\scriptscriptstyle\mathcal{S},\mathcal{S}'}^0$ given by (\ref{drift}) .

\paragraph{Isotropy and anisotropy}
As we said at the beginning of the section, the necessary and sufficient conditions for the isotropy of the diffusion tensor are not the same as the ones considered by \cite{Othmer_Hillen.00}, as the transition probability does not depend on the incoming velocity. This implies that the mean outgoing velocity for a given incoming velocity is always zero and, therefore, the adjoint persistence is zero.
In our case, we have that $\mathbb{D}_{\scriptscriptstyle \mathcal{S},\mathcal{S}'}^0$ is isotropic if and only if $T_0[\mathcal{S},\mathcal{S'}]$ is isotropic as a function of $\hat{\vb}$, $\ie$,  if $T_0[\mathcal{S},\mathcal{S'}]$ is invariant with respect to rotations of $\frac{\pi}{2}$ as a function of $\hat{\vb}$. This also implies that $\Ub_{\scriptscriptstyle \mathcal{S},\mathcal{S}'}^0$ is zero and that the directional persistence is zero. Viceversa, if $\Ub_{\scriptscriptstyle \mathcal{S},\mathcal{S}'}^0$ is not zero, $\mathbb{D}_{\scriptscriptstyle \mathcal{S},\mathcal{S}'}^0$ is anisotropic.

\begin{table}
\begin{tabular}{|l|c|c|c|c|c|}
\hline\noalign\\[-10pt]
Biological aspect & Section & $\mathcal{J}$ & ${\rm Macro}\atop{\rm limit}$ &  Application & Figure \\ 
\hline
{\bf Orientation}  &&&&&\\ 
\hline 
Local sensing & \ref{Local_sensing} & \eqref{Jlocal} & \eqref{difflocal} & & \\ 
\hline 
${\rm Encounter\ rate\ depending}\atop{\rm  on\ the\ incoming\ velocity}$ & \ref{Encounterratedependenttaxis} & \eqref{Jturning} & \eqref{rate.dep} & &   \\ 
\hline 
Non-local sensing  & \ref{Nonlocal_sensing} & \eqref{J_r} &   &  
Chemotaxis &\ref{mono.simu.1}, \ref{mono.simu.1bis}, \ref{mono.simu.2}   \\
&  &  &   &  Cell adhesion &\ref{fig:adhesion}   \\
&  &  &   &  Durotaxis &\ref{durotaxis}   \\
 (small sensing limit) & 			& \eqref{J_r.loc} & \eqref{diff.adv.mono}--\eqref{chinonlocalsensing}   & & \\
\hline 
Comparative sensing  & \ref{Comparative} & \eqref{Jcomparative} &  &  
Chemotaxis &\ref{comp} \\
 (small sensing limit) & & \eqref{comp.sens} & \eqref{diff.adv.mono}--\eqref{chicomparativesensing} &   & \\ 
\hline 
{\bf Speed}&&&&&
\\ \hline 
Random polarization  & \ref{isotropicpol} & \eqref{TSR0} & \eqref{convdiffrandom} & ECM steryc hindrance & \ref{matrix}\\ 
(Non-local speed sensing)  & & & & & \\
\hline 
{\bf Orientation and Speed}&&&&&
\\ \hline  
Non-local orientation  &  \ref{doublechemo} & \eqref{nonlocalchemo} & \eqref{nonlocalchemodiff} &  Chemotaxis and  & \ref{mat3}\\ 
and speed sensing& & & & ECM steryc hindrance& \\
 \hline

\end{tabular}
\caption{Summary of models}
\label{table}
\end{table}

\subsection{Boundary conditions}
We shall consider the following biological no-flux condition \citep{Plaza}
\begin{equation}\label{noflux}
\int_{V_p} p(\tau,\boldsymbol{\xi},\vb_p)\vb\cdot {\bf n}(\boldsymbol{\xi})d\vb_p=0, \quad \forall \boldsymbol{\xi} \in \partial \Omega, \quad \tau>0
\end{equation}
being ${\bf n}(\boldsymbol{\xi})$ the outer normal to the boundary $\partial \Omega$ in the point $\boldsymbol{\xi}$, and we remind that $\vb_p=(v,\hat{\vb})$. This choice is motivated by the fact that in experiments in vitro there is no exchange of biological material between the system and its environment.
Equation \eqref{noflux} implies that there is no normal mass flux across the boundary \citep{Lemou}.
At the macroscopic level \eqref{noflux} gives \citep{Plaza}
\[
\Big(  \mathbb{D}_{\scriptscriptstyle\mathcal{S},\mathcal{S}'}\nabla\rho -\rho \Ub_{\scriptscriptstyle\mathcal{S},\mathcal{S}'}^1\Big)\cdot {\bf n}(\boldsymbol{\xi})=0, \quad \rm{on} \quad \partial \Omega.
\]
for the diffusive limit, whilst for the hyperbolic limit the corresponding boundary condition is 
\[
\Ub^0(\boldsymbol{\xi})\cdot {\bf n}=0, \quad \rm{on} \quad \partial \Omega.
\]
Two important classes of kinetic boundary conditions which satisfy \eqref{noflux} are the regular
reflection boundary operators and the non-local (in velocity) boundary operators of diffusive types defined by \cite{Palc} (see also the work by \cite{Lods}). Let us denote the boundary operator
\[
\mathcal{R}[p](\tau,\boldsymbol{\xi}, v,\hat{\vb})=p(\tau,\boldsymbol{\xi}, v',\hat{\vb}').
\]
Two main examples of the regular reflection boundary conditions are 
\begin{itemize}
\item bounce back reflection condition
\begin{equation}\label{bounce.back}
p(\tau,\boldsymbol{\xi}, v',\hat{\vb}')=p(\tau,\boldsymbol{\xi}, v,-\hat{\vb}) \quad \rm{if} \quad \hat{\vb}\cdot {\bf n}\ge 0
\end{equation} 
and $p(\tau,\boldsymbol{\xi}, v',\hat{\vb}')=p(\tau,\boldsymbol{\xi}, v,\hat{\vb})$ elsewhere.
\item specular reflection boundary condition
\begin{equation}
p(\tau,\boldsymbol{\xi}, v',\hat{\vb}')=p(\tau,\boldsymbol{\xi}, v, \hat{\vb}-(\hat{\vb}\cdot{\bf n}){\bf n})\quad \rm{if} \quad \hat{\vb}\cdot {\bf n}\ge 0
\end{equation}
and $p(\tau,\boldsymbol{\xi}, v',\hat{\vb}')=p(\tau,\boldsymbol{\xi}, v,\hat{\vb})$ elsewhere.
\end{itemize}
Diffusive boundary conditions are prescribed in the form
\[
p(\tau,\boldsymbol{\xi},v',\hat{\vb}')=\int_{ \hat{\vb}^*\cdot {\bf n}\ge 0} h(\boldsymbol{\xi},\vb_p,\vb_p^*)p(\tau,\boldsymbol{\xi},v^*,\hat{\vb}^*)|\vb^*\cdot {\bf n}|d\vb_p^* 
\]
where $h=h(\boldsymbol{\xi},\vb_p,\vb_p^*)$ is a non-negative measurable function called the Gaussian equilibrium function satisfying
\[
\int_{ \hat{\vb}^*\cdot {\bf n}\le 0 }h(\boldsymbol{\xi},\vb_p,\vb_p^*)|\vb^*\cdot {\bf n}|d\vb^*_p=1.
\]
A linear combination of a regular reflection with a diffusive boundary operator is
called a Maxwell-type boundary operator and reads
\begin{equation}\label{Maxwell}
p(\tau,\boldsymbol{\xi}, v',\hat{\vb}')=\alpha(\boldsymbol{\xi})p(\tau,\boldsymbol{\xi}, v,\mathcal{V}(\hat{\vb}))+(1-\alpha(\boldsymbol{\xi}))M(\boldsymbol{\xi},v,\hat{\vb})\int_{ \hat{\vb}^*\cdot {\bf n}\ge 0} p(\tau,\boldsymbol{\xi},v^*,\hat{\vb}^*)|\vb^*\cdot {\bf n}|d\vb_p^*,
\end{equation}
where $M(\boldsymbol{\xi},v,\hat{\vb})$ is the Maxwellian of the wall and $\mathcal{V}(\hat{\vb})=-\hat{\vb}$ for the bounce back reflection condition and $\mathcal{V}(\hat{\vb})=\hat{\vb}-(\hat{\vb}\cdot{\bf n}){\bf n}$ for the specular reflection.
We shall consider regular reflection boundary conditions for the one-dimensional simulations and Maxwell type boundary conditions for the two-dimensional simulations (see section \ref{Numerics}).
\section{Directional bias}
In this section we will specialize the above kinetic models and macrosopic limits to several situations of biological interest, that are listed in Table \ref{table} in order to guide the reader through them.
For sake of clarity, for the moment we will forget about the $\mathcal{S}'$ field, assuming that once the cell is polarized its speed is determined through a distribution function over the speed
$\psi(\x,v)$. 
The case in which the distribution function depends on a signaling cue $\mathcal{S}'$ will be denoted as double bias and will be treated in Section \ref{double.bias}.
Therefore, $\psi(\x,v)$ describes the random unbiased probability for a cell at position $\x$ of having modulus $v$ after a random re-orientation. 
In this case we expect a factorized turning probability, as the distributions for the (biased) direction and for the speed are independent.\\
We will apply the model to several non-local sensing dynamics, comparing the results with other existing models.  
\subsection{Directional turning probability} \label{dir.tur.prob}
Let us introduce the quantities
\begin{equation}\label{dir.transition}
B [\mathcal{S}](\x,\hat{\vb})=\displaystyle \int_{\mathbb{R}_+}\gamma_{\scriptscriptstyle \mathcal{S}}(\lambda) \mathcal{T}_{\lambda}^{\hat{\vb}}[\mathcal{S}](\x)  \, d\lambda \,,
\end{equation}
and
\begin{equation}
\bar B[\mathcal{S}](\x,\hat{\vb})=c_2(\x) B[\mathcal{S}](\x,\hat{\vb})\,,
\end{equation}
where $c_2(\x)=\dfrac{1}{\displaystyle \int_{\mathbb{S}^{d-1}} \int_{\mathbb{R}_+}\gamma_{\scriptscriptstyle \mathcal{S}}(\lambda) \mathcal{T}_{\lambda}^{\hat{\vb}}[\mathcal{S}](\x) d\lambda \, d\hat{\vb}}$ is the normalization constant.\\
In this case, the turning probability  can then be factorized as
\begin{equation}
T[\mathcal{S}](\x,\vb_p) = \psi(\x,v) \bar B[\mathcal{S}](\x,\hat{\vb})\,. 
\end{equation}
We also introduce
\begin{equation}
\bar{U}(\x)=\displaystyle\int_0^{U} \psi (\x,v) v \, dv\,,
\end{equation}
that is the mean speed and
\begin{equation}
D(\x)=\displaystyle\int_0^{U} \psi (\x,v) (v-\bar{U})^2 \, dv\,,
\end{equation}
that is the variance of the distribution function of the speed. 
The final expression of the operator $\mathcal{J}[p]$ then becomes
\begin{equation}\label{Jr_intermed}
\mathcal{J}[p](t,\x,\vb_p) = \mu (\x) \,  \Big( \rho(t,\x) \psi(\x,v)\bar B[\mathcal{S}](\x,\hat{\vb})  - p(t,\x,\vb_p) \Big) \,.
\end{equation}
The distribution which makes (\ref{Jr_intermed}) vanish corresponds to a macroscopic velocity
\begin{equation}
\Ub_{\scriptscriptstyle\mathcal{S}}(\x)=\bar{U}(\x)\displaystyle\int_{\mathbb{S}^{d-1}}\bar B[\mathcal{S}](\x,\hat{\vb})\hat{\vb} \, d\hat{\vb} \, 
\end{equation}
and to a diffusion tensor 
\begin{equation}
\mathbb{D}_{\scriptscriptstyle\mathcal{S}}(\x)=\displaystyle\int_{V_p}\psi(\x,v)\bar B[\mathcal{S}](\x,\hat{\vb})(\vb-\Ub_{\scriptscriptstyle\mathcal{S}})\otimes (\vb-\Ub_{\scriptscriptstyle\mathcal{S}}) \, d\vb_p.
\end{equation}
The expansion (\ref{T.expanded}) of $T$ reflects now on the expansion of $\bar B$
$$
\bar B[\mathcal{S}](\boldsymbol{\xi},\hat{\vb})= \bar B[\mathcal{S}]_0(\boldsymbol{\xi},\hat{\vb})+\epsilon \bar B[\mathcal{S}]_1(\boldsymbol{\xi},\hat{\vb})\,,
$$
with
\begin{equation}
\begin{array}{lr}
\Ub_{\mathcal{S}}^i(\boldsymbol{\xi})=\bar{U}(\boldsymbol{\xi})\displaystyle\int_{\mathbb{S}^{d-1}} \bar B [\mathcal{S}]_i(\boldsymbol{\xi}, \hat{\vb})\hat{\vb}  \, d\hat{\vb}\,,
\end{array}
\end{equation}
and
\begin{equation}\label{tensor}
\mathbb{D}_{\scriptscriptstyle\mathcal{S}}^i({\bf \boldsymbol{\xi}}) =\int_{V_p} \psi(\x,v) \bar B [\mathcal{S}]_i(\boldsymbol{\xi}, \hat{\vb})(\vb-\Ub_{\scriptscriptstyle\mathcal{S}}^i) \otimes (\vb-\Ub_{\scriptscriptstyle\mathcal{S}}^i) \, d\vb_p \, .
\end{equation}
We remark that if (\ref{even}) holds true ($\ie$, if $\Ub_{\scriptscriptstyle\mathcal{S}}^0$ vanishes)
\begin{equation}
\mathbb{D}_{\scriptscriptstyle\mathcal{S}}^0(\boldsymbol{\xi})=D(\boldsymbol{\xi})\displaystyle\int_{\mathbb{S}^{d-1}}\bar B_0[\mathcal{S}](\boldsymbol{\xi},\hat{\vb})\hat{\vb}\otimes \hat{\vb} \, d\hat{\vb}.
\end{equation}
The diffusive limit in the case in which the turning frequency does not depend on the microscopic velocity, $\ie$ Eq. \eqref{macro.diff}, now reads
\begin{equation}\label{macro.diff.mono}
\dfrac{\partial \rho}{\partial \tau}+\nabla \cdot \Big(\rho \Ub_{\scriptscriptstyle\mathcal{S}}^1 \Big) =\nabla \cdot \Big( \dfrac{1}{\mu} \nabla \cdot \big(\mathbb{D}_{\scriptscriptstyle\mathcal{S}}^0\rho \big)\Big)\, ,
\end{equation}
while the diffusive limit in the case in which $\mu_1$ depends on the microscopic velocity, $\ie$ Eq. \eqref{macro.diff.t}, now reads
\begin{equation}\label{macro.diff.t.bis}
\begin{array}{lr}
\dfrac{\partial \rho}{\partial \tau}+\nabla \cdot\Big(\rho \big(\Ub_{\scriptscriptstyle\mathcal{S}}^1
+\Ub_{\scriptscriptstyle\mathcal{S}}^{1,\mu_1}\big)\Big)= \nabla \cdot\Big( \dfrac{1}{\mu_0} \nabla \cdot \big(  \mathbb{D}_{\scriptscriptstyle\mathcal{S}}^0 \rho\big) \Big)\,, 
\end{array}
\end{equation}
where 
\begin{equation}\label{U1S}
\begin{array}{lr}
\Ub_{\scriptscriptstyle\mathcal{S}}^{1,\mu_1}
=\dfrac{1}{\mu_0}\displaystyle\int_{V_p}\mu_1(\vb_p)\psi(v)\bar B[\mathcal{S}]_0(\hat{\vb}) \vb d\vb_p,
\end{array}
\end{equation}
while the hyperbolic limit is
\begin{equation}\label{macro.drift.bis}
\dfrac{\partial \rho}{\partial \tau}+\nabla \cdot \big( \rho \Ub_{\scriptscriptstyle\mathcal{S}}^0 \big)=0 \, .
\end{equation}

\subsection{Examples}\label{examples}
 The function $\mathcal{T}^{\lambda}_{\hat{\vb}}$, introduced to account for a bias in the choice of the direction, has to be chosen according to the specific taxis process to be considered. In particular, as the cell scouts and detects the signal around itself, the functional $\mathcal{T}^{\lambda}_{\hat{\vb}}$ will depend on $\mathcal{S}$ through the quantity $\mathcal{S}(\x+\lambda \hat{\vb})$ as introduced by \cite{Othmer_Hillen.02} and later treated by \cite{Hillen_Painter_Schmeiser.06}, with $\lambda =R_{\scriptscriptstyle\mathcal{S}}$. For us, $\lambda \in [0,R_{\scriptscriptstyle\mathcal{S}}]$, where $R_{\scriptscriptstyle\mathcal{S}}$ accounts for the size of the sensing neighborhood and the information is weighted through $\gamma_{\scriptscriptstyle \mathcal{S}} (\lambda)$. This quantity will be considered in order to define for every $\x$ the probability of going in direction $\hat{\vb}$, which is $\bar B[\mathcal{S}(\x)](\hat{\vb})$. According to the characteristics of this probability, we will recover a macroscopic model, depending on the value of
\begin{equation}
\int_{\mathbb{S}^{d-1}} \bar B[\mathcal{S}]_0(\boldsymbol{\xi}, \hat{\vb}) \hat{\vb} d\hat{\vb}
\end{equation} 
which discriminates between a diffusion driven or a drift driven phenomena, according to whether (\ref{even}) is satisfied or not.

\subsubsection{Local sensing}\label{Local_sensing}
In the first example we consider a transition rate that depends only on the information given by the control factor at that site, $\ie , 
\gamma_{\scriptscriptstyle \mathcal{S}}(\lambda)=\delta (\lambda -0)$. 
In this case, the term $\bar B[\mathcal{S}]$ in the transition probability is simply
\begin{equation}
\bar B[\mathcal{S}](\x,\hat{\vb})=\dfrac{1}{|\mathbb{S}^{d-1}|} \, ,
\end{equation}
where $| \cdot |$ denotes the measure of the set.
Then the turning operator reads
\begin{equation}\label{Jlocal}
\mathcal{J}[p](t,\x,\vb_p) =  \mu (\x) \,  \Big( \dfrac{1}{|\mathbb{S}^{d-1}|} \rho(t, \x)  \psi(\x,v)- p(t,\x,\vb_p) \Big)  \, .
\end{equation}
This is what we expect, as the measure of $\mathcal{S}$ in $\x$ does not affect the choice of the direction. In this model, $\mathcal{S}(\x)$ may only affect the frequency of turning.
As (\ref{even}) is satisfied and $B[\mathcal{S}]_1=0$, we perform a diffusive limit. 
 In addition, since 
\begin{equation}
\int_{\mathbb{S}^{d-1}} \hat{\vb}\otimes \hat{\vb} \, d\hat{\vb} =\frac{|\mathbb{S}^{d-1}|}{d}\mathbb{I},
\end{equation}
$\mathbb{D}_{\scriptscriptstyle\mathcal{S}}({\bf \boldsymbol{\xi}})$ is isotropic
 \begin{equation}
 \mathbb{D}_{\scriptscriptstyle\mathcal{S}}(\boldsymbol{\xi})=D(\boldsymbol{\xi})\mathbb{I}, 
 \end{equation}
 and, if $\mu \sim \mathcal{O}(1)$, we have that 
\begin{equation}\label{difflocal}
\dfrac{\partial \rho}{\partial \tau}(\tau,{\bf \boldsymbol{\xi}}) =\nabla \cdot \Big(\dfrac{1}{\mu({\bf \boldsymbol{\xi}})} \nabla \big( D(\boldsymbol{\xi})\rho (\tau,{\bf \boldsymbol{\xi}})  \big) \Big)  \, .
\end{equation}
Therefore, there is no directional bias which comes from the transition probability, coherently with the fact that no non-local information is taken into account. 

\subsubsection{Encounter rate dependent taxis}\label{Encounterratedependenttaxis}
In order to consider situations in which organisms are too small to perform a non-local measurement, like E.Coli,
following the ideas by \cite{Berg_Block_Segall} and \cite{Othmer_Hillen.02}, we allow here the encounter rate to weakly depend on a signal  $S$, still  working with local models. For sake of simplicity,  
 we assume a time-independent signal, or however changes over a time scale much larger than the free-fly time.
In fact,  E.Coli can measure a pathwise gradient in time, $\ie$, $\dfrac{D \mathcal{S}}{Dt}=\dfrac{\partial}{\partial t}\mathcal{S}+\vb\cdot \nabla \mathcal{S}$. 
\cite{Berg_Block_Segall} show that the movement of E.Coli may be modelled with a Poisson process with turning frequency depending on the temporal gradient of $\mathcal{S}$
\begin{equation}\label{freq.weak}
\mu(\x,\vb_p)=\mu_r(\x)\exp\left[ -f(\mathcal{S})\left(\dfrac{\partial}{\partial t}\mathcal{S}+\vb\cdot \nabla \mathcal{S}\right)\right]
\end{equation}
being $\mu_r$ the turning rate in absence of the external signal.
The turning operator reads
\begin{equation}\label{Jturning}
\begin{array}{rl}
&\mathcal{J}[p](t,\x,\vb_p)=-\mu_r\exp\left[ -f(\mathcal{S})\left(\dfrac{\partial}{\partial t}\mathcal{S}+\vb\cdot \nabla \mathcal{S}\right)\right]p(t,\x,\vb_p)\\[12pt]
&+\dfrac{1}{|\mathbb{S}^{d-1}|}\psi(\boldsymbol{\xi}, v)\displaystyle\int_{V_p}\mu_r\exp\left[ -f(\mathcal{S})\left(\dfrac{\partial}{\partial t}\mathcal{S}+\vb'\cdot \nabla \mathcal{S}\right)\right] p(t,\x,\vb_p') d \vb_p' \,.
\end{array}
\end{equation}
For instance, one can take the function proposed by \cite{Lapidus} $\ie$ $f(\mathcal{S})=\dfrac{C_1 K_D}{(K_D+\mathcal{S})^2}$ where $K_D$ is the dissociation constant for the attractant. 
Hence, in the spirit of the work by \cite{Othmer_Hillen.02} by a parabolic scaling, we can expand (\ref{freq.weak}) as 
$$
\mu(\boldsymbol{\xi})=\mu_r(\boldsymbol{\xi})\Big( 1-\epsilon f(\mathcal{S}) \vb\cdot \nabla \mathcal{S} +\mathcal{O}(\epsilon^2) \Big)\,,
$$
and set
\begin{equation}
\mu_0(\boldsymbol{\xi})=\mu_r(\boldsymbol{\xi})\,,
\end{equation}
and
\begin{equation}
\mu_1(\boldsymbol{\xi},\vb_p)=-\mu_r(\boldsymbol{\xi}) f(\mathcal{S})\vb\cdot \nabla \mathcal{S}.
\end{equation}
We then obtain $\Ub_{\mathcal{S}}^1={\bf 0}$ because $B[\mathcal{S}]_1 =0$ and, recalling \eqref{U1S},
$$\Ub_{\mathcal{S}}^{1,\mu_1}=-\frac{-\mu_r}{\mu_r} f(\mathcal{S})\frac{1}{|\mathbb{S}^{d-1}|}\int_{0}^U\psi(\boldsymbol{\xi},v)vdv\int_{\mathbb{S}^{d-1}}\hat{\vb}\otimes \hat{\vb} d\hat{\vb}\,\nabla \mathcal{S}= \frac{\bar{U}(\boldsymbol{\xi})f(\mathcal{S})}{d} \nabla \mathcal{S}(\boldsymbol{\xi})$$
and the macroscopic equation reads
\begin{equation}\label{rate.dep}
\begin{array}{lr}
\dfrac{\partial \rho}{\partial \tau}(\tau,{\bf \boldsymbol{\xi}}) + \nabla \cdot \Big(\rho (\tau,\boldsymbol{\xi})\frac{\bar{U}(\boldsymbol{\xi})f(\mathcal{S})}{d} \nabla \mathcal{S}(\boldsymbol{\xi}) \Big) =\nabla \cdot \Big(\dfrac{1}{\mu_r({\bf \boldsymbol{\xi}})} \nabla \big( D(\boldsymbol{\xi})\rho (\tau,{\bf \boldsymbol{\xi}}) \big)\Big).
\end{array}
\end{equation}
that is a chemotaxis model also in the case of local sensing, when an organism is too small for measuring a gradient, but it is able to measure the temporal gradient of $\mathcal{S}$ along its path. 

\subsubsection{Non-local sensing average}\label{Nonlocal_sensing}
We now consider the case in which the choice of the new velocity depends on a suitable average of the value of the signal $\mathcal{S}$ perceived through transmembrane receptors by a cell that extending its protrusions can scout its neighborhood.\\
We shall consider in this case
$$
\mathcal{T}_{\lambda}^{\hat{\vb}}[\mathcal{S}](\x)=b\big(\mathcal{S}(\x+\lambda \hat{\vb})\big)
$$
that models the fact that in order to decide its new direction, the cell measures $\mathcal{S}$ in $\x+\lambda \hat{\vb}$ and evaluates an average of a quantity which is linked to this measure ($b$).
Therefore, recalling \eqref{Jr_intermed} the operator for biased random motion including taxis is
\begin{equation}\label{J_r}
\begin{array}{lr}
\mathcal{J}[p](t,\x,\vb_p)=\\[12pt]
\phantom{a}\mu(\x) \,  \left( \rho(t,\x)  \psi(\x,v) c(\x) \displaystyle{\int_{\mathbb{R}_+}} b(\mathcal{S}(\x+\lambda \hat{\vb})) \gamma_{\scriptscriptstyle \mathcal{S}} (\lambda) \, d\lambda \, - p(t,\x,\vb_p) \right) \hspace{-2pt} \, ,
\end{array}
\end{equation}
if the turning frequency does not depend on the microscopic velocity.
In order to understand what to expect from the integration of (\ref{J_r}) that will be performed in Section \ref{numerics}, we can look at the asymptotic limit.\\
For a general $b$ and $R_{\scriptscriptstyle\mathcal{S}}$, the macroscopic velocity of order zero is
\begin{equation}\label{solv.barrier.b}
\Ub_{\mathcal{S}}^0(\boldsymbol{\xi})=\bar{U}(\boldsymbol{\xi})\dfrac{\displaystyle\int_{\mathbb{S}^{d-1}}  \left( \int_{\mathbb{R}^+} b(\mathcal{S}(\boldsymbol{\xi}+\lambda \hat{\vb}))\gamma_{\scriptscriptstyle \mathcal{S}} (\lambda)  \, d\lambda \right) \, \hat{\vb} \, d\hat{\vb}}{\displaystyle\int_{\mathbb{S}^{d-1}}  \left( \int_{\mathbb{R}^+} b(\mathcal{S}(\boldsymbol{\xi}+\lambda \hat{\vb}))\gamma_{\scriptscriptstyle \mathcal{S}} (\lambda)  \, d\lambda \right) d\hat{\vb}}.
\end{equation}
As, in this case, (\ref{even}) is generally not satisfied, we have a non-zero drift term and a hyperbolic limit needs be performed, leading to the integro-differential equation
\begin{equation}\label{Drift.adv.mono}
\dfrac{\partial \rho}{\partial \tau}+\nabla \cdot \big( \rho \Ub_{\mathcal{S}}^0 \big)=0 \, .
\end{equation}
In particular, if $b(\mathcal{S})=\mathcal{S}$,
this drift term measures the dominant direction in the extracellular factor in the neighbourhood of the cell as
\begin{equation}
\Ub_{\mathcal{S}}^0(\boldsymbol{\xi})=\bar{U}(\boldsymbol{\xi})\dfrac{\displaystyle\int_{\mathbb{S}^{d-1}} \displaystyle \left( \int_{\mathbb{R}^+} \mathcal{S}(\boldsymbol{\xi}+\lambda \hat{\vb}))\gamma_{\scriptscriptstyle \mathcal{S}} (\lambda)  \, d\lambda \right) \, \hat{\vb} \, d\hat{\vb}}{\displaystyle\int_{\mathbb{S}^{d-1}}  \left( \int_{\mathbb{R}^+} \mathcal{S}(\boldsymbol{\xi}+\lambda \hat{\vb})\gamma_{\scriptscriptstyle \mathcal{S}} (\lambda)  \, d\lambda \right)  d\hat{\vb}}.
\end{equation}
However, it is possible to simplify the turning operator \eqref{J_r}, if $R_{\scriptscriptstyle\mathcal{S}}$ is much smaller than the characteristic length $l_{\mathcal{S}}$ of variation of $\mathcal{S}$, $\eg$, $l_{\mathcal{S}}=1/\max \frac{|\nabla \mathcal{S}|}{\mathcal{S}}$. In fact, we can expand $b$ in a Taylor series 
\begin{equation}\label{bapprox}
b(\mathcal{S}(\x+\lambda \hat{\vb}))=b(\mathcal{S}(\x))+\lambda b'(\mathcal{S}(\x))\nabla \mathcal{S}(\x)\cdot \hat{\vb} +\mathcal{O}(\lambda ^2).
\end{equation}
Therefore, if $b(\mathcal{S}(\x))$ does not vanish, the operator may be approximated by
\begin{equation}\label{Bbar.grad.b}
\bar B[\mathcal{S}](\x,\hat{\vb})=\dfrac{1}{|\mathbb{S}^{d-1}|} \left[ 1+\Lambda\dfrac{b'(\mathcal{S}(\x))}{b(\mathcal{S}(\x))} \nabla \mathcal{S}(\x)\cdot \hat{\vb}\right]
\end{equation}
where
\begin{equation}
\Lambda=\dfrac{\displaystyle \int_{\mathbb{R}_+} \gamma_{\scriptscriptstyle \mathcal{S}} (\lambda) \lambda d \lambda}{\displaystyle \int_{\mathbb{R}_+} \gamma_{\scriptscriptstyle \mathcal{S}} (\lambda)  d \lambda}.
\end{equation}
For instance, if only the signals at a membrane having a distance $R_{\scriptscriptstyle\mathcal{S}}$ from $\x$ is taken into account, $\ie$ $\gamma_{\scriptscriptstyle \mathcal{S}}(\lambda)=\delta (\lambda -R_{\scriptscriptstyle\mathcal{S}})$, then 
$\Lambda=R_{\scriptscriptstyle\mathcal{S}}$.
If instead all the signals between the cell center and its membrane are uniformly mediated, $\ie$ $\gamma_{\scriptscriptstyle \mathcal{S}}(\lambda)=H (R_{\scriptscriptstyle\mathcal{S}}-\lambda)$, then $\Lambda=\dfrac{R_{\scriptscriptstyle\mathcal{S}}}{2}$. \\
In this case, the turning operator is
\begin{equation}\label{J_r.loc}
\mathcal{J}[p](t,\x,\vb_p)=\mu(\x)   \left[ \rho(t,\x)  \dfrac{\psi(\x,v)}{|\mathbb{S}^{d-1}|} \Big( 1+\Lambda\dfrac{b'(\mathcal{S}(\x))}{b(\mathcal{S}(\x))} \nabla \mathcal{S}(\x)\cdot \hat{\vb}\Big)- p(t,\x,\vb_p) \right] \hspace{-2pt} \, .
\end{equation}
We remark that the smallness of $R_{\scriptscriptstyle\mathcal{S}}$ and the approximation \eqref{bapprox} localizes   the non-local integral model (\ref{J_r}) into \eqref{J_r.loc}. Under these limit assumptions it is possible to perform a parabolic scaling  
that gives 
\begin{equation}\label{BbarS}
\bar B[\mathcal{S}](\boldsymbol{\xi}, \hat{\vb})=\dfrac{1}{|\mathbb{S}^{d-1}|}\left( 1 + \epsilon\Lambda \dfrac{b'(\mathcal{S}(\boldsymbol{\xi}))}{b(\mathcal{S}(\boldsymbol{\xi}))}\nabla \mathcal{S}(\boldsymbol{\xi})\cdot \hat{\vb}\right)  \, .  
\end{equation}
Hence, $\bar B[\mathcal{S}]_0(\boldsymbol{\xi}, \hat{\vb})=\dfrac{1}{|\mathbb{S}^{d-1}|}$ and $\bar B[\mathcal{S}]_1(\boldsymbol{\xi}, \hat{\vb})=\dfrac{1}{|\mathbb{S}^{d-1}|}\Lambda \dfrac{b'(\mathcal{S}(\boldsymbol{\xi}))}{b(\mathcal{S}(\boldsymbol{\xi}))}\nabla \mathcal{S}(\boldsymbol{\xi})\cdot \hat{\vb}$.
In this case  recalling (\ref{macro.diff.mono}) we obtain
\begin{equation}\label{diff.adv.mono}
\dfrac{\partial \rho}{\partial \tau}(\tau,{\bf \boldsymbol{\xi}})+\nabla \cdot \Big(\rho(\tau,\boldsymbol{\xi}) \chi(S(\boldsymbol{\xi}))\nabla \mathcal{S}(\boldsymbol{\xi}) \Big)  =\nabla \cdot \Big( \dfrac{1}{\mu({\bf \boldsymbol{\xi}})}\nabla \big(D({\bf \boldsymbol{\xi}})\rho(\tau,\boldsymbol{\xi})\big)\Big)  \, ,
\end{equation}
where 
\begin{equation}\label{chinonlocalsensing}
\chi(S(\boldsymbol{\xi}))=\frac{\Lambda\bar{U}(\boldsymbol{\xi}) }{d}\dfrac{b'(\mathcal{S}(\boldsymbol{\xi}))}{b(\mathcal{S}(\boldsymbol{\xi}))}\,.
\end{equation}
So, the chemotactic sensitivity $\chi$ takes into account of the kinetic response through $\bar{U}$ and of the sensing capability through $\gamma_{\scriptscriptstyle \mathcal{S}}(\lambda)$ contained in $\Lambda$. More importantly, different signal-dependent sensitivity models can be obtained according to the choice of $b$. 
Viceversa, any relation on the chemotactic sensitivity can be obtained by a proper $b$ given by
$$b(\mathcal{S})=\exp\left[\dfrac{d}{\Lambda\bar{U}}\int \chi (\mathcal{S})\,d\mathcal{S}\right]\,.$$
Trivially, if $b$ is independent of $\mathcal{S}$, one has no chemotaxis. If $b$ is proportional to $\mathcal{S}$ 
(and $\mathcal{S} \ne 0$, always), then 
\begin{equation}\label{chemo.sensitivity}
\chi(S(\boldsymbol{\xi}))=\Lambda \frac{\bar{U}(\boldsymbol{\xi})}{d\mathcal{S}(\boldsymbol{\xi})} \, .
\end{equation}

Let us now introduce the parameter
\begin{equation}\label{eta}
\eta=\dfrac{R_{\scriptscriptstyle \mathcal{S}}}{l_{\scriptscriptstyle \mathcal{S}}}.
\end{equation}
We have seen that if the sensing radius is smaller then the characteristic length of variation of the chemotactic field, $\ie \, \eta \ll 1$, then \eqref{J_r} can be simplified to \eqref{J_r.loc}, whereas this is not possible if the sensing radius is larger then $l_{\scriptscriptstyle \mathcal{S}}, \ie \, \eta \gg 1$. This leads to different macroscopic limits \eqref{diff.adv.mono} and \eqref{Drift.adv.mono} respectively. This different macroscopic behavior and the proper choice of the scaling may be justified thanks to a nondimensionalization argument.  
We shall rescale the variables as
$$
\boldsymbol{\xi}=\dfrac{\x}{l_{\scriptscriptstyle \mathcal{S}}}, \quad \tilde{\vb}_p=\dfrac{\vb_p}{U_{ref}}, \quad\tau=\dfrac{t}{\sigma_t}, 
$$
where we shall choose
\[
U_{ref}=R_{\scriptscriptstyle \mathcal{S}}\bar{\mu}
\]
being $\bar \mu$ a reference frequency.
The time scale $\sigma_t$ can be chosen as a diffusion time $T_{Diff}$ scale or a drift time scale $T_{Drift}$. In general we may write \citep{Othmer_Hillen.00}
$$
T_{Diff}=\dfrac{l_{\scriptscriptstyle \mathcal{S}}^2}{D},\quad D=\dfrac{U_{ref}^2}{\bar{\mu}}, \qquad T_{Drift}=\dfrac{l_{\scriptscriptstyle \mathcal{S}}}{U_{ref}}
$$
The regime is diffusive - and we will choose $\sigma_t=T_{Diff}$ - if the frequency $\bar{\mu}$ is very large with respect to the reference time scale $\sigma_t$, $\ie$ if we can find a small parameter $\epsilon$ such that 
\[
\bar{\mu}=\dfrac{\epsilon^{-2}}{\sigma_t}.
\]
The latter is equivalent to
$$
l_{\scriptscriptstyle \mathcal{S}}=\mathcal{O}\left(\dfrac{U_{ref}}{\epsilon}\right)
$$
which implies the hierarchy
\begin{equation}
T_{Drift} \ll T_{Diff}. 
\end{equation}\label{hierarhy}
In the present case this is equivalent to 
\[
\eta=\dfrac{R_{\scriptscriptstyle \mathcal{S}}}{l_{\scriptscriptstyle \mathcal{S}}} <1.
\]
On the other hand, the macroscopic regime is hyperbolic, and we choose a faster time scale if
\[
\bar{\mu}=\dfrac{\epsilon^{-1}}{\sigma}.
\]
In this case  the appropriate choice will be 
\[
\sigma_t=T_{Drift}=\dfrac{l_{\scriptscriptstyle \mathcal{S}}}{U_{ref}}=\dfrac{\eta}{\mu}
\]
as the hierarchy \eqref{hierarhy} does not hold anymore. Hence, the following relation holds
\[
\eta \sim \dfrac{1}{\epsilon} \gg 1.
\]
With respect to \eqref{chemo.sensitivity}, different chemotactic coefficients can be obtained by different choices of the bias function $b(\mathcal{S})$: for instance, if $b(\mathcal{S})=k+\mathcal{S}^n$ with $k>0$, then we have a saturating chemotactic sensitivity
\begin{equation}\label{saturatingsensitivity}
\chi(\mathcal{S}(\boldsymbol{\xi}))=\frac{\Lambda\bar{U}(\boldsymbol{\xi})}{d} \frac{S^{n-1}(\boldsymbol{\xi})}{k+\mathcal{S}^n(\boldsymbol{\xi})}.
\end{equation} 
Another example is the receptor binding process. In this case we may consider a saturating dependence on the signal concentration 
\begin{equation}\label{b.s}
b\big(\mathcal{S}(\x + \lambda\hat{\vb})\big) = \dfrac{\mathcal{S}^n(\x + \lambda\hat{\vb})}{k + \mathcal{S}^n(\x +\lambda \hat{\vb})} \,,
\end{equation}
yielding
\begin{equation}
\chi(\mathcal{S}(\boldsymbol{\xi}))=\frac{\Lambda\bar{U}(\boldsymbol{\xi})}{d} \frac{kn}{\mathcal{S}(\boldsymbol{\xi})[k+\mathcal{S}^n(\boldsymbol{\xi})]}.
\end{equation} 
We notice that in this case, if we choose $n=1$ and $\gamma_{\scriptscriptstyle \mathcal{S}}(\lambda)=\delta(\lambda-R_{\scriptscriptstyle\mathcal{S}})$, $\ie$, a membrane sensing, we recover the model proposed by \cite{Hillen_Painter_Schmeiser.06}. \\
Generally speaking, $b$  increasing with $\mathcal{S}$ will give rise to chemoattraction, while $b$  decreasing with $\mathcal{S}$ will lead to a repulsive effect.  For instance, if 
$$b(\mathcal{S}(\x + \lambda\hat{\vb}))=\frac{1}{k+\mathcal{S}^n(\x +\lambda \hat{\vb})}\,,$$ 
we have \eqref{saturatingsensitivity} but with the opposite sign, $\ie$, alignment in the opposite direction with respect to the gradient of $\mathcal{S}$.

At this stage we did not specify whether $\mathcal{S}$ represents a diffusing or a matrix bound chemical. Even more, one can deal with {\it durotaxis} in a completely analogous way. In this case the signal $\mathcal{S}$ is the perceived stiffness of the ECM and  the sensing of the mechanical properties of the ECM around the cell will induce motion  toward stiffer region of the ECM if $b$ is an increasing function of $\mathcal{S}$, as we shall see as last application in Section \ref{numerics}.

\subsubsection{Comparative sensing}\label{Comparative}

In some cases the signals perceived by the cell in differently localized receptors on its membrane is amplified by a polarization dynamics involving PTEN and the phosforillation of PIP2 into PIP3 \citep{Ambrosi.04}. This causes the formation of a ``head" and a ``tail" in the cell that chooses the direction accordingly. In order to mimick this phenomenon, we assume here that the turning rate depends on what is sensed in $\x+\lambda \hat{\vb}$ and in $\x-\lambda \hat{\vb}$, $\ie$,
\begin{equation}
\mathcal{T}_{\hat{\vb}}^{\lambda}[\mathcal{S}(\x)] = \alpha + \beta b\big(\mathcal{S}(\x + \lambda\hat{\vb}),\mathcal{S}(\x-\lambda \hat{\vb})\big)\,, \quad \alpha >\beta >0
\end{equation}
where $b\in(-1,1)$ is a quantity comparing the values $\mathcal{S}(\x-\lambda\hat{\vb})$ and $\mathcal{S}(\x + \lambda\hat{\vb})$ in a way that keeps $\mathcal{T}_{\hat{\vb}}^{\lambda}$ always positive.
Hence,
\begin{equation}
\begin{array}{lr}\label{Jcomparative}
\mathcal{J}[p](t,\x,\vb_p)=\\[12pt]
\mu(\x) \,  \left[ \rho(t,\x)  \psi(\x,v) c(\x)  \left(\alpha \Lambda_0 + \beta 
\displaystyle{\int_{\mathbb{R}_+}} b\big(\mathcal{S}(\x + \lambda\hat{\vb}),\mathcal{S}(\x-\lambda \hat{\vb})\big) \gamma_{\scriptscriptstyle \mathcal{S}} (\lambda) \, d\lambda\right) - p(t,\x,\vb_p) \right] \hspace{-2pt} \, ,
\end{array}
\end{equation}
being $\Lambda_0=\displaystyle \int_{\mathbb{R}^+} \gamma_{\scriptscriptstyle \mathcal{S}} (\lambda) d\lambda$.
In the attractive case, the cell is most likely to go where there is a larger concentration of chemical, and therefore we might take
\begin{equation}\label{b.grad.attr}
b\big(\mathcal{S}(\x+\lambda\hat{\vb}),\mathcal{S}(\x - \lambda\hat{\vb})\big) = \dfrac{\mathcal{S}(\x+\lambda \hat{\vb}) -\mathcal{S}(\x-\lambda \hat{\vb})}{2k+\mathcal{S}(\x+\lambda \hat{\vb}) +\mathcal{S}(\x-\lambda \hat{\vb})}\,.
\end{equation}
On the other hand, in the repulsive case, the cell tends to go where there is a smaller concentration of chemical, and therefore we might take
\begin{equation}\label{b.grad.rep}
b\big(\mathcal{S}(\x+\lambda\hat{\vb}),\mathcal{S}(\x - \lambda\hat{\vb})\big) =  \dfrac{\mathcal{S}(\x-\lambda \hat{\vb}) -\mathcal{S}(\x+\lambda \hat{\vb})}{2k+\mathcal{S}(\x+\lambda \hat{\vb}) +\mathcal{S}(\x-\lambda \hat{\vb})}\,.
\end{equation}
As done in the previous section, if we can assume that the size of the neighborhood providing information through signaling is small,   one can perform a Taylor expansion of the function $\mathcal{T}_{\lambda}^{\hat{\vb}}$ in $\lambda=0$ and write
\begin{equation}\label{F.expansion}
\mathcal{T}_{\lambda}^{\hat{\vb}}= \alpha +\beta \dfrac{\lambda\nabla \mathcal{S}(\x)\cdot \hat{\vb}}{k+\mathcal{S}(\x)} \,
\end{equation}
in the attractive case (the repulsive case is similar with a minus sign, or equivalently a negative $\beta$).
The biased transition probability becomes
\begin{equation}\label{B.bar.calculated.cell}
\bar B[\mathcal{S}](\boldsymbol{\xi}, \hat{\vb}) = \dfrac{1}{|\mathbb{S}^{d-1}|}\left( 1 + \frac{\beta}{\alpha}\Lambda \dfrac{\nabla \mathcal{S}(\x)\cdot \hat{\vb}}{k+\mathcal{S}(\x)}\right)\,,
\end{equation}
that for a logarithmic dependence of $b$ from $\mathcal{S}$ has the same structure as \eqref{BbarS}.
In this limit the operator for biased random motion including taxis becomes
\begin{equation}\label{comp.sens}
\mathcal{J}[p]=\mu \,  \left( \rho(t,\x)\psi(\x,v) \dfrac{1}{|\mathbb{S}^{d-1}|}\left( 1 + \dfrac{\beta}{\alpha}\Lambda
\dfrac{\nabla \mathcal{S}(\x)\cdot \hat{\vb}}{k+\mathcal{S}(\x)}\right)  - p(t,\x,\vb) \right) \hspace{-2pt} \, .
\end{equation}
 Equation (\ref{B.bar.calculated.cell}) satisfies (\ref{even}) and, therefore, a diffusive limit can be performed. As, $\bar B[\mathcal{S}]_0(\boldsymbol{\xi}, \hat{\vb})=\psi({\bf \boldsymbol{\xi}};v)\dfrac{1}{|\mathbb{S}^{d-1}|}$ and $\bar B[\mathcal{S}]_1(\boldsymbol{\xi}, \hat{\vb})=\dfrac{\beta}{\alpha}\Lambda \dfrac{\nabla \mathcal{S}(\x)\cdot \hat{\vb}}{k+\mathcal{S}(\x)}$, 
we get a drift-diffusion equation \eqref{diff.adv.mono} with 
\begin{equation}\label{chicomparativesensing}
\chi(\mathcal{S})=\frac{\beta\bar{U}\Lambda}{d\alpha} \,\dfrac{1}{ k+\mathcal{S}} \, .
\end{equation}

Even if in the limit of small sensing radii the comparative sensing is almost the same as the non-local sensing introduced in the previous section (in the sense that the diffusive limit are similar), it allows to add some details. In fact, the ratio of the coefficients $\beta$ and $\alpha$ measures the different weight of the diffusive and the advective (chemotactic) contributions. Further, we may include in $\beta$ a dependence on other substances like $\mathcal{S}$ itself or auto-inducers to model quorum sensing. For example in the case of the cellular
slime mold Dictyostelium discoideum, the response to the auto-inducer Netrin-1 may be attractive in high concentrations of cAMP, whilst it may be decreasing in case of low levels of cAMP \citep{Deery}. \cite{Painter_Hillen.02} consider a response in the form $\beta (w) =1-\frac{w}{w^*}$, being $w^*$ a saturation level. 
In addition, when the sensing radius is not small the kinetic models are different, reflecting the fact that the mechanisms are fundamentally different, with the comparative sensing giving rise to a stronger response, especially when considering that the reception of the signal can be amplified by feedback mechanisms arising from the activation of proper protein cascades inside the cells, giving rise, for instance to the segregation of PIP2 and PIP3 in different parts of the cell.

\subsection{Numerical simulations}\label{numerics}
We simulate the kinetic model with turning operator given by \eqref{J_r} with $b(\mathcal{S})=\mathcal{S}$ both in one and two dimensions. In order to do that, we use the numerical scheme proposed by \cite{Filbet} in which  a kinetic model for chemotaxis is simulated in two-dimensions using a van Leer scheme for the space transport. 
\subsubsection{Numerical resolution of the kinetic model}\label{Numerics}
We consider a computational domain that will be in the form $[x_{min},x_{max}]\times[0,U]$ in the one dimensional case and $[x_{min},x_{max}]\times[y_{min},y_{max}]\times V_p$ in the two dimensional case.
The computational domain is discretized with a Cartesian mesh ${\bf X}_h\times {\bf V}_{p_h}$, where ${\bf X}_h$ and ${\bf V}_{p_h}$ are defined by (in two dimensions)
\[
\begin{cases}
{\bf X}_{h} =\lbrace \x_{i,j}=(x_i,y_j)=(x_{min}+i\Delta x, y_{min}+j\Delta y), \quad 0 \le i \le n_x, \quad 0 \le j \le n_y \rbrace \\[20pt]
{\bf V}_{p_h} =\lbrace \vb_{l,k}=v_k(\cos \theta_l,\sin \theta_l),\quad \theta_l=(l+1/2)\Delta \theta, 
\quad 0\le l\le n_{ang}-1, v_k=v_0+k\Delta v,\quad 0 \le k \le n_v \rbrace
\end{cases}
\] 
where $\Delta x=\dfrac{x_{max}-x_{min}}{n_y}, \Delta y=\dfrac{y_{max}-y_{min}}{n_y}, \Delta v=\dfrac{U}{n_v}, \Delta \theta =\dfrac{2\pi}{n_{ang}}$.
Denoting by $p^n_{i,j,l,k}$ an approximation of the distribution function $p(t^n,\x_{i,j},\vb_{p_{l,k}})$, where $\vb_{p_{l,k}}=(v_k,\hat{\vb}_l)$.
We introduce the first order splitting
\[
\begin{cases}
\dfrac{p_{i,j,l,k}^{n+1/2}-p_{i,j,l,k}^n}{\Delta t}+\vb \cdot \nabla_{\x,h}p_{i,j,l,k}^n=0 \qquad \rm{(transport \quad step)}\\[20pt]
\dfrac{p_{i,j,l,k}^{n+1}-p_{i,j,l,k}^{n+1/2}}{\Delta t}=\mu\Big(\rho_{i,j}^{n+1}T[\mathcal{S},\mathcal{S}']_{i,j,l,k}-p_{i,j,l,k}^{n+1}\Big) \qquad \rm{(relaxation \quad step)}
\end{cases}
\]
where $h=(\Delta t, \Delta x, \Delta y)$, $\vb \cdot \nabla_{\x,h}p_{i,j,l,k}^n$ is an approximation of the transport operator $\vb \cdot \nabla p$ computed with a Van Leer scheme. It is a high resolution monotone, conservative scheme which is second order if the solution is smooth and first order near the shocks. $T[\mathcal{S},\mathcal{S'}]_{i,j,l,k}$ is the discretization of the transition probability $T$. We observe that as the turning operator preserves mass and the turning probability is known and does not depend on $p$, the relaxation step may be implicit and we may consider the density at time $n+1$. In particular the density is computed by using a trapezoidal rule
\[
\rho_{i,j}^n=\Delta v \Delta \theta\sum_{k=0}^{n_v}\sum_{l=0}^{n_{ang}-1}p_{i,j,l,k}^n.
\]
Concerning boundary conditions, in the one dimensional case we consider regular reflecting conditions. In one dimension, in a domain $[0,L]$, the bounce-back and the specular reflection boundary conditions coincide, that is $p(t,x=0,v)=p(t,x=0,-v)$ and $p(t,x=L,-v)=p(t,x=L,v)$. We do not consider Maxwell type conditions as only the outgoing speed would be affected.
In the two dimensional case, the regular reflection is biologically unrealistic, as cells do not bounce back nor they collide with the wall as hard spheres. Therefore, Maxwell type boundary conditions are more realistic, and we shall consider for the Maxwellian to the wall
\[
M(\x,\vb_p)=T[\mathcal{S},\mathcal{S}'](\x,\vb_p),
\]
being $T$ the asymptotic equilibrium of the system with this class (no-flux) of boundary conditions.

\paragraph{Chemotactic motion}\label{sim_chemo}
\hspace{-2pt}

\begin{figure}[htbp!]
\subfigure[]{\includegraphics[width=0.47\textwidth]{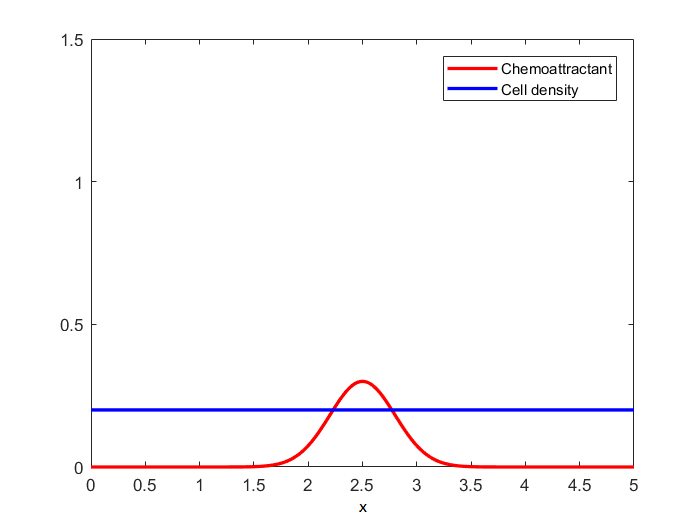}}
\subfigure[]{\includegraphics[width=0.47\textwidth]{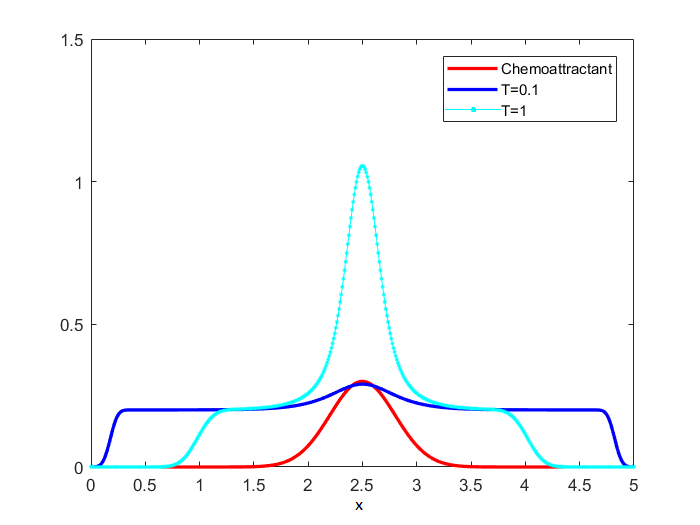}}
\centering
\subfigure[]{\includegraphics[scale=0.5]{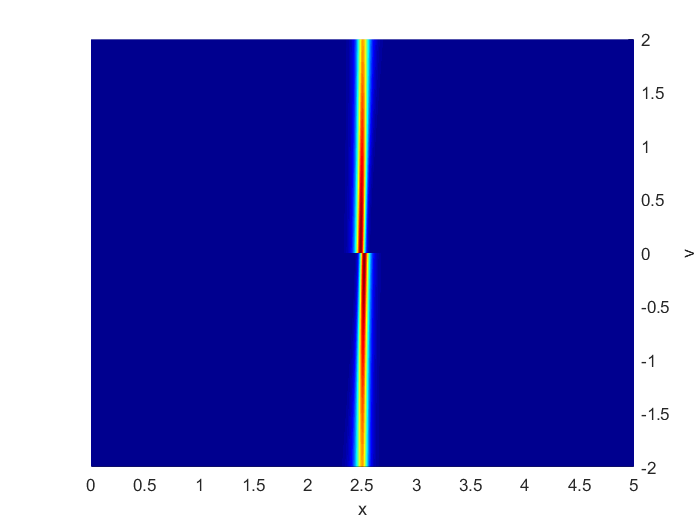}}

\caption{Evolution of a constant macroscopic cell density in presence of a gaussian distribution of chemoattractant ((a) and (b)). The chemotactic sensitivity is $b(\mathcal{S})=\mathcal{S}$, $\psi(v)=\dfrac{1}{U}, U=2$, and $\mu=50$. The sensing radius is $R_{\scriptscriptstyle \mathcal{S}}=0.2$ and $\gamma_{\scriptscriptstyle \mathcal{S}}=\delta(\lambda-R_{\scriptscriptstyle \mathcal{S}})$. The stationary distribution function is given in (c), while the stationary density in Fig. \ref{mono.simu.1bis} (a).}
\label{mono.simu.1}
\end{figure}

\begin{figure}
\subfigure[]{\includegraphics[width=0.47\textwidth]{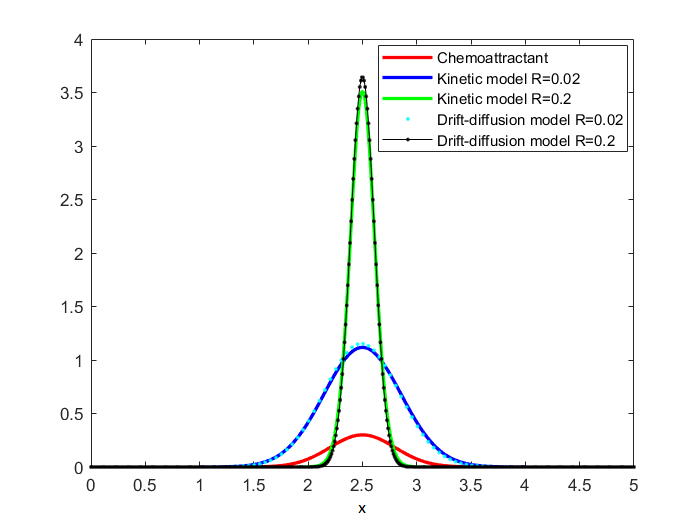}}
\subfigure[]{\includegraphics[width=0.47\textwidth]{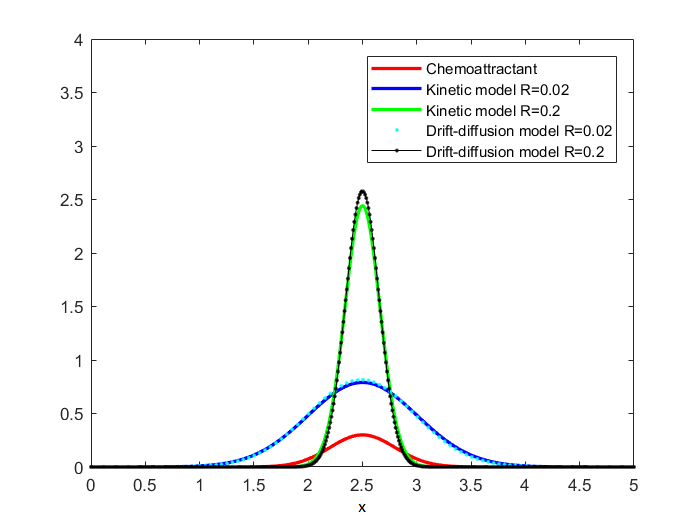}}

\subfigure[]{\includegraphics[width=1.0\textwidth]{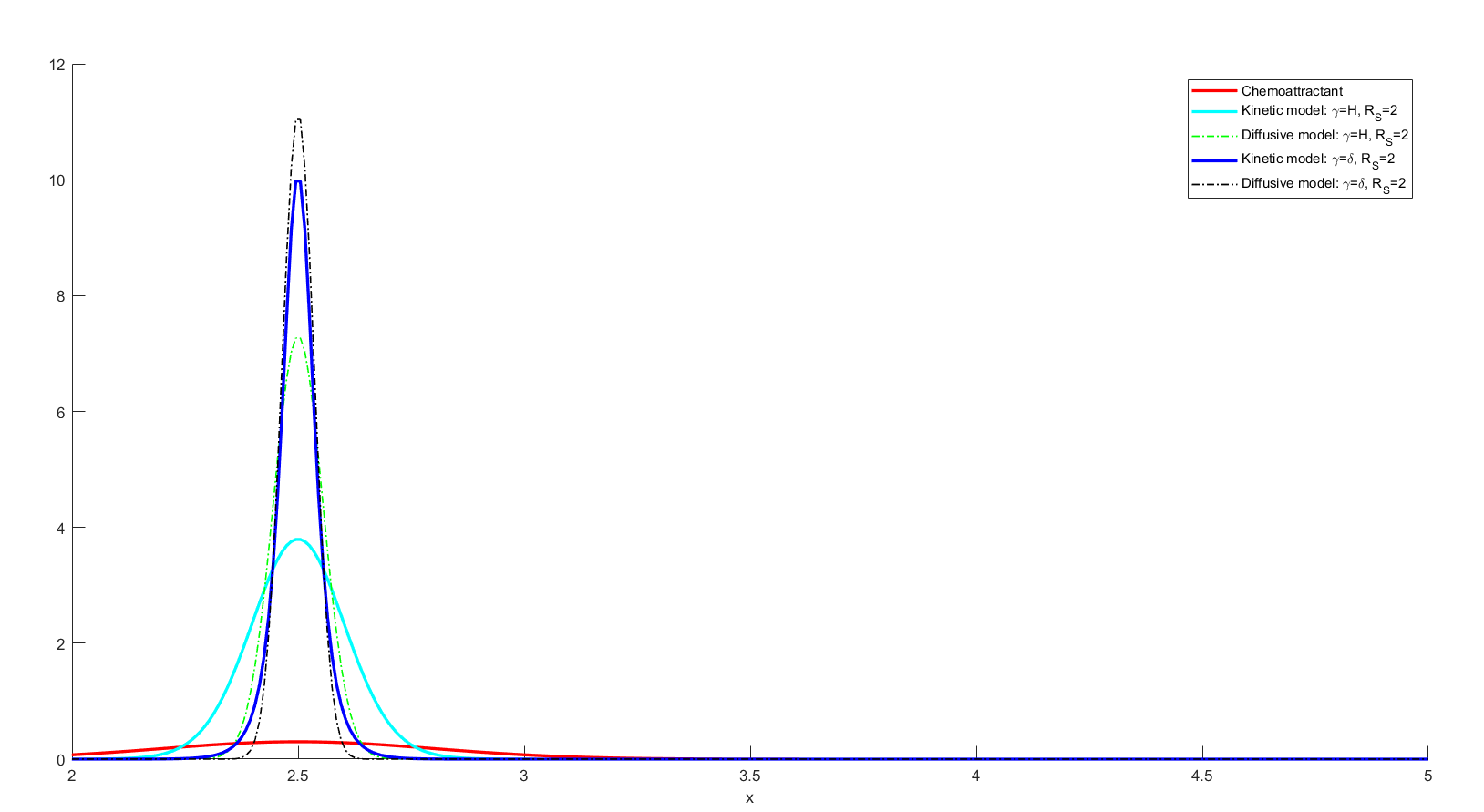}}

\caption{Comparison of the stationary density of the kinetic model with the stationary solution \eqref{stat.state} of the diffusion limit \eqref{diff.adv.mono} for (a) $\gamma_{\scriptscriptstyle \mathcal{S}}=\delta(\lambda-R_{\scriptscriptstyle \mathcal{S}})$ and (b) $\gamma_{\scriptscriptstyle \mathcal{S}}=H(\lambda-R_{\scriptscriptstyle \mathcal{S}})$. In (c) the solutions for $R_{\scriptscriptstyle \mathcal{S}}=2$ are given showing  discrepancies between the density of the solution of the kinetic model and of the diffusive limit (\ref{diff.adv.mono}) obtained when $\eta\ll 1$.}
\label{mono.simu.1bis}
\end{figure}

\noindent In the first simulation we consider a fixed gaussian distribution of chemoattractant 
$$\mathcal{S}(x)=0.3\exp\left[\dfrac{(x-2.5)^2}{0.3}\right]\,,$$
and a constant initial condition for the cell population, as shown as in Figure \ref{mono.simu.1} (a). From Figures \ref{mono.simu.1}(b), \ref{mono.simu.1bis}, we can observe that cells tend to assume a profile similar to that of the chemoattractant (see also Figure \ref{mono.simu.1bis}). 
This is not surprising since  for a turning rate $\mu$ and distribution function $\psi(v)$ independent of $x$ (and therefore constant $\bar U$ and $D$) in one dimension the stationary solution of \eqref{diff.adv.mono} is given by
\begin{equation}\label{rhostaz}
\rho(x)=C [b(\mathcal{S}(x))]^m\quad{\rm with}\ \ m=\dfrac{\Lambda \mu\bar{U}}{D}\,,
\end{equation}
and the constant $C$ given by the initial mass, $\ie$,
\begin{equation}\label{stat.state}
\rho(x)=\int_\Omega\rho_0(x)\,dx\, \dfrac{[b(\mathcal{S}(x))]^m}
{\displaystyle{\int_\Omega{[b(\mathcal{S}(x))]^m}dx}}\,.
\end{equation}
The maximum of the cell stationary state in response to the chemoattractant depends, as shown in Figure  \ref{mono.simu.1bis}, both on the size of the sensing neighborhood and on the kind of sensing. A larger sensing radius leads to a stronger motility of cells and, therefore, to a higher peak in the steady state. This is because a bigger sensing radius allows to scout a bigger neighborhood. Furthermore, the sensing of a volume ($\gamma_{\scriptscriptstyle \mathcal{S}}=H$) leads to an average of $\mathcal{S}$ over a bigger region, with respect to $\gamma_{\scriptscriptstyle \mathcal{S}}=\delta$. This trend is also confirmed by the diffusive limit. In fact, the maximum density depends on the power $m$ defined in 
\eqref{rhostaz}. Having set $\mu=50$ and chosen a uniform distribution in $[0,2]$, so that $\bar U=1$ and $D=\dfrac{1}{3}$, then $m=150\Lambda$ with 
$\Lambda=0.02$ for  $\gamma_{\scriptscriptstyle \mathcal{S}}=\delta(\lambda-R_{\scriptscriptstyle \mathcal{S}})$ and 
$\Lambda=0.01$ for $\gamma_{\scriptscriptstyle \mathcal{S}}=H(\lambda-R_{\scriptscriptstyle \mathcal{S}})$.

\begin{table}[htbp]
\centering

\begin{tabular}{|l|c|c|c|c|c|c|c|}
\hline\noalign\\[-10pt]
$\quad l_{\scriptscriptstyle \mathcal{S}}$& $R_{\scriptscriptstyle \mathcal{S}}$& $\gamma_{\scriptscriptstyle \mathcal{S}}$& $\Lambda$& $\eta$& relative $L_{\infty}$ error  \\ 
\hline
17.964& 0.02& $\delta$& 0.02& 0.001&  0.03\\
\hline
17.964& 0.02& H& 0.01& 0.001&   0.03\\
\hline
17.964& 0.2& $\delta$& 0.2& 0.01&   0.03\\
\hline
17.964& 0.2& H& 0.1& 0.01&  0.05\\
\hline
17.964& 2& $\delta$& 2& 0.1& 0.13\\
\hline
17.964& 2& H& 1& 0.1& 0.53\\
\hline
\end{tabular}
\caption{Chemotactic motion.}
\label{table2}
\end{table}

Referring to  Table \ref{table2}, when $R_{\scriptscriptstyle \mathcal{S}}=0.02\ll l_\mathcal{S}$, the collision operator \eqref{J_r} is well approximated by \eqref{J_r.loc} and we can compare the solution between the kinetic model and Equation (\ref{diff.adv.mono}) with $\chi(S)=\Lambda \frac{\bar{U}}{\mathcal{S}}$. 
As shown  in Figures  \ref{mono.simu.1bis}(a),(b), the macroscopic density of the stationary solution of the kinetic model with turning operator given by \eqref{J_r} is very close to the analytic solution \eqref{rhostaz} of the diffusive limit.
 In addition, one can observe that  a larger sensing radius gives rise to a stronger sensitivity as reflected by a larger $\Lambda$ and therefore a larger $\chi(\mathcal{S})$ in the advection-diffusive equation \eqref{diff.adv.mono} and a larger $m$ in \eqref{rhostaz}.
The comparison between the two solutions remains quite good for $R_{\scriptscriptstyle \mathcal{S}}=0.2$, which corresponds to  $\eta=0.01$.
Instead, as shown in Figure  \ref{mono.simu.1bis} and Table \ref{table2}, an $R_{\scriptscriptstyle \mathcal{S}}=2$ is not much smaller than $l_{\scriptscriptstyle \mathcal{S}}$, yielding to larger discrepancies.  
The  trends  obtained for $\gamma_{\scriptscriptstyle \mathcal{S}}=\delta$ and $\gamma_{\scriptscriptstyle \mathcal{S}}=H$ are similar.
However, the transient time is larger if the sensing function is the Heavyside function.

\begin{figure}[htbp]
\begin{center}
\includegraphics[scale=0.35]{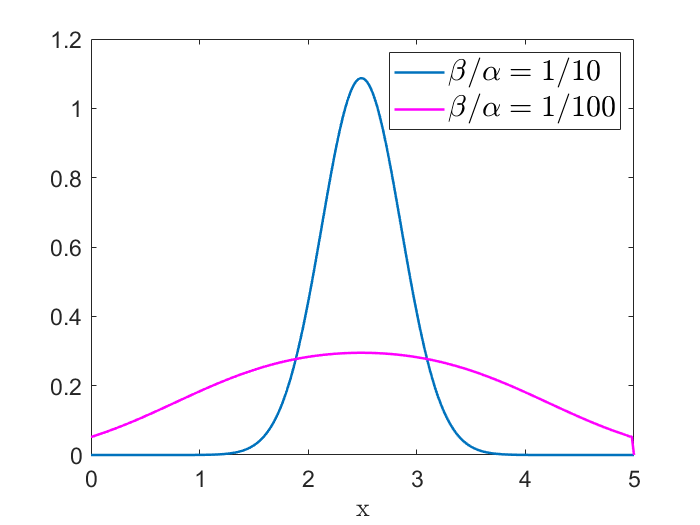}
\end{center}
\caption{Stationary density of the kinetic model with a comparative sensing kernel for $R_{\scriptscriptstyle \mathcal{S}}=0.02$ and 
$\gamma_{\scriptscriptstyle \mathcal{S}}=\delta(\lambda-R_{\scriptscriptstyle \mathcal{S}})$.} 	
\label{comp}
\end{figure}

In the second simulation reported in Figure \ref{comp}, we use a comparative sensing kernel (\ref{comp.sens}). Here $R_{\scriptscriptstyle \mathcal{S}}=0.02$ and we consider different values of $\alpha$ and $\beta$. We may observe that if $\alpha$ is much larger than $\beta$, then the diffusive dynamics is dominant. 

\begin{figure}[htbp]
\begin{center}
 	\includegraphics[width=0.42\textwidth]{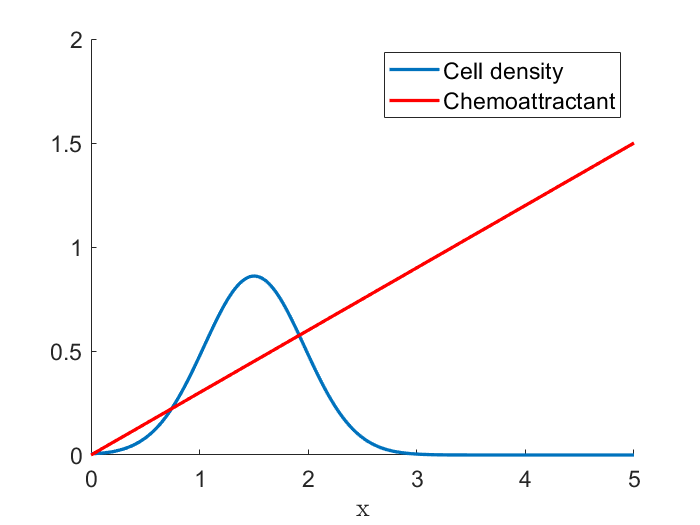}
 	\includegraphics[width=0.42\textwidth]{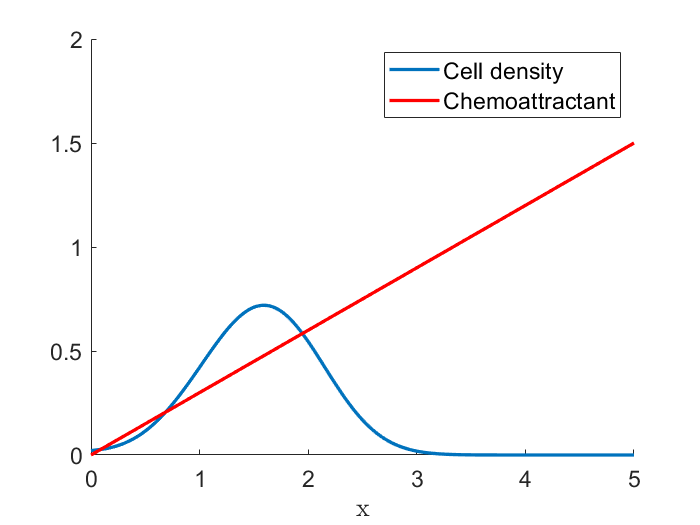}
 	
 	\includegraphics[width=0.42\textwidth]{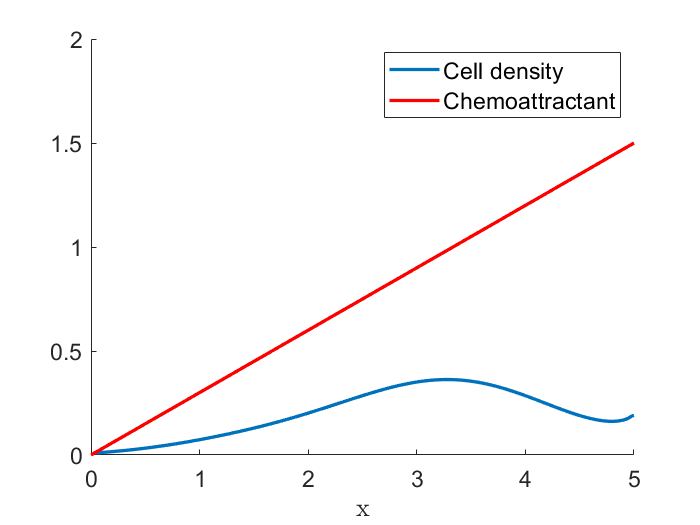}
 	\includegraphics[width=0.42\textwidth]{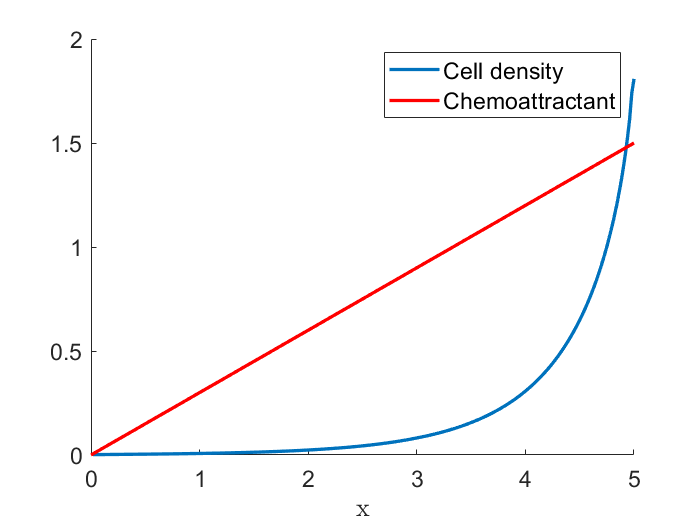}
\end{center}

\caption{Evolution of a gaussian macroscopic distribution of cell density  for $t=0,0.5,5,10$ in presence of a linear chemoattractant distribution  $\mathcal{S}(x)=0.3x$ shown in red. The sensing radius $R_{\scriptscriptstyle \mathcal{S}}=0.02$ and other parameters are the same as in Figure \ref{mono.simu.1}.}
\label{mono.simu.2}
\end{figure}

Finally, in the set of simulations shown in Figure \ref{mono.simu.2}, we  start from a macroscopic gaussian distribution for cells, that moves due to the perception of the chemoattractant which is distributed linearly. Cells gradually move to the right until they reach the right boundary where the specular reflection boundary condition, corresponding to a no-flux condition for the macroscopic density, keeps them in the domain.

\paragraph{Cell-cell adhesion}
\hspace{-2pt}

\begin{figure}[!htbp]
\begin{center}
\includegraphics[scale=0.5]{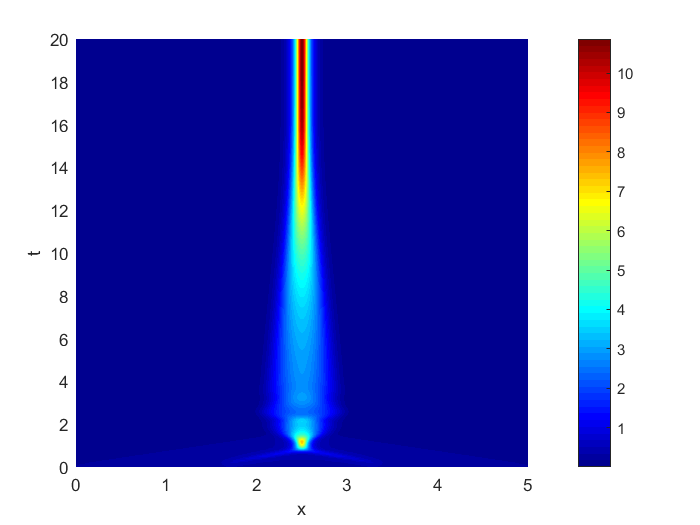}

\end{center}

\caption{Temporal evolution of the macroscopic density starting from a perturbed constant distribution under the action of cell-cell adhesion. Here $\gamma_{\rho}=\delta(\lambda-R_{\rho})$, with $R_{\rho}=1$.}
\label{fig:adhesion}
\end{figure}

\noindent 
Using the same framework, we may also consider cellular adhesion by considering $\mathcal{S}=\rho$ as a biasing signal.
In this case, taking for instance $b(\rho)=\rho$ and $\gamma_{\rho}$ a Dirac delta, the turning operator reads
$$
\begin{array}{lr}
\mathcal{J}[p](t,\x,\vb_p)=
\phantom{a}\mu(\x) \,  \left[ \rho(t,\x)  \psi(\x,v) c(\x)  \rho(\x+R_{\rho}\hat{\vb})  - p(t,\x,\vb_p) \right] \hspace{-2pt} \, ,
\end{array}
$$
As initial condition we take $p(0,x,\vb_p)=\dfrac{0.2}{U}(1+0.09\sin x)$ moving both to the right and to the left, that corresponds to a small perturbation of the constant initial condition with unitary mass that would stay constant. However, cell-cell adhesion triggers an instability so that the small perturbation amplifies, reaching a peaked distribution in the center of the domain as shown in Figure \ref{fig:adhesion}.

In two dimensions we show cell-cell adhesion for a four peaks distribution. Boundary conditions are of Maxwell-type, with $\alpha=0.5$ in \eqref{Maxwell}. We may observe that at the beginning the distributions peak up. Then the sensing radius leads to aggregation of the single peaks in a nearly circular crown structure. Peaks  get closer and eventually aggregate. 
\begin{figure}[!htbp]
\begin{center}
\subfigure[$t=0$]{\includegraphics[scale=0.19]{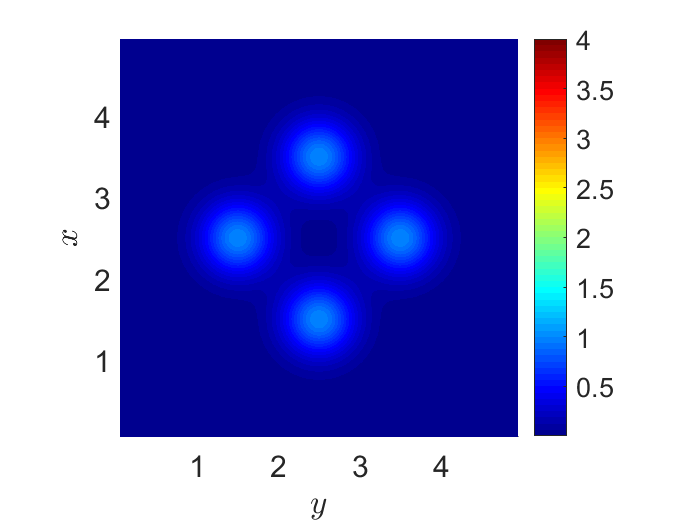}}
\subfigure[$t=2.5$]{\includegraphics[scale=0.19]{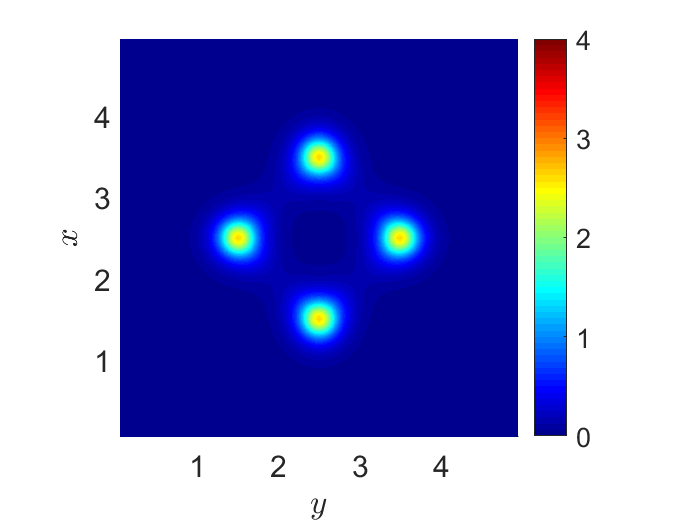}}
\subfigure[$t=7.5$]{\includegraphics[scale=0.19]{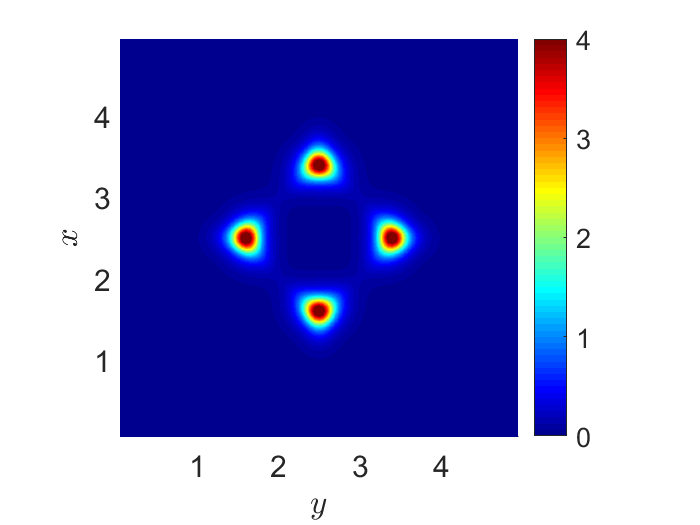}}
\subfigure[$t=12.5$]{\includegraphics[scale=0.19]{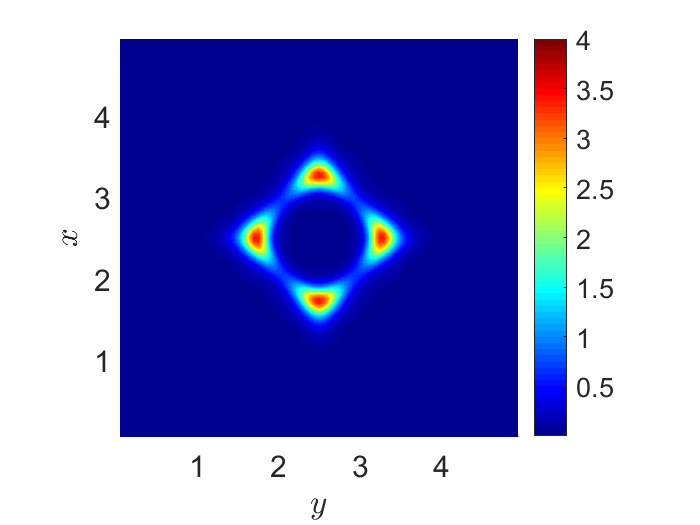}}

\subfigure[$t=15$]{\includegraphics[scale=0.19]{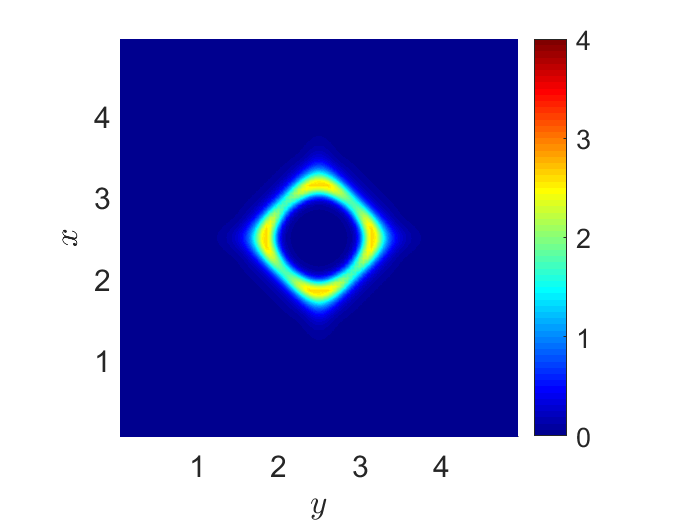}}
\subfigure[$t=17.5$]{\includegraphics[scale=0.19]{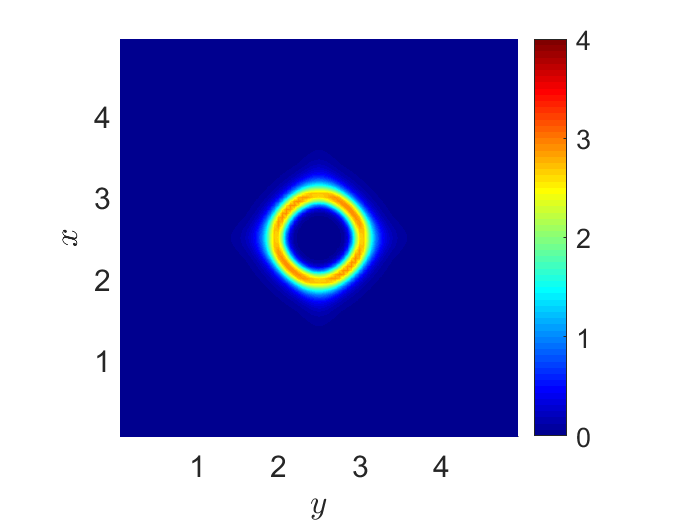}}
\subfigure[$t=20$]{\includegraphics[scale=0.19]{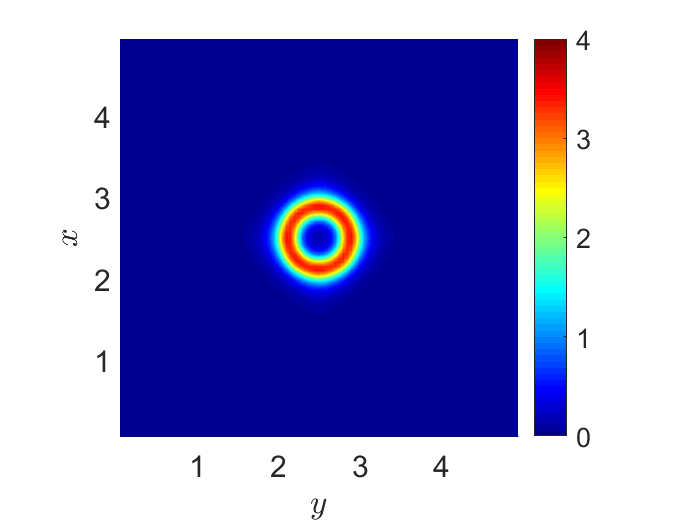}}
\subfigure[$t=50$]{\includegraphics[scale=0.19]{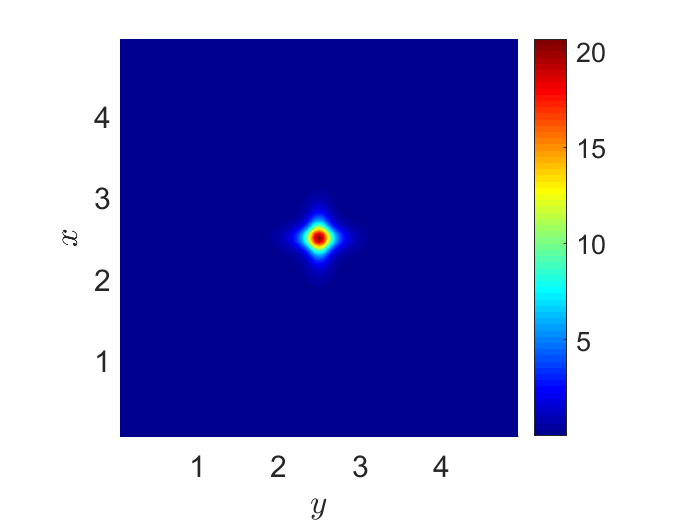}}

\end{center}

\caption{Temporal evolution of the macroscopic density starting from the four peaks  distribution in (a) under the action of cell-cell adhesion. Here $\gamma_{\rho}=\delta(\lambda-R_{\rho})$, with $R_{\rho}=0.25$.}
\label{fig:adhesion2d}
\end{figure}
\paragraph{Durotaxis}
\hspace{-2pt}

\begin{figure}[!htbp]
\begin{center}
\subfigure[]{\includegraphics[scale=0.38]{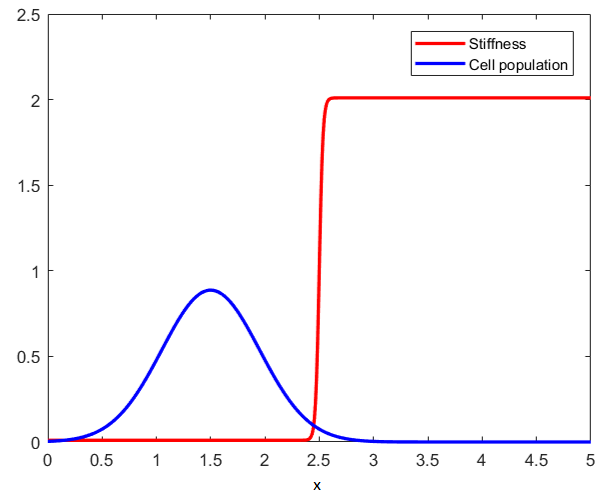}}
\subfigure[]{\includegraphics[scale=0.305]{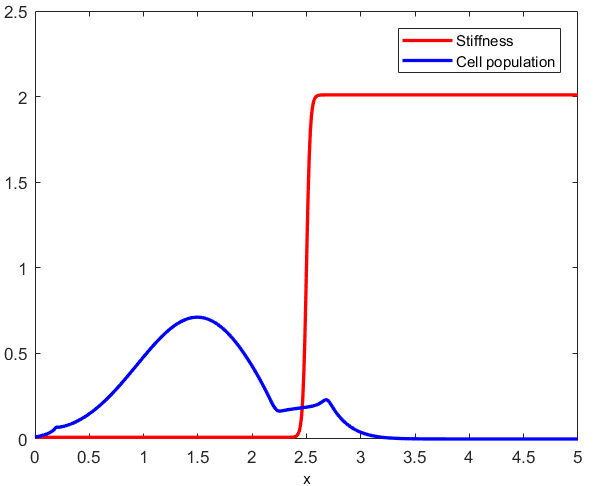}}
\subfigure[]{\includegraphics[scale=0.38]{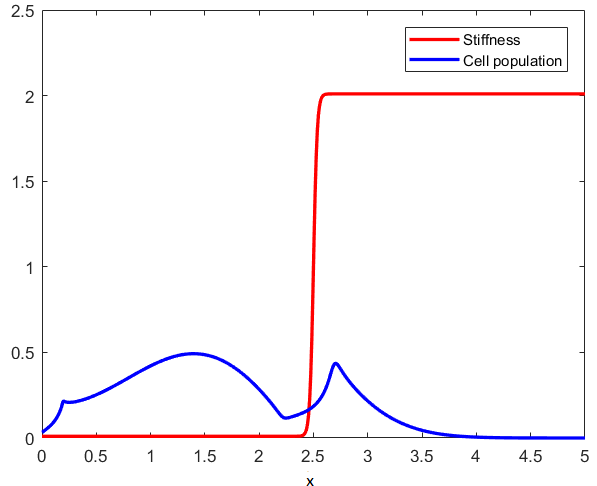}}
\end{center}
\begin{center}
\subfigure[]{\includegraphics[scale=0.4]{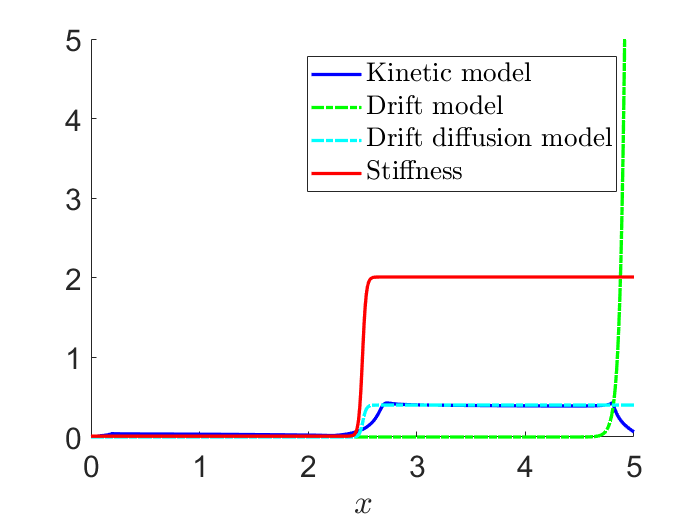}}
\subfigure[]{\includegraphics[scale=0.4]{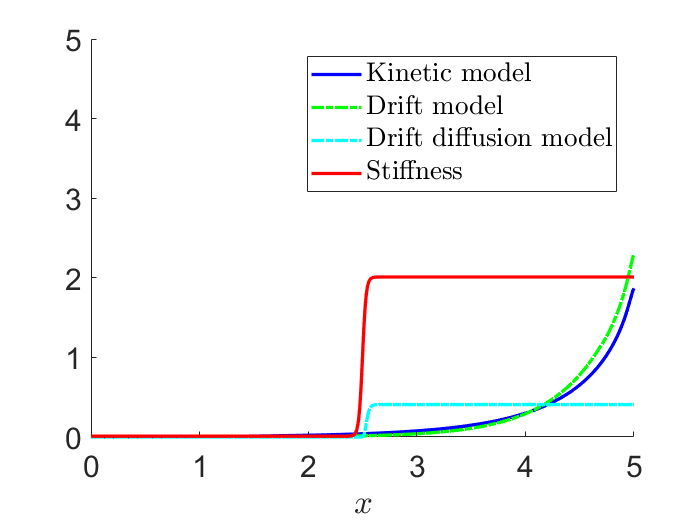}}
\end{center}
\caption{Cell migration under the action of durotaxis. The red line refers to the stiffness of the extracellular matrix. The initial condition is uniformly ditributed in the velocity space and has a macroscopic density which is a gaussian centered in $1.5$ (a). In the top row we show the evolution of the cell density for $t=0,1,10$ and for $R_{\scriptscriptstyle \mathcal{S}}=0.02$ and in the bottom row the stationary states for $R_{\scriptscriptstyle \mathcal{S}}=0.02$ (d) and $R_{\scriptscriptstyle \mathcal{S}}=2.5$ (e) and the comparison with the solutions of the two macroscopic models (respectively, diffusive and hyperbolic). }
\label{durotaxis}
\end{figure}

\noindent 
We now consider the case in which cells are guided by the rigidity of the extracellular matrix moving towards stiffer regions.
We consider a stiffness profile as in Fig. \ref{durotaxis} (red line) and $b(\mathcal{S})=\mathcal{S}$. The sensing function is again a Dirac delta and the sensing radius in the simulation shown in the top row of Figure \ref{durotaxis} (red line) is $R_{\scriptscriptstyle \mathcal{S}}=0.02$. So, the cells within that radius start moving to the stiffer region. Eventually, a stationary state is reached that depends on the sensing radius.
In particular, Figure \ref{durotaxis} (bottom row) clearly shows  that for a small radius $R_{\scriptscriptstyle \mathcal{S}}=0.02$ (which corresponds to $\eta=0.02$) the analytic solution of the advection-diffusion equation (\ref{stat.state}) is similar to the macroscopic density of the solution to the kinetic model. If, instead, $R_{\scriptscriptstyle \mathcal{S}}=2.5$ (which corresponds to $\eta=3$), we have that (\ref{stat.state}) fails in approximating the solution to the kinetic equation, while the solution to the hyperbolic macroscopic equation well approximates the profile of the cell density.

\begin{figure}[!htbp]
\begin{center}
\subfigure[]{\includegraphics[scale=0.4]{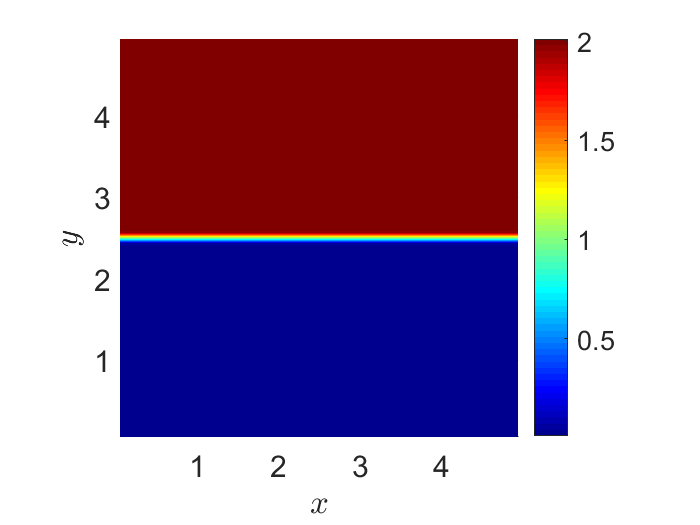}}
\subfigure[$T=0$]{\includegraphics[scale=0.5]{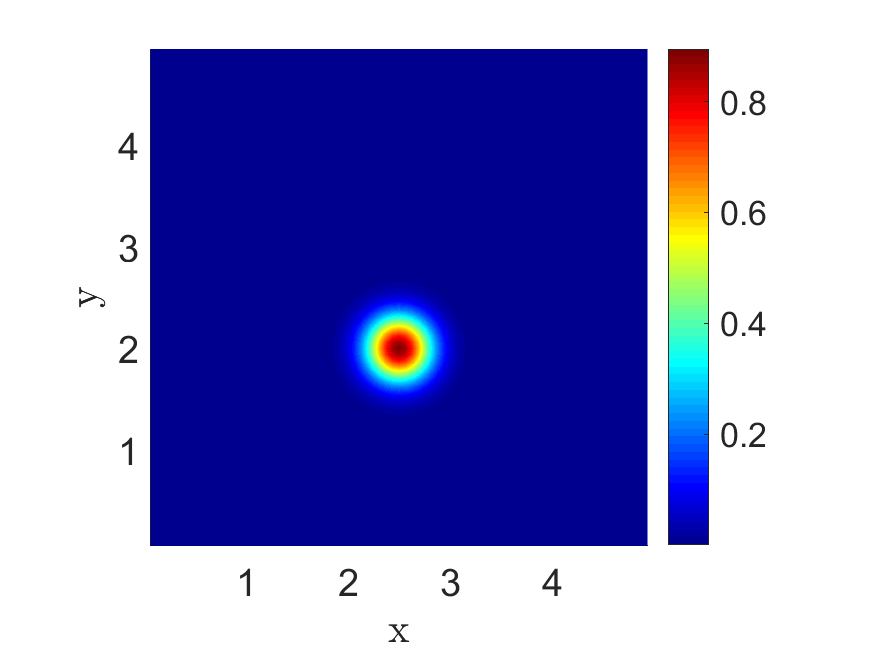}}
\end{center}
\begin{center}
\subfigure[$t=0.6$]{\includegraphics[scale=0.33]{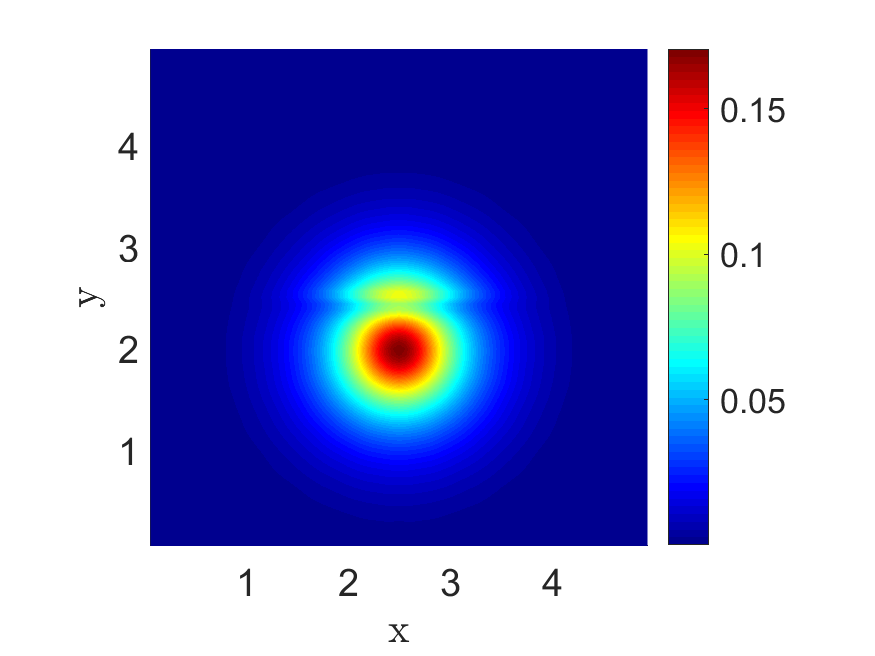}}
\subfigure[$t=6$]{\includegraphics[scale=0.33]{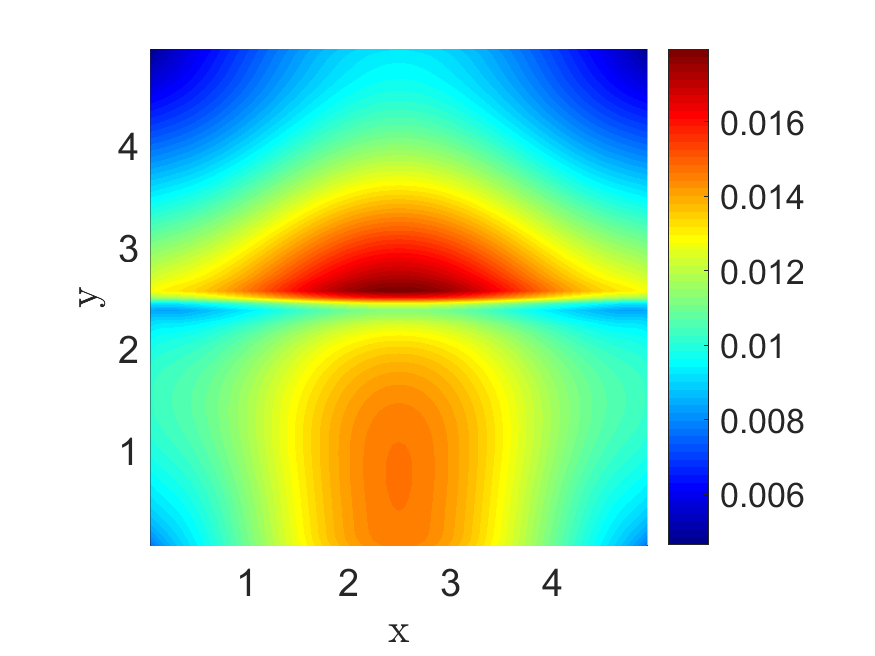}}
\subfigure[$t=22$]{\includegraphics[scale=0.33]{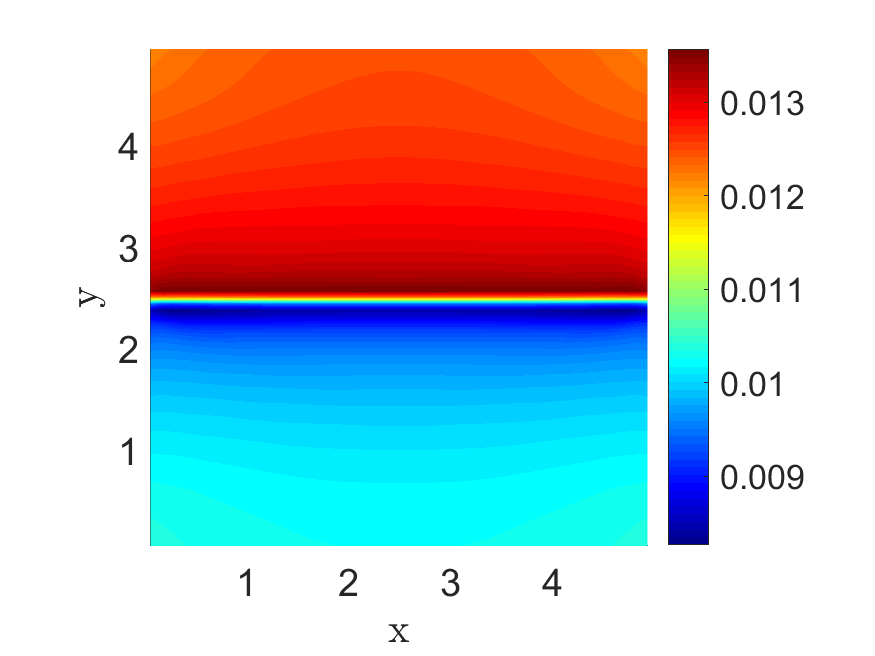}}
\end{center}
\begin{center}
\subfigure[$t=0.6$]{\includegraphics[scale=0.33]{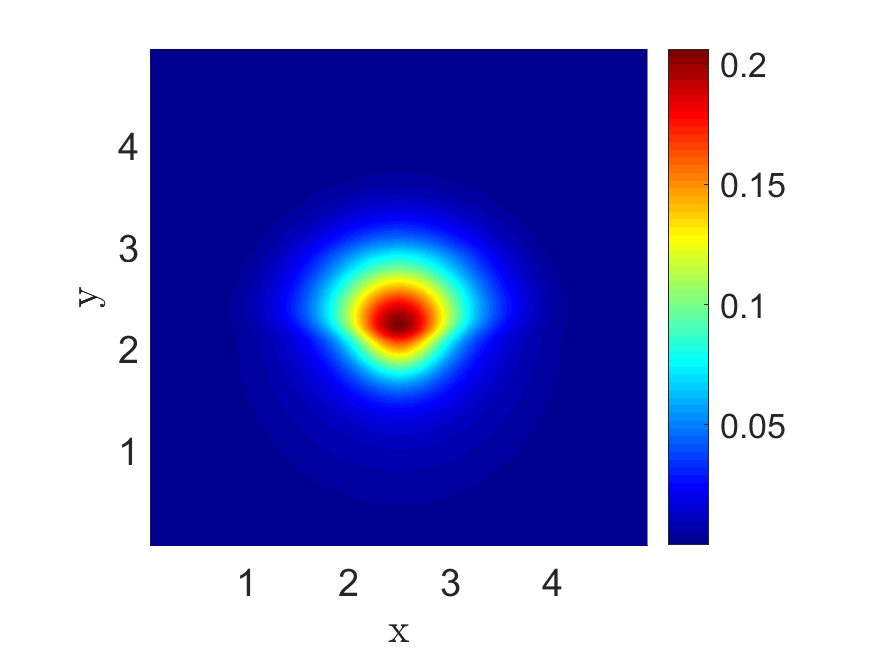}}
\subfigure[$t=3$]{\includegraphics[scale=0.33]{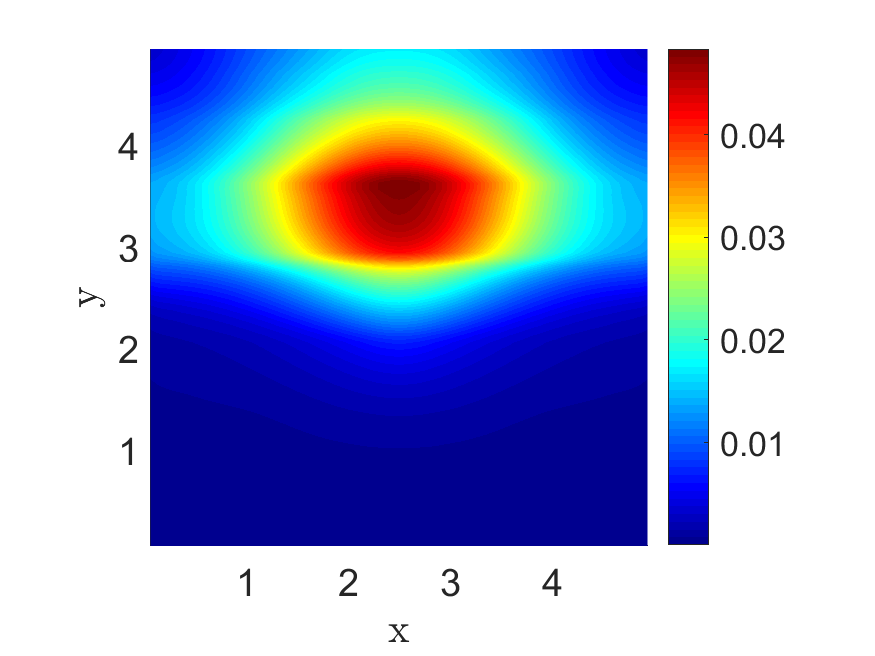}}
\subfigure[$t=22$]{\includegraphics[scale=0.33]{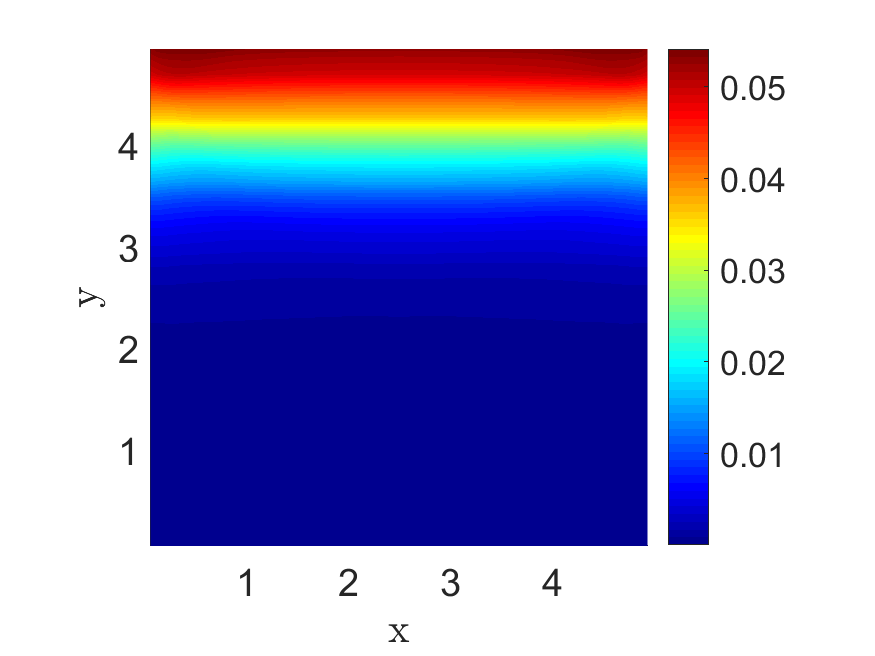}}
\end{center}
\caption{Evolution of an initial distribution given in (b) in presence of a heterogeneous environment with stiffness given in (a). In (c)-(f) $R_{\scriptscriptstyle \mathcal{S}}=0.02$, while in (f)-(h) $R_{\scriptscriptstyle \mathcal{S}}=2.5$. }
\label{durotaxis.2D}
\end{figure}

In Figure \ref{durotaxis.2D} we present the corresponding simulations in two dimensions. We may observe that the qualitative behavior is the same. Here we show the results with Maxwell type boundary conditions ($\alpha=0$ in \eqref{Maxwell}). We checked that the macroscopic stationary state is the same with regular reflection boundary conditions ($\alpha=1$ in \eqref{Maxwell}).

\section{Double bias}\label{double.bias}

In the previous section we have assumed that the external signal only affects the choice of the turning direction keeping the distribution function $\psi$ determining the speed always the same and in particular unaffected by environmental cues or external chemical signals. Instead, in cell motility one should distinguish polarization mechanisms from the ability to move as they can depend on different factors. In fact, as stated in the introduction, a cell can even be polarized but unable to move. 
For this reason, in this section we consider a double bias, due to a second factor $\mathcal{S}'$ evaluated in a non-local way that influences the speed, once the direction $\hat{\vb}$ is chosen according to the bias ruled by $\mathcal{S}$. These can be represented not only by different external free and bound chemical factors, or subcellular cues involved in cell motility, but also by cell and ECM density. 

In order to do that, we allow the distribution function $\psi({\bf x},v)$ to depend on the level of the signal $\mathcal{S}'$.  
In order to stress this dependence, we use the following notations
$\psi=\psi(\x,v|\mathcal{S}'(\x+\lambda'\hat{\vb}))$ for the distribution function, $\bar{v} (\x|\mathcal{S}'(\x+\lambda'\hat\vb))$ for its mean, and $D (\x|\mathcal{S}'(\x+\lambda'\hat\vb))$ for its variance, where $|\mathcal{S}'$ reads as "given $\mathcal{S}'$ in a point $\x+\lambda'\hat{\vb}$ in the neighborhood of the cell". Then, the cell polarized along $\hat\vb$ determines its speed averaging the signal over the sensing radius $R_{\scriptscriptstyle \mathcal{S}'}$ through a weight function $\gamma_{\scriptscriptstyle \mathcal{S}'}$.
\\
In the same spirit as Eq. (\ref{dir.transition}), introducing also 
\begin{equation}
\Psi[\mathcal{S}'](\x,v|\hat{\vb})=  \int_{\mathbb{R}_+}\gamma_{\scriptscriptstyle \mathcal{S}'}(\lambda') \psi(\x,v|\mathcal{S}'(\x+\lambda'\hat{\vb})) \, d\lambda ' \, ,
\end{equation}
the turning probability in \eqref{J_r.mu.S} or \eqref{J_r.S} reads
\begin{equation}\label{distribution.g.r.double}
T[\mathcal{S},\mathcal{S}'](\x,\vb_p)=c(\x)B[\mathcal{S}](\x,\hat{\vb})
\Psi[\mathcal{S}'](\x,v|\hat{\vb}).
\end{equation}
The normalization term $c$ can be factorized as
$$
c(\x)=c_1(\x)c_2(\x),
$$
with 
$$c_1(\x)=
\dfrac{1}{\displaystyle{\int_{\mathbb{R}_+}\gamma_{\scriptscriptstyle \mathcal{S}'}(\lambda') \, d\lambda '}}=\dfrac{1}{\Gamma_0'},$$ 
and 
$$c_2(\x)=\dfrac{1}{\displaystyle \int_{\mathbb{S}^{d-1}} \int_{\mathbb{R}_+}\gamma_{\scriptscriptstyle \mathcal{S}}(\lambda) \mathcal{T}_{\lambda}^{\hat{\vb}}[\mathcal{S}](\x) d\lambda \, d\hat{\vb}}.$$
In this way, 
\begin{equation}
\bar B[\mathcal{S}](\x,\hat{\vb})= c_2(\x)\displaystyle\int_{\mathbb{R}_+}\gamma_{\scriptscriptstyle \mathcal{S}}(\lambda) \mathcal{T}_{\lambda}^{\hat{\vb}}[\mathcal{S}](\x) d\lambda \,
\end{equation} and 
\begin{equation}\label{bar.Psi}
\bar \Psi[\mathcal{S}'](\x,v|\hat{\vb})= \dfrac{1}{\Gamma_0'}\displaystyle\int_{\mathbb{R}_+}\gamma_{\scriptscriptstyle \mathcal{S}'}(\lambda') \psi(\x,v|\mathcal{S}'(\x+\lambda'\hat{\vb})) \, d\lambda ' \, 
\end{equation} 
are both distribution functions and the turning probability reads $T=\bar B\bar{\Psi}$.\\
Going back to the general case, the turning operator reduces to
\begin{equation}\label{turning.operator.2}
\mathcal{J}[p](\x,\vb_p) = \mu(\x) \, \Big( \rho(t,\x)c(\x)B[\mathcal{S}](\x,\hat{\vb})
\Psi[\mathcal{S}'](\x,v|\hat{\vb})  - p(t,\x,\vb_p) \Big) \,
\end{equation}
and the distribution function which nullifies the turning operator is
$$
p=\rho(t,\x) c(\x) B[\mathcal{S}](\x,\hat{\vb})
\Psi[\mathcal{S}'](\x,v|\hat{\vb}).  
$$\\
Now, ${\bf U}_{\scriptscriptstyle\mathcal{S},\mathcal{S}'}$ in Eq. (\ref{mean.Vr1}) reads
\begin{equation}
{\bf U}_{\scriptscriptstyle\mathcal{S},\mathcal{S}'}(\x)=
\dfrac{c(\x)}{c_1(\x)} \displaystyle \int_{\mathbb{S}^{d-1}}
B[\mathcal{S}](\hat{\vb})\,  
\bar{U}_{\scriptscriptstyle \mathcal{S}'}(\x|\hat{\vb})
 \hat{\vb} \, d\hat{\vb}
\end{equation}
where
\begin{equation}
\bar{U}_{\scriptscriptstyle \mathcal{S}'}(\x|\hat{\vb})=\displaystyle  c_1(\x)\int_{\mathbb{R}_+}\bar{v} (\x|\mathcal{S}'(\x+\lambda'\hat\vb))\gamma_{\scriptscriptstyle \mathcal{S}'}(\lambda')\, d\lambda' 
\end{equation}
and the diffusion tensor is
\begin{equation}
\mathbb{D}_{\scriptscriptstyle\mathcal{S},\mathcal{S}'}(\x)=c(\x) \displaystyle \int_{V_p}
B[\mathcal{S}](\hat{\vb}) \,  
\Psi[\mathcal{S}'](v|\hat{\vb})
 (\vb-{\bf U}_{\scriptscriptstyle\mathcal{S},\mathcal{S}'}) \otimes (\vb-{\bf U}_{\scriptscriptstyle\mathcal{S},\mathcal{S}'})\, d\vb_p.
\end{equation}

Let us now assume that, like $B$, $\Psi[\mathcal{S}']$ may be written, up to re-scaling, as
$$\Psi[\mathcal{S}']=\Psi[\mathcal{S}']_0+\epsilon\Psi[\mathcal{S}']_1 +\mathcal{O}(\epsilon^2).$$  
The means and variances of $\Psi[\mathcal{S}']_0$ and $\Psi[\mathcal{S}']_1$ will be, respectively, denoted by $\bar U_{\scriptscriptstyle \mathcal{S}'}^0$, $\bar U_{\scriptscriptstyle \mathcal{S}'}^1$, $\bar{D}_{\scriptscriptstyle \mathcal{S}'}^0$, $\bar{D}_{\scriptscriptstyle \mathcal{S}'}^1$. 

Up to re-scaling, the re-scaled turning probability becomes $T[\mathcal{S},\mathcal{S}'](\vb_p)=T[\mathcal{S},\mathcal{S}']_0(\vb_p)+\epsilon T[\mathcal{S},\mathcal{S}']_1(\vb_p)$, where, now
\begin{equation}
T[\mathcal{S},\mathcal{S}']_0(\vb_p)=c(\x)B[\mathcal{S}]_0(\hat{\vb})\Psi[\mathcal{S}']_0(v|\hat{\vb})
\end{equation}
and
\begin{equation}
T[\mathcal{S},\mathcal{S}']_1(\vb_p)=c(\x)B[\mathcal{S}]_0(\hat{\vb})\Psi[\mathcal{S}']_1(v|\hat{\vb})+c(\x)B[\mathcal{S}]_1(\hat{\vb})\Psi[\mathcal{S}']_0(v|\hat{\vb})\,.
\end{equation}
In this case, Eqs.(\ref{diff.cond.1}) and (\ref{diff.cond.2}) are satisfied if
$$\displaystyle \int_{0}^U \Psi[\mathcal{S}']_0 dv=\displaystyle \int_{0}^U \Psi[\mathcal{S}'] dv \qquad \displaystyle \int_{0}^U \Psi[\mathcal{S}']_i dv=0 \quad \forall i\geq 1.$$
Therefore, the macroscopic velocity of order 0 is
\begin{equation}\label{macro.drift.double}
{\bf U}_{\scriptscriptstyle\mathcal{S},\mathcal{S}'}^0(\boldsymbol{\xi})=
\frac{c(\boldsymbol{\xi})}{c_1(\boldsymbol{\xi})} \int_{\mathbb{S}^{d-1}}
B[\mathcal{S}]_0(\hat{\vb}) \,  
\bar{U}_{\scriptscriptstyle \mathcal{S}'}^0(\boldsymbol{\xi}|\hat{\vb})
\hat{\vb} \, d\hat{\vb}
\end{equation}
while the macroscopic velocity of order 1 is
\begin{equation}
{\bf U}_{\scriptscriptstyle\mathcal{S},\mathcal{S}'}^1(\boldsymbol{\xi})=
c(\boldsymbol{\xi})\int_{\mathbb{S}^{d-1}} \Big(
B[\mathcal{S}]_0(\hat{\vb})\Psi_1(v|\hat{\vb})+B[\mathcal{S}]_1(\hat{\vb})\Psi_0(v|\hat{\vb})
\Big)\hat{\vb} \, d\hat{\vb}
\end{equation}
and the equilibrium diffusion tensor
\begin{equation}
\mathbb{D}_{\scriptscriptstyle\mathcal{S},\mathcal{S}'}^0(\boldsymbol{\xi})=c(\boldsymbol{\xi})\int_{V_p}
B[\mathcal{S}]_0(\hat{\vb}) \,  
\bar{D}_{\scriptscriptstyle \mathcal{S}'}^0(\boldsymbol{\xi}|\hat\vb)
 (\vb-{\bf U}_{\scriptscriptstyle\mathcal{S},\mathcal{S}'}^0) \otimes (\vb-{\bf U}_{\scriptscriptstyle\mathcal{S},\mathcal{S}'}^0)\, d\vb_p.
\end{equation}
The diffusion tensor is in general anisotropic, as $\bar{D}^0$ may be not isotropic in $\hat{\vb}$.\\
If (\ref{even}) is satisfied, the equilibrium tensor is
\begin{equation}
\mathbb{D}_{\scriptscriptstyle\mathcal{S},\mathcal{S}'}^0(\boldsymbol{\xi})=c(\boldsymbol{\xi}) \displaystyle \int_{\mathbb{S}^{d-1}}
B[\mathcal{S}]_0(\hat{\vb}) \,  
\bar{D}_{\scriptscriptstyle \mathcal{S}'}^0(\boldsymbol{\xi}|\hat\vb)
 \hat{\vb}\otimes \hat{\vb}\, d\hat{\vb}.
\end{equation}
where
$$\bar{D}_{\scriptscriptstyle \mathcal{S}'}^0(\boldsymbol{\xi}|\hat{\vb})=\displaystyle \int_{\mathbb{R}_+} D^0(\boldsymbol{\xi}|\mathcal{S}'(\x+\lambda'\hat\vb))\gamma_{\scriptscriptstyle \mathcal{S}'}(\lambda')\, d\lambda' .$$ 
In this case, we get as a diffusive limit Eq. (\ref{macro.diff}), or (\ref{macro.diff.t}) if the frequency depends weakly on the polarization. Otherwise, if (\ref{even}) is not satified, we get (\ref{macro.drift}) as a hyperbolic limit.

\subsection{Examples}

First of all, we observe that if $\gamma_{\scriptscriptstyle \mathcal{S}'}(\lambda)=\gamma_{\scriptscriptstyle \mathcal{S}}(\lambda)=\delta(\lambda-0)$ are Dirac deltas, then the local model with $T[\mathcal{S},\mathcal{S}']=\frac{1}{|\mathbb{S}^{d-1}|}\psi(\x,v)$ is recovered. Hence $\Ub_{\scriptscriptstyle \mathcal{S},\mathcal{S}'}$ is zero and $\bar{U}_{\scriptscriptstyle\mathcal{S}'}(\x|\hat\vb)$ and $\bar{D}_{\scriptscriptstyle\mathcal{S}'}(\x|\hat\vb)$ are the mean speed and variance of $\psi(\x,v)$ as defined in Section \ref{dir.tur.prob}.  

In this section we shall at first consider $\gamma_{\scriptscriptstyle \mathcal{S}}(\lambda)=\delta(\lambda-0)$, $\ie$, when there is no bias in the decision of the direction of motion. We shall do this in order to analyze the influence of the 
signal $\mathcal{S}'$. In particular, we will see that if, for instance, $\mathcal{S'}$ is the ECM density then in the limit we recover a model describing the steryc hindrance of motion. Finally, we will analyse models where both $\mathcal{S}$ and $\mathcal{S}'$ have a non-local influence in determining cell motion.

\subsubsection{Random polarization}\label{isotropicpol}

In order to understand the meaning of the transport model we here focus on the particular case in which there is no bias in the decision of the direction of motion. This does not mean that the situation is isotropic, because the speed can have different distribution functions in the randomly chosen direction of motion because the distribution of $\mathcal{S}'$ sensed ahead may be different.

In this case the turning operator (\ref{distribution.g.r.double}) simplifies to 
\begin{equation}\label{TSR0}
T[\mathcal{S},\mathcal{S}'](\x,\vb_p)=\bar \Psi[\mathcal{S}'](\x,v|\hat{\vb}) \, ,
\end{equation}
where $\bar{\Psi}$ is defined in (\ref{bar.Psi}),
the macroscopic velocity of the transition probability is 
$$\Ub_{\scriptscriptstyle\mathcal{S}'}=\dfrac{1}{\displaystyle |\mathbb{S}^{d-1}|\Gamma_0'} \displaystyle \int_{\mathbb{S}^{d-1}} \bar{U}_{\scriptscriptstyle\mathcal{S}'} \hat{\vb} \, d\hat{\vb},$$ and the variance-covariance matrix is $$\mathbb{D}_{\scriptscriptstyle \mathcal{S}'}=\int_{V_p}\bar \Psi[\mathcal{S}'](\x,v|\hat{\vb})(\vb-{\bf U}_{\scriptscriptstyle\mathcal{S}'}) \otimes (\vb-{\bf U}_{\scriptscriptstyle\mathcal{S}'})\, d\hat{\vb}.$$
Up to re-scaling, we obviously have that $$\bar\Psi[\mathcal{S}']=\bar{\Psi}[\mathcal{S}']_0+\epsilon\bar{\Psi}[\mathcal{S}']_1+\mathcal{O}(\epsilon^2),$$
and we can consider the following cases:
\begin{description}
\item[$\bullet$]
If $\bar\Psi[\mathcal{S}']_0(\boldsymbol{\xi},v|\hat{\vb})$ is even as a function of $\hat{\vb}$, $\ie$ if $\bar{U}$ is even as a function of $\hat{\vb}$, we have that $\Ub_{\mathcal{S}'}^0(\boldsymbol{\xi})=\boldsymbol{0}$. We can therefore perform a diffusive limit to get
$$
\dfrac{\partial \rho}{\partial \tau} =\nabla \cdot \left[\dfrac{1}{\mu} \nabla \cdot \big(\mathbb{D}_{\scriptscriptstyle\mathcal{S}'}^0\rho \big)\right]\, ,
$$
if $\bar\Psi[\mathcal{S}']_1$ is zero and
\begin{equation}\label{convdiffrandom}
\dfrac{\partial \rho}{\partial \tau}+\nabla \cdot \Big(\rho \Ub_{\scriptscriptstyle\mathcal{S}'}^1 \Big) =\nabla \cdot \left[\dfrac{1}{\mu} \nabla \cdot \big(\mathbb{D}_{\scriptscriptstyle\mathcal{S}'}^0\rho\big) \right] \, ,
\end{equation}
otherwise. The tensor $\mathbb{D}_{\scriptscriptstyle\mathcal{S}'}^0$ may be anisotropic, as $\bar{U}$ and $\bar{D}$ may be anisotropic as functions of $\hat{\vb}$.
\item[$\bullet$]
If $\bar\Psi[\mathcal{S}']_0(\boldsymbol{\xi},v|\hat{\vb})$ is not even as a function of $\hat{\vb}$, $\ie$ if $\bar{U}$ is not even as a function of $\hat{\vb}$, we can perform a hyperbolic limit to get
$$
\dfrac{\partial \rho}{\partial \tau}+\nabla \cdot \big( \rho \Ub_{\scriptscriptstyle\mathcal{S}'}^0 \big)=0\,.
$$
\end{description}
Therefore, we see that, even if there is no directional sensing, as the speed sensing is conditioned by the direction, there may be a drift driven phenomenon or an anisotropic diffusive one.

\subsubsection{ECM steryc hindrance}
We shall now focus on the effect of the density $M(\x)$ of extracellular matrix. Following the experimental results by \cite{Palecek}, \cite{DiMilla}, \cite{Wolf_Friedl.13}, we take into account that there is an optimal ECM density or pore cross section for cell migration. In fact, for lower densities cell motion results hampered for the lack of adhesion sites that the cell can use to exert traction. 
On the other hand, for higher densities cell motion is hampered  as well by excessive adhesion and lack of space among the fibers. 
Actually, it is known that there is a maximum value of ECM density, or better a minimum value of pore area cross section, above which the fiber network is too tight to be penetrated (the so-called physical limit of migration \citep{Wolf_Friedl.13, Scianna_Preziosi.13, Giverso.18}. It is more controversial whether there is a minimum value of ECM density  denoted in the following by $M_0$ below which cells are unable to migrate because of the absence of ECM fibers on which to crawl. The existence of these  physical limits of migration will be examined in more detail in a future work. 

One can then take such experimentally determined functions as an indication of the dependence of the 
mean  $\bar v$ of the distribution function $\psi$ from the density $M$ of extracellular matrix.
In the following simulation, to mimick such a behaviour we assume that 
\begin{equation}\label{barvM}
\bar{v}(M)=v_{Max} \dfrac{M}{M_{Max}}\exp\left[\dfrac{M_{max}-M}{M_{max}}\right]\,,
\end{equation}
and choose $\psi$ to be 
$$
\psi(v)=\dfrac{4v}{\bar v^2}e^{-2v/\bar v}\,.
$$

\begin{figure}[!htbp]

\centering
\subfigure[]{\includegraphics[scale=0.3]{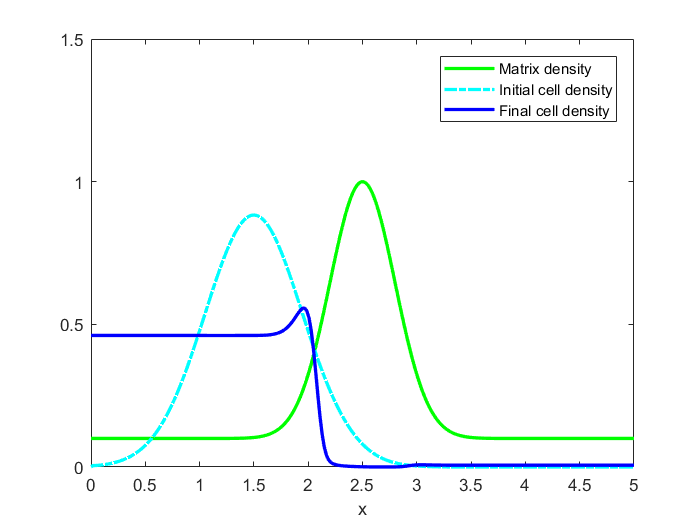}}
\subfigure[]{\includegraphics[scale=0.3]{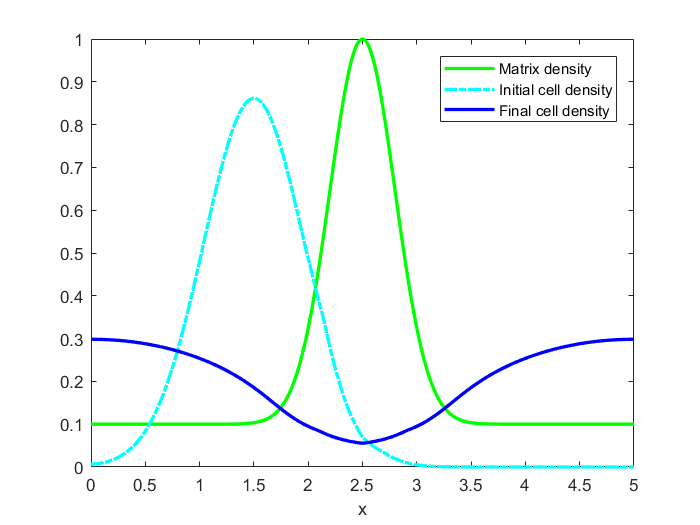}}

\subfigure[]{\includegraphics[scale=0.3]{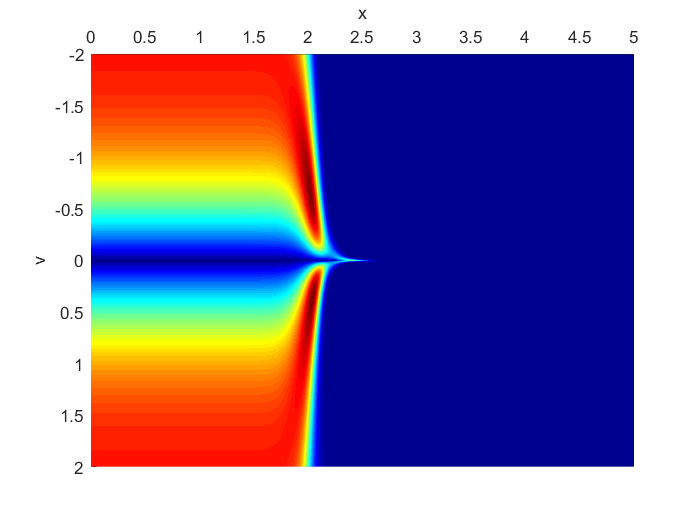}}
\subfigure[]{\includegraphics[scale=0.3]{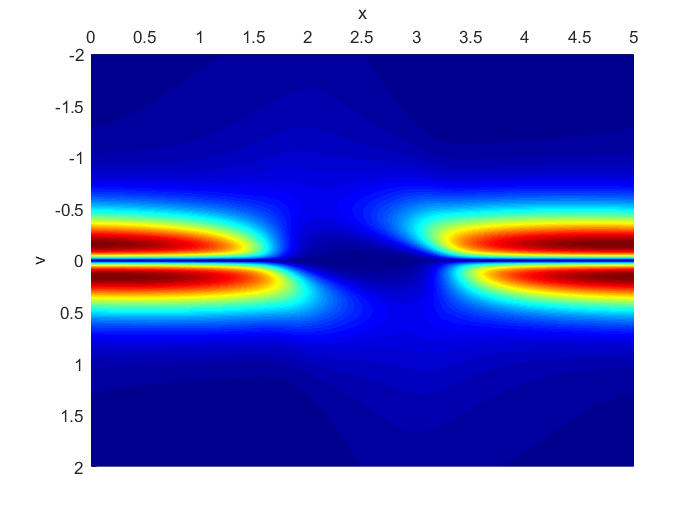}}
\caption{Initial condition (cyan line) and  density at $t=200$ (full line) for $M_{max}=1/10$ (a) and $M_{max}=1/0.3$ (b) in the presence of an heterogeneous distribution of ECM. In both cases $R_{\scriptscriptstyle M}=0.04$ and $\gamma_{\mathcal{S}}$ is a Heavyside function. The corresponding final distribution functions at $t=200$ are given in (c) and (d).}
\label{matrix}
\end{figure}

We consider an initial condition which is uniformly distributed in the velocity space and with macroscopic density a gaussian centered in $x=1.5$ and the heterogeneous distribution of ECM density ranging in $[M_0,M_1]=[0.1,1]$ shown in Fig. \ref{matrix} (a) and (b)(green line). 
We take here $\gamma_{\scriptscriptstyle M}(\lambda')=H_M (R_{\scriptscriptstyle M}-\lambda')$ as sensing function, meaning that the speed is determined by uniformly averaging the density of ECM in the direction $\hat\vb$ looking ahead up to a distance $R_{\scriptscriptstyle M}$ that in the simulations is $0.04$.

If a matrix density distributed as in Fig. \ref{matrix} (a) and (b) is present, cells change their speed accordingly. In particular, if $M_{max}<M_0=0.1$ (as in the simulations reported in Fig. \ref{matrix} (a)) cell mean speed is on the decreasing branch of \eqref{barvM}, i.e., always smaller than $v_{max}\frac{M_0}{M_{max}} \exp\left[-(M_0-M_{max})/M_{max}\right]$. Hence, they are strongly slowed down by the ECM. On the other hand,  if $M_{max}>M_1=1$ (as in the simulations reported in Fig. \ref{matrix} (b)) cell mean speed is on the increasing branch
of \eqref{barvM}, i.e., always larger than the value given above.
So,  they progressively invade the whole spatial domain and crawl across of the ECM.

\begin{figure}[!htbp]

\begin{center}
\subfigure[]{\includegraphics[scale=0.26]{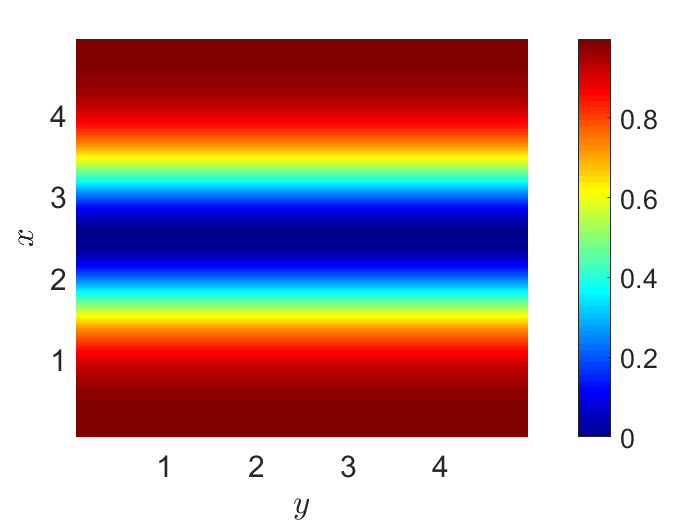}}
\subfigure[$t=0$]{\includegraphics[scale=0.26]{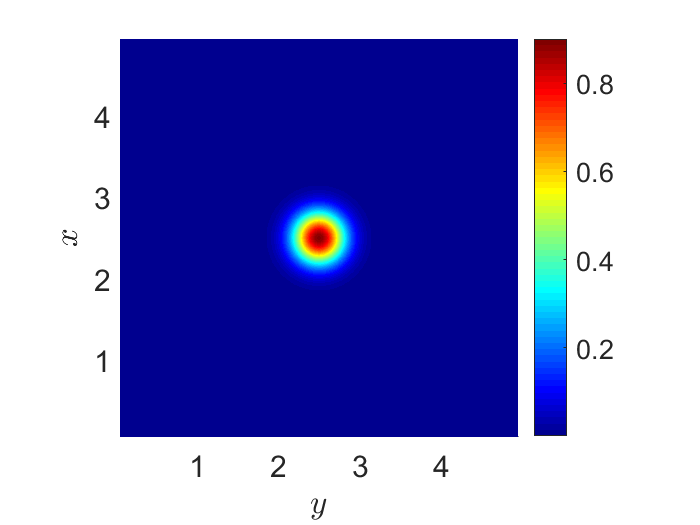}}
\subfigure[$t=0.5$]{\includegraphics[scale=0.26]{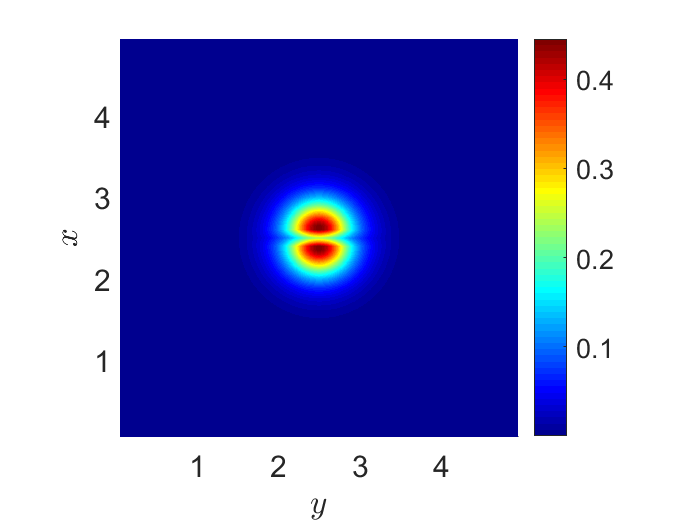}}
\end{center}

\begin{center}
\subfigure[$t=1$]{\includegraphics[scale=0.258]{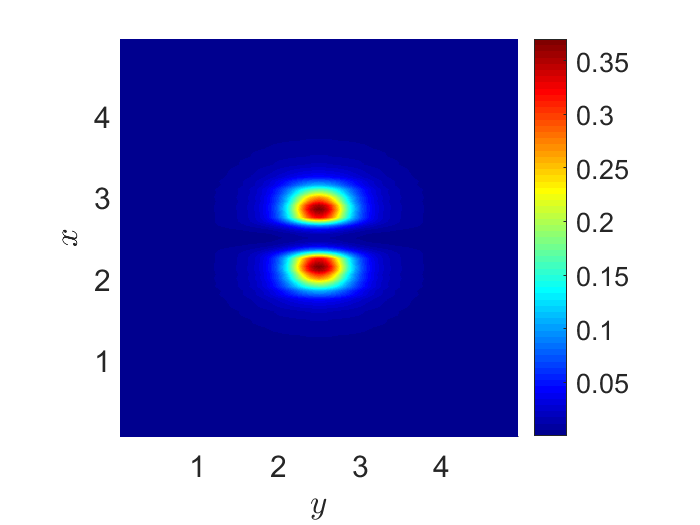}}
\subfigure[$t=10$]{\includegraphics[scale=0.258]{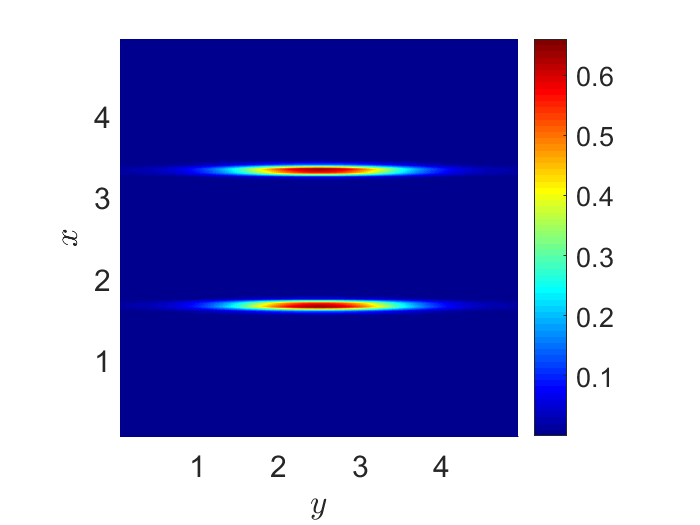}}
\subfigure[$t=60$]{\includegraphics[scale=0.258]{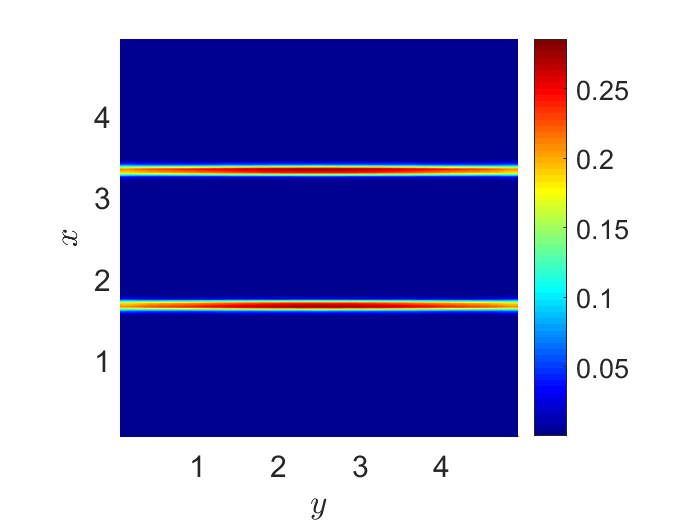}}
\end{center}
\caption{Matrix density, with $M_{max}=1$ (a). The sensing radius is $R_{\scriptscriptstyle M}=0.25$ and $\gamma_{\mathcal{S}}$ is a Dirac delta. The time evolution of the two dimensional macroscopic density is given in (b)-(f).}
\label{matrix2d}
\end{figure}

In two dimensions, if a matrix density is distributed as in Fig. \ref{matrix2d} (a), we observe that cells tend to go where the value of the matrix density is  $0.5*M_{max}$, $\ie$ where the speed is the largest. Upon reaching this zone with maximum mean speed (see Fig.\ref{matrix2d}(e)) diffusion looks anisotropic because cells tend to follow the region with optimal ECM density staying away from the regions where matrix density gets closer to $M_{max}$ where motility is hampered.

\subsubsection{ECM steryc hindrance and non-local chemotaxis}\label{doublechemo}
We shall now consider a real double bias, which means that there is a non-local sensing of the field $\mathcal{S}$ for the directional bias, $\ie$, $\gamma_{\scriptscriptstyle \mathcal{S}}(\lambda)=\delta(\lambda-R_{\scriptscriptstyle\mathcal{S}})$ or $\gamma_{\scriptscriptstyle \mathcal{S}}(\lambda)=H(R_{\scriptscriptstyle\mathcal{S}}-\lambda)$. In addition we take $\mathcal{T}_{\lambda}^{\hat{\vb}}[\mathcal{S}](\x)=b\big( \mathcal{S}(\x+\lambda\hat{\vb})\big)$. Then the turning operator reads
\begin{equation}\label{nonlocalchemo}
\begin{array}{lr}
\mathcal{J}[p](t,\x,\vb_p)=\\[8pt]
\mu (\x) \Big(\dfrac{c(\x)}{\Gamma_0'}\Psi(\x,v|\mathcal{S}',\hat{\vb}) \displaystyle \int_{\mathbb{R}_+} b(\mathcal{S}(\x+\lambda \hat{\vb})) \gamma_{\scriptscriptstyle \mathcal{S}} (\lambda) d\lambda - p(t,\x,\vb_p) \Big)
\end{array}
\end{equation}
As (\ref{even}) is hardly satisfied, the macroscopic limit is advective
\begin{equation}\label{nonlocalchemodiff}
\dfrac{\partial \rho}{\partial \tau}+\nabla \cdot \left[ \rho  \int_{\mathbb{S}^{d-1}}  \left( \bar{U}_{\scriptscriptstyle\mathcal{S'}}^0(\boldsymbol{\xi}|\hat{\vb}) \int_{\mathbb{R}_+} b(\mathcal{S}(\boldsymbol{\xi}+\lambda \hat{\vb}))\gamma_{\scriptscriptstyle \mathcal{S}} (\lambda)  \, d\lambda \right) \, \hat{\vb} \, d\hat{\vb}\right]=0 \, .
\end{equation}
On the other hand, if $R_{\scriptscriptstyle\mathcal{S}}$ can be considered small, the turning operator reads
$$
\mathcal{J}[p](t,\x,\vb_p)=
\mu (\x) \left[\dfrac{1}{\displaystyle |\mathbb{S}^{d-1}|\Gamma_0' }\Psi[\mathcal{S}'](\x,v|\hat{\vb}) \left( 1+ \Lambda
\frac{b'}{b}\nabla \mathcal{S}\cdot \hat{\vb}\right) - p(t,\x,\vb_p) \right]\,.
$$
In this case
$$
T[\mathcal{S},\mathcal{S}']_0=\dfrac{1}{\displaystyle |\mathbb{S}^{d-1}|\Gamma_0' }\Psi[\mathcal{S}']_0
$$
and
$$
T[\mathcal{S},\mathcal{S}']_1=\dfrac{1}{\displaystyle |\mathbb{S}^{d-1}|\Gamma_0'}\left(\Psi[\mathcal{S}']_0\Lambda
\frac{b'}{b}\nabla \mathcal{S}\cdot \hat{\vb} +\Psi[\mathcal{S}']_1 \right)
$$
Condition (\ref{even}) is satisfied if $\Psi[\mathcal{S}']_0$ is even as a function of $\hat{\vb}$ for all $\x$ and the diffusive limit is (\ref{macro.diff}) with advective velocity
$$
{\bf U}_{\scriptscriptstyle\mathcal{S},\mathcal{S}'}^1(\x)=\dfrac{1}{\displaystyle |\mathbb{S}^{d-1}|\Gamma_0'}\left(\int_{\mathbb{S}^{d-1}}
\bar{U}_{\scriptscriptstyle \mathcal{S}'}^1\hat{\vb} \, d\hat{\vb}+\Lambda\frac{b'}{b} \mathbb{T}\nabla \mathcal{S} \right),
$$
being
\begin{equation}
\mathbb{T}=\int_{\mathbb{S}^{d-1}}\bar U_{\scriptscriptstyle \mathcal{S}'}^0\hat{\vb}\otimes
\hat{\vb} \, d\hat{\vb}.
\end{equation}
It may be anisotropic as, in general, the mean speed is anisotropic.

\noindent If $\Psi[\mathcal{S}']_0$ is not even as a function of $\hat{\vb}$, we must perform a hyperbolic limit that gives
\begin{equation}
\dfrac{\partial \rho}{\partial \tau}+\nabla \cdot \big( \rho \Ub_{\scriptscriptstyle\mathcal{S},\mathcal{S'}}^0 \big)=0 \quad {\rm with}\quad
\Ub_{\scriptscriptstyle\mathcal{S},\mathcal{S'}}^0=\int_{\mathbb{S}^{d-1}} \bar{U}_{\scriptscriptstyle\mathcal{S'}}^0(\x|\hat{\vb}) \hat{\vb} \, d\hat{\vb}. 
\end{equation} 
This shows that even if the directional sensing gives rise to an even distribution function of order zero in $\hat{\vb}$ and, then, to a diffusive behavior, the speed sensing may change this tendency and we may have a drift driven dynamics.

As an application we shall now consider the case of cell migration in the extra-cellular matrix under the action of a chemoattractant. 
We then integrate the transport equation (\ref{transport.general}) with the operator \eqref{nonlocalchemo} with $b(\mathcal{S})=\mathcal{S}$, the mean $\bar{v}$ given by \eqref{barvM}. We take here the same sensing function as before, $\ie,$ $\gamma_{\scriptscriptstyle M}(\lambda')=H_M (R_{\scriptscriptstyle M}-\lambda')$. Also $\gamma_{\scriptscriptstyle \mathcal{S}}(\lambda)=H (R_{\scriptscriptstyle \mathcal{S}}-\lambda)$. 

\begin{figure}[htbp]
\begin{center}
\subfigure[]{\includegraphics[scale=0.3]{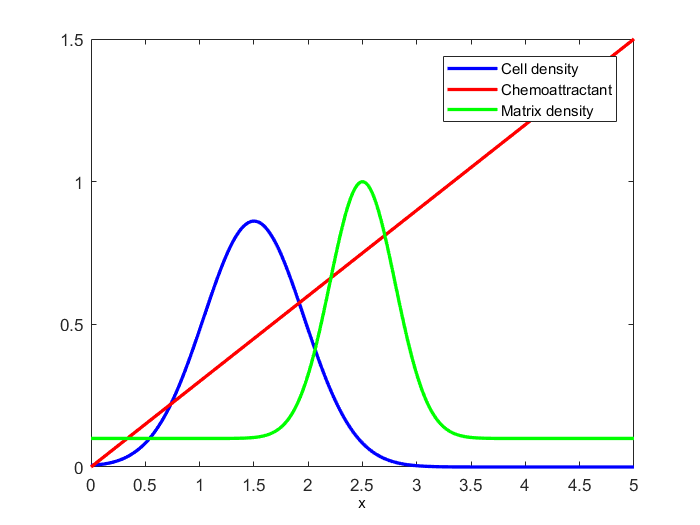}}
\subfigure[]{\includegraphics[scale=0.4]{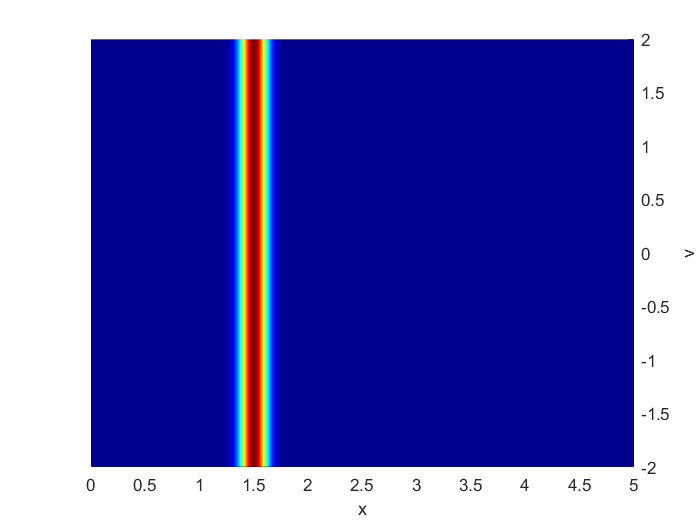}}
\end{center}
\subfigure[$t=5$]{\includegraphics[scale=0.2]{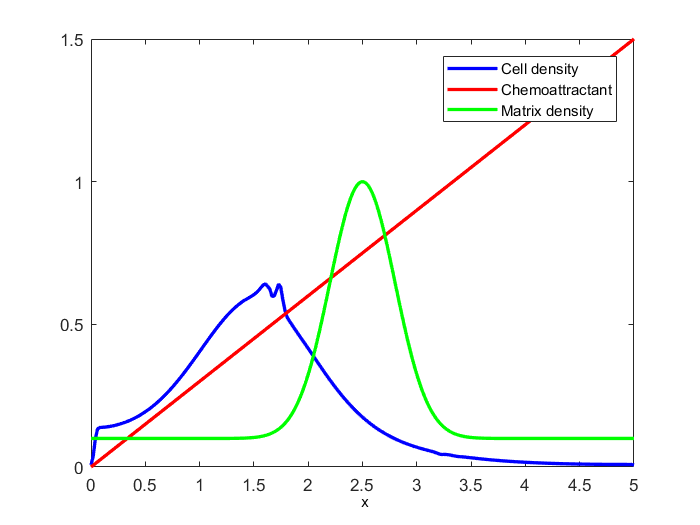}}
\subfigure[$t=15$]{\includegraphics[scale=0.2]{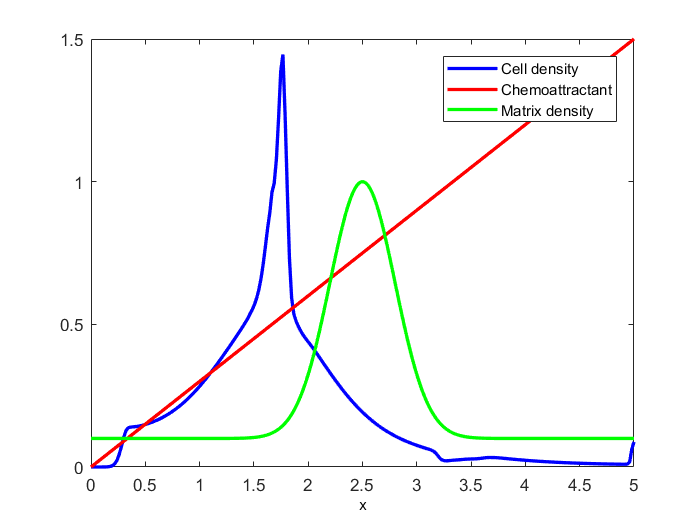}}
\subfigure[$t=150$]{\includegraphics[scale=0.2]{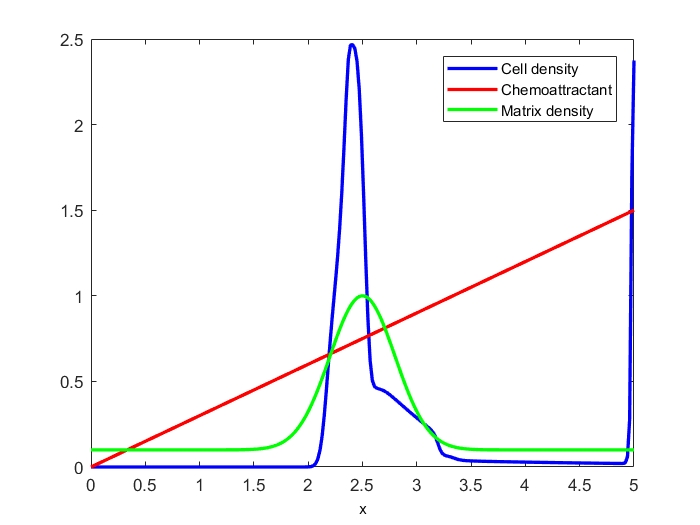}}

\subfigure[$t=5$]{\includegraphics[scale=0.2]{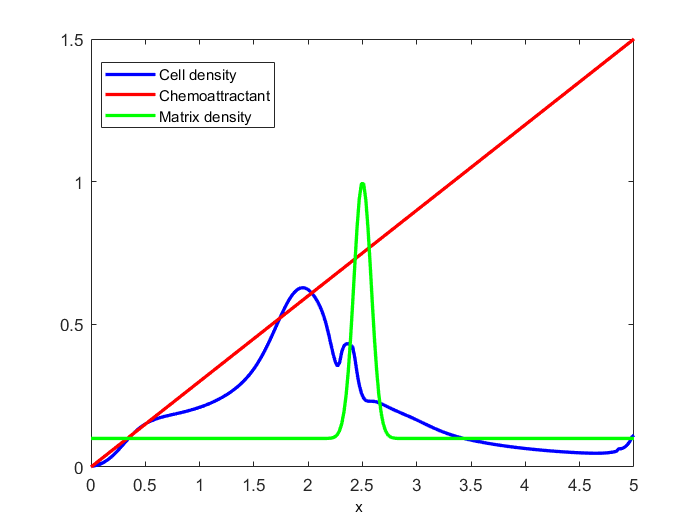}}
\subfigure[$t=15$]{\includegraphics[scale=0.2]{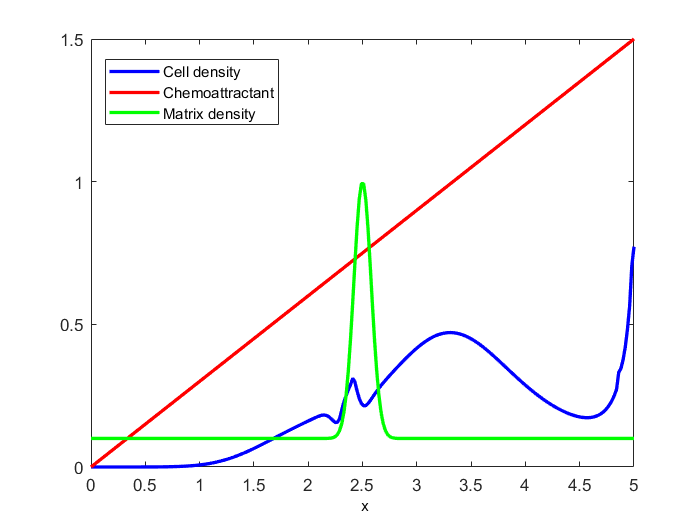}}
\subfigure[$t=200$]{\includegraphics[scale=0.2]{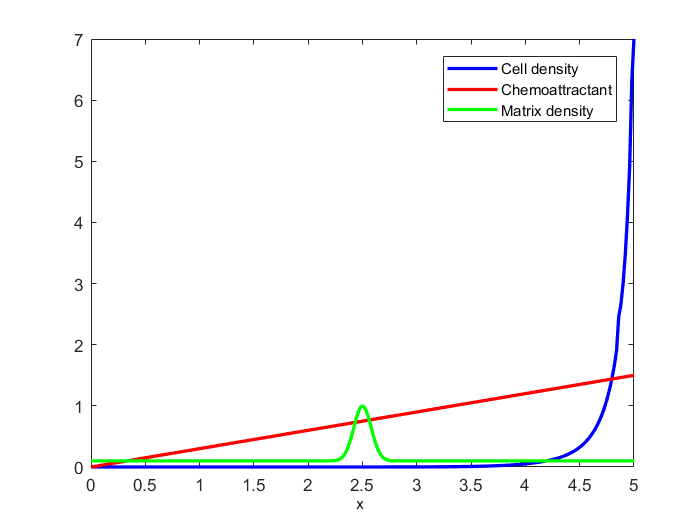}}

\subfigure[]{\includegraphics[scale=0.265]{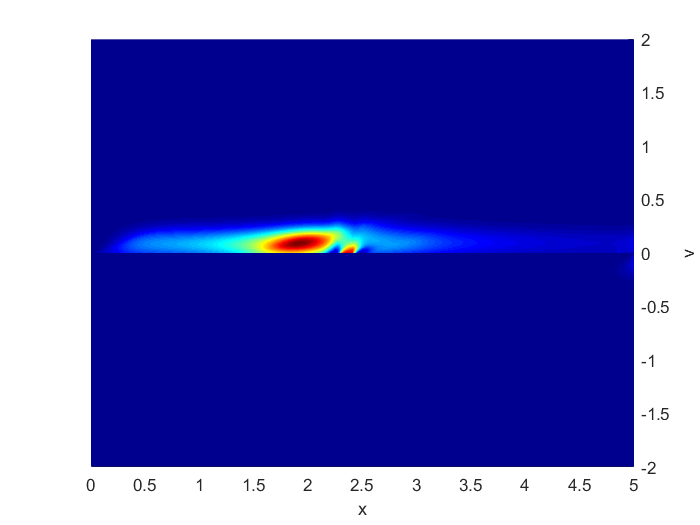}}
\subfigure[]{\includegraphics[scale=0.265]{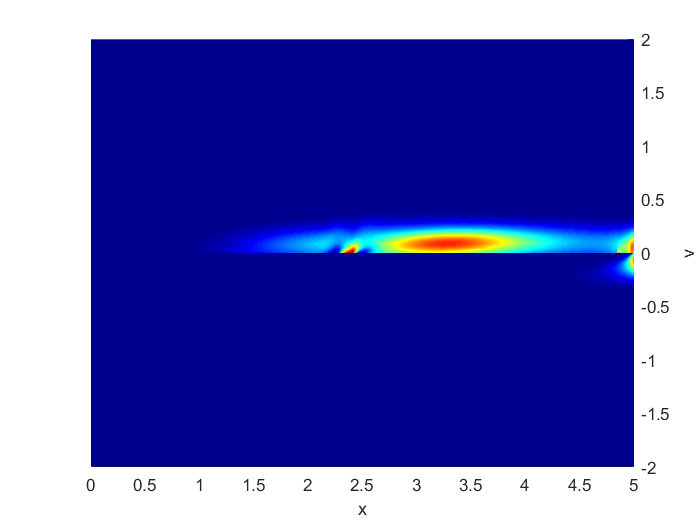}}
\subfigure[]{\includegraphics[scale=0.265]{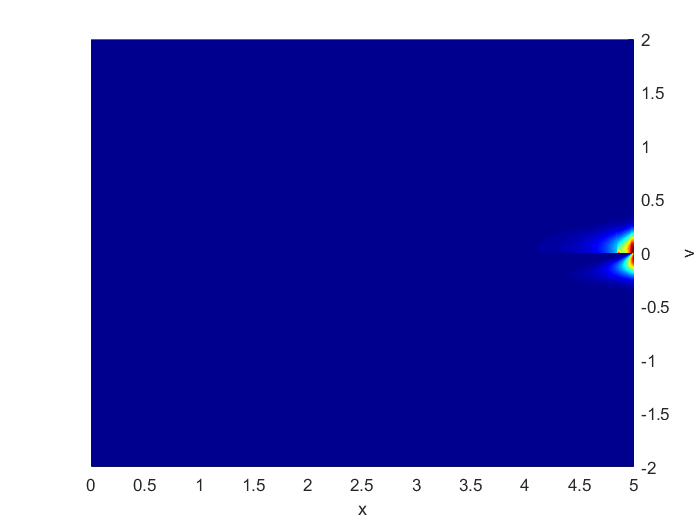}}

\caption{Evolution of the macroscopic cell density in response to a chemoattractant and an heterogeneous distribution of extracellular matrix. (a): initial macroscopic density; (b): distribution function. (c,d,e): $M_{max}=1/10, R_{\scriptscriptstyle \mathcal{S}}=1$. (f,g,h): $M_{max}=1/0.3, R_{\scriptscriptstyle \mathcal{S}}=0.04$. (i,j,k): evolution of the distribution function corresponding respectively to (f,g,h).}
\label{mat3}
\end{figure}

In Figure \ref{mat3}, the initial gaussian distribution of cells, then, starts travelling to the right under the action of the linearly distibuted chemoattractant. However, at variance with what happend in Figure \ref{mono.simu.2}, when cells encounter the denser area of ECM they slow down. If $M_{max}$ is large ($M_{max}=1/0.3$), the mean speed is large and the cells are driven by the chemoattractant, even if the radius of the direction  sensing is small ($R_{\scriptscriptstyle \mathcal{S}}=0.04$). So, eventually the cells succeed in going through the ECM as shown in the third row of Figure 8. The evolution of the distribution function is also given in the bottom row. If $M_{max}$ is small ($M_{max}=1/10$), even if the influence of the chemoattractant is very strong ($R_{\scriptscriptstyle \mathcal{S}}=1$), cells are strongly hampered by the  ECM as shown in the second row of Figure 8.

\section{Conclusion}

The kinetic model developed in this paper is based on several observations, all related to a particular feature of the model. Specifically, 
\begin{enumerate}
\item in order to move cells react to external stimuli by polarizing and deciding their direction of motion, forming a  ``head'' and ``tail''. Then they try to move in that direction forming adhesion sites and polimerizing actin at the head and removing adhesion sites at the tail. It is therefore convenient to write the kinetic model with the orientation and speed, separately, as independent variables; 
\item Both orientation and speed depend on cell sensing of their environment over a finite radius related, for instance, to the length of their protrusion. This gives the kinetic model a non-local character;
\item Orientation and speed can be related to different chemical and mechanical cues, with the former influencing mainly orientation and the latter mainly speed (though this is not always the case). This induces the inclusion of different cues determining the changes of orientation and speed in the turning operator;
\item While chemical cues are measured by the cells extending their protrusions, many mechanical cues are related to the fact that the nucleus, which represents the stiffest organelle of the cell, has difficulty in passing through narrow pores in the ECM or through densely packed cells. So, the sensing radii involved in the identification of the orientation and of the speed might be different  with the former being larger than the latter.
\end{enumerate}

A particular attention was devoted to the identification of the proper macroscopic limit according to the properties of the turning operator. However, all simulations integrate the kinetic model. In some cases, the solution is compared with the solution of the related macroscopic equations. The model was in fact applied to several sensing strategies, including when the signal is perceived locally, averaged over a region, or perceived  at the cell ``head'' and ``tail'' and then compared. Several applications were also considered as chemotaxis (or equivalently haptotaxis), durotaxis, cell-cell adhesion. The effect of the presence of an heterogeneous ECM was also considered in absence of any cue affecting cell orientation, then yielding a random polarization, and in presence of a chemical cue.

We have left to a coming paper the case in which the density of cell or of ECM is so large that they represent physical limits for migration. This introduces a dependence of the sensing radius determining the speed that depends on the density of cells or of ECM. In fact, for instance, if the characteristic pore size of the ECM is very small, then cell can protrude its cytoskeleton across the dense ECM  and scout for chemical signals and polarize, but the cell cannot penetrate the ECM because the nucleus is stuck. This holds true even in the case in which the layer of dense ECM is very thin, like in basal membrane or in cell lining, e.g., endothelial or mesothelial lining. The cell would have no difficulties in moving after passing through the membrane but cannot cross it. One should then address the question of deducing a kinetic model capable of dealing with such situations that are of great interest because of their relationship with the spread of metastasis.

\break
\newpage
\section*{Aknowledgements}
This work was partially supported by Istituto Nazionale di Alta Matematica, Ministry of Education, Universities and Research, through the ‘‘MIUR grant Dipartimenti di Eccellenza 2018-2022’’ and Compagnia San Paolo that finances NL PhD scholarship.


\bibliography{references}

\end{document}